\documentclass[a4paper,11pt]{article}
\pdfoutput=1        

\usepackage{ulem}
\usepackage{jheppub}
\usepackage[T1]{fontenc}
\usepackage{hyphenat}
\usepackage{multicol}
\usepackage{xcolor}
\usepackage{tcolorbox}
\usepackage{slashed}
\usepackage{float}
\usepackage{bbold}
\usepackage{bm}
\usepackage[compat=1.1.0]{tikz-feynman}
\usepackage{tikz,pgfplots,pgfplotstable}
\graphicspath{ {figures/} }

\let\emph\textit

\title{The Genuine Type-V Seesaw Model: Phenomenological Introduction}
\author{Saiyad Ashanujjaman\href{https://orcid.org/0000-0001-5643-2652}{\includegraphics[scale=0.4]{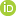}}}
\author{and Kirtiman Ghosh}
\affiliation[]{\footnotesize Institute of Physics, Bhubaneswar, Sachivalaya Marg, Sainik School Post, Bhubaneswar 751005, India}
\affiliation[]{\footnotesize Homi Bhabha National Institute, Training School Complex, Anushakti Nagar, Mumbai 400085, India}
\emailAdd{saiyad.a@iopb.res.in}
\emailAdd{kirti.gh@gmail.com}

\abstract{We study a model which generates Majorana neutrino masses at tree-level via low-energy effective operator with mass-dimension-9. Introduction of such a higher dimensional operator brings down the lepton number violating mass scale to TeV making such model potentially testable at present or near future colliders. This model possesses several new $SU(2)_L$ fermionic multiplets, in particular, three generations of triplets, quadruplets and quintuplets, and thus a rich phenomenology at the LHC. As the lepton flavour violation arises very naturally in such setup, we put constraints on the Yukawa couplings and heavy fermion masses from the current experimental bounds on lepton flavour violating processes. We also obtain 95\% CL lower bounds on the masses of the triplets, quadruplets and quintuplets using a recent CMS search for multilepton final states with 137 inverse femtobarn integrated luminosity data at 13 TeV center of mass energy. The possibility that the heavy fermions could be long-lived leaving disappearing charge track signatures or displaced vertex at the future colliders like LHeC, FCC-he, MATHUSLA, etc. is also discussed.}

\keywords{Lepton Flavour Violation, Generalised Type-III Seesaw Mechanism, Quintuplet / Type-V Seesaw Model, Multilepton final states search}

\begin{document} 

\preprint{
\begin{flushright} 
IP/BBSR/2020-10
\end{flushright} 
}

\maketitle

\flushbottom

\section{Introduction}
\label{sec:intro}
The Standard Model (SM) of particle physics, based upon gauge and Lorentz symmetries, has been surpassingly successful in predicting a wide range of crucial phenomena with eloquent experimental confirmation of most of its theoretical predictions at a persuasive level of accuracy and precision --- discovery of the Higgs boson at the Large Hadron Collider (LHC), which attests the spontaneous symmetry breaking mechanism advocated by Brout, Englert, and Higgs (BEH), being the latest entry in that list. The SM at present provides our best fundamental understanding of particle physics phenomenology. Notwithstanding its mighty success, SM falls short at offering explanations about quite a few theoretical arguments as well as some experimental observations. Theoretical arguments including some philosophical and aesthetic displeasures are fermion mass and mixing hierarchies, the electroweak vacuum instability, the strong CP problem, naturalness problem, {\it etc.} Experimental inconsistencies include issues like neutrino masses and mixings, the anomalous magnetic moment of electron and muon, baryon asymmetry and dark matter. These inadequacies cry out for new physics beyond the Standard Model (BSM) scenarios. In this work, we study a particular BSM scenario which only addresses the issue of non-zero neutrino masses and mixings.

The neutrino oscillation experiments \cite{nu_osci_expt_1,nu_osci_expt_2,nu_osci_expt_3,nu_osci_expt_4,nu_osci_expt_5,nu_osci_expt_6} performed over the last few decades provided imperative evidences for neutrino flavour oscillation, and hence nonzero neutrino masses and mixings \cite{nu_osci_param}. A Dirac mass term for the SM neutrinos can be constructed by extending the SM by right-handed neutrinos via the usual BEH mechanism. However, this requires very tiny Yukawa couplings causing a hierarchy among the Yukawa couplings. This hierarchy persuades philosophical and aesthetic displeasure making this scheme unappealing. The tiny neutrino masses also can be accommodated by introducing new degrees of freedom to the SM, preferably at a high scale which, after being integrated out, lead to high-dimensional\footnote{Whenever the word `dimension' is used, we mean `mass-dimension'.} gauge-invariant and non-renormalisable operator(s) involving the SM fields only. These operators violate lepton number by two units $(\Delta L =2)$ in the effective theory of the Standard Model extension (SME). After the electroweak symmetry is broken, the Majorana neutrino masses are generated. The Majorana masses are suppressed by the lepton number violating (LNV) mass scale. The large LNV mass scale involved in such a scenario explains the smallness of the observed neutrino masses readily. This makes such a scenario not only plausible but also very natural, and hence an appealing one. Such lowest dimensional operator is popularly known as Weinberg operator \cite{seesaw_gut_2} and the corresponding mechanism of neutrino mass generation is known as seesaw mechanism:
\begin{equation*}
\mathcal{O}^{d=5} \sim \frac{1}{\Lambda}LLHH~,
\end{equation*}
where $\Lambda$ is the LNV mass scale. In minimal extensions of the SM, there are exactly three UV completions corresponding to the Weinberg operator at tree level \cite{seesaw_gut_5}. These are popularly known as type-I \cite{seesaw1_1,seesaw1_2,seesaw1_3,seesaw1_4,seesaw1_5}, type-II \cite{seesaw2_1,seesaw2_2,seesaw2_3,seesaw2_4,seesaw2_5,seesaw2_6} and type-III \cite{seesaw3_1} seesaw. Type-I and type-III seesaw employ $\rm{SU(2)_L}$ singlet and triplet right-handed neutrinos, respectively, whereas type-II employs $\rm{SU(2)_L}$ scalar triplet. These have been discussed very widely in the literature. Also, type-I and type-III seesaw have been discussed in different grand unified theories (GUT) framework \cite{seesaw_gut_1,seesaw_gut_2,seesaw_gut_3,seesaw_gut_4,seesaw_gut_5,seesaw_gut_6,seesaw_gut_7,seesaw_gut_8}.

The most general gauge-invariant and non-renormalisable neutrino mass operator with $\Delta L=2$ of dimension $d$ ($d \geq 5$), which can be constructed using the SM fields, can be written as \cite{high_dim_1,high_dim_2}
\begin{equation}
\mathcal{O}^d \sim \frac{1}{\Lambda^{d-4}} LLHH\left(H^\dagger H\right)^{(d-5)/2}~.
\label{eq:eff_operator_d}
\end{equation}
While writing down the above operator, flavor, spinor and gauge indices are kept implicit for brevity. The above operator can be realized both at the tree and loop level. In case the neutrino masses are generated radiatively, each loop integrals will yield a suppression of $1/16\pi^2$. The general form of Majorana neutrino mass generated via the above operator is given by \cite{high_dim_1,high_dim_2,high_dim_3,high_dim_4,high_dim_5,high_dim_6}
\begin{equation*}
m_\nu \sim \epsilon \times \left(\frac{1}{16\pi^2} \right)^n \times \left(\frac{v}{\sqrt{2}\Lambda}\right)^{d-5} \times \frac{v^2}{2\Lambda}~,
\end{equation*}
where $v$ is the SM vacuum expectation value (VEV), $n$ is the number of loops at which neutrino masses are induced, and $\epsilon$ denotes additional LNV suppression which could arise in particular models, for instance, in inverse seesaw scenarios, R parity-violating SUSY models, {\it etc}. The neutrino masses, $m_\nu$, scale as $\frac{v^2}{2\Lambda}$ if they are generated via the so-called Weinberg operator, $\mathcal{O}^{d=5}$, at tree level; whereas they are further suppressed by the factor $\left(v/\sqrt{2}\Lambda\right)^{d-5}$ if they are generated via $d>5$ dimensional operator at tree level. In order to explain the observed sub-eV neutrino masses, for classical seesaw, Yukawa couplings of order $\mathcal{O}(1)$ require $\Lambda$ to be at GUT scale; whereas for $\Lambda$ to be at TeV scale, one has to assume Yukawa couplings of order $\mathcal{O}(10^{-6})$. At this point, it is worth mentioning that although the fields which mediate neutrino masses in classical seesaw are naturally motivated to have very high scale masses, nothing precludes them to have masses at the TeV scale. Assuming the absence of any additional suppression or loop suppression, if one expects Yukawa couplings to be large enough $\left[\mathcal{O}(1)-\mathcal{O}(10^{-1})\right]$ as well as $\Lambda$ to be at TeV scale, one has to consider the scenarios where neutrino masses are generated via $d\geq 9$ or higher dimensional operator at tree level. The LNV scale or new physics scale involved in such a scenario is very naturally connected to the physics at the TeV scale making the same potentially testable, and hence falsifiable at the LHC. A model is said to be `genuine' at a given dimension $d$, if all the lower dimensional contributions to neutrino masses are automatically absent without the need for any additional $U(1)$ or $\mathbb{Z}_n$ symmetry. The particle contents of a genuine model is such that it does not allow UV completions of any lower-dimensional operator. Note that the very same $d$-dimensional operator in Eq.~\eqref{eq:eff_operator_d} always leads to lower-dimensional loop models. For the $d$-dimensional contribution to the neutrino masses at tree level to dominate over that of the $(d-2)$-dimensional one at 1-loop, one requires $\Lambda \lesssim 2$ TeV\footnote{See Ref.~\cite{high_dim_6} for detail discussion.}.

In this paper, we study a genuine model at $d=9$ which was introduced for the first time in Ref.~\cite{high_dim_6}. Ref.~\cite{high_dim_7} includes a short discussion on the phenomenology of long-lived particle for this model. This model contains, in addition to the SM fields, many exotic fermionic fields transforming as $3,4$ and $5$ representations of $SU(2)_L$ --- vector-like triplet(s) with hypercharge 1, vector-like quadruplet(s) with hypercharge $1/2$ and chiral quintuplet(s) with hypercharge 0. If one follows the convention of labeling a seesaw model with the size of the representation of the fermionic field (or more precisely, the fermionic field which yields lepton number violation) it employs, {\it i.e.}, if one labels a seesaw model employing a fermion transforming as $n$-representation of $SU(2)_L$ as type-$n$ or $n$-plet seesaw model, then this model would be called a Type-V or quintuplet seesaw model\footnote{Note that, in the literature, there are a few so-called Type-V seesaw models, {\it e.g.}, the model studied in Ref.~\cite{kumericki,picek} which employs quintuplet fermions with zero hypercharge or the model studied in Ref.~\cite{kumericki1} which employs quintuplet fermions with non-zero hypercharge, but none of these models are `genuine'.}. In the framework of this model, the lepton flavour violation arises very naturally. Hence, the model parameters will be constrained by the current experimental bounds from the lepton flavour violating (LFV) decays. On the other hand, the explanation of the neutrino oscillation data ({\it i.e.,} the tiny neutrino masses and mixings) requires the exotic fermions to be at TeV scale. Owing to the non-trivial representations of the exotics under the $SU(2)_L$ gauge group, they can be pair produced copiously at the LHC, and their subsequent decays will lead to several interesting signatures at the LHC. In this article, we provide a phenomenological introduction of the `genuine' Type-V seesaw model and obtain bounds from the LFV processes as well as collider experiments.

In the next section (section \ref{sec:model}), we introduce the model and present the relevant Lagrangian for neutrino mass generation. The observed neutrino masses and mixings can dexterously be accommodated in this model as discussed in section \ref{sec:nu_mass_mixing}. In section \ref{sec:lfv}, calculation for the decay rates of lepton flavour violating magnetic transitions $\mu \to e\gamma$ and $\tau \to \ell \gamma$ ($\ell=e,\mu$) are presented. We also put constraints on the Yukawa couplings and heavy fermion masses using the current experimental bounds on LFV processes. Production and subsequent decays of the exotic fermions are discussed in section \ref{sec:production} and \ref{sec:decay} respectively. In section \ref{sec:collider_signals}, multilepton final states at the LHC are discussed. $95\%$ CL upper limits on the total production cross sections of heavy fermionic pairs are estimated in section \ref{sec:multilepton} using a recent search based on multilepton final states by CMS \cite{cms_multilepton_137}. In section \ref{sec:displaced}, we discuss the possibility that the heavy fermions can also be long-lived leaving disappearing track signatures or displaced vertex inside detector. We summarise the discussion in section \ref{sec:conclusion}. Mass matrices for differently charged leptons and their diagonalisation are presented in Appendix \ref{app:mass_matrices}. Relevant gauge interactions for both production and decay of the exotics are presented in Appendix \ref{app:gauge_Lag}.

\section{Model}
\label{sec:model}
Keeping the symmetry of the SM gauge group $SU(3)_C \times SU(2)_L \times U(1)_Y$ unaltered, in the model to be presented here, the fermionic sector of the SM is extended by vector-like $SU(2)_L$ triplets ($\Sigma_{L,R}$), vector-like $SU(2)_L$ quadruplets ($\Delta_{L,R}$) and chiral $SU(2)_L$ quintuplets ($\Phi_R$). Note that one generation of these exotic fermion multiplets are not sufficient to explain more than one non-zero neutrino masses, and, hence, we include three copies of each\footnote{Note that one copy of vector-like quadruplet, one copy of chiral quintuplet (vector-like triplet) and three copies of vector-like triplets (chiral quintuplet) will suffice to generate two non-vanishing neutrino masses.}. The particle contents along with their quantum numbers under the SM gauge group are shown in Table~\ref{table:model_contents}.
\begin{table}[h]
\scalebox{0.9}{
\begin{tabular}{|c|}
\hline 
\hline
\textbf{Gauge group: $SU(3)_{C} \times SU(2)_{L} \times U(1)_{Y}$}\\  
\hline
\hline 
\\
\begin{minipage}{0.35\textwidth}
\textbf{SM baryon fields}:\\ 
$Q_L={\begin{pmatrix} u \\ d \end{pmatrix}}_L\sim(3,2,\frac{1}{6})$
\\ \\
$u_R\sim (3,1,\frac{2}{3})$
\\ \\
$d_R \sim (3,1,-\frac{1}{3})$
\end{minipage}
\begin{minipage}{0.35\textwidth}
\textbf{SM lepton fields}:\\ 
$L={\begin{pmatrix} \nu_\ell \\ \ell \end{pmatrix}}_L\sim (1,2,-\frac{1}{2})$
\\ \\
$\ell_R\sim (1,1,-1)$
\\ \\ 
\end{minipage}
\begin{minipage}{.27\textwidth}
\textbf{SM scalar field}:\\ 
$H={\begin{pmatrix} H^{+} \\ H^{0} \end{pmatrix}}\sim(1,2,\frac{1}{2})$ 
\\
with $H^0=\frac{1}{\sqrt{2}}(v+h+i\eta)$
\\ \\ \\
\end{minipage}
\\
\\
\hline
\\
\textbf{Exotic fermionic fields:}\\
\begin{minipage}{.37\textwidth}
$\Sigma_{L,R}={\begin{pmatrix} \Sigma^{++} \\ \Sigma^{+} \\ \Sigma^0 \end{pmatrix}}_{L,R}\sim (1,3,1),~~~~$
\end{minipage}%
\begin{minipage}{.375\textwidth}
$\Delta_{L,R}={\begin{pmatrix} \Delta^{++}\\ \Delta^+ \\ \Delta^0 \\ \Delta^- \end{pmatrix}}_{L,R}\sim(1,4,\frac{1}{2}),~~~$
\end{minipage}%
\begin{minipage}{.3\textwidth}
$\Phi_R ={\begin{pmatrix} \Phi^{++}\\ \Phi^+ \\ \Phi^0 \\ \Phi^- \\ \Phi^{--} \end{pmatrix}}_R\sim(1,5,0)$
\end{minipage}
\\
\\
\hline 
\hline
\end{tabular}
}
\caption{\label{table:model_contents} Particle contents of the model with their gauge quantum numbers. Superscripts of the fields denote the corresponding electric charges. We adopt the convention $Q = T_3 + Y$. The generation indices are suppressed for the sake of brevity, and will be suppressed throughout this article.}
\end{table}

The relevant parts of the gauge invariant and renormalisable Lagrangian relevant for the generation of tiny non-vanishing neutrino masses and mixings are given by
\begin{eqnarray}
\label{eq:yukawa_int}
-{\cal L}_{\rm Yukawa} &=& Y_\ell~ \overline{L}_i H^i \ell_R + Y_{23}~\big(\overline{\Sigma_L}\big)^i_j~H^j~\widetilde L_i ~+~ Y_{34}~\big(\overline{\Delta_R} \big)_{ijk}~(\Sigma_L)^i_{j^\prime}~H^*_{k^\prime}~\epsilon^{jj^\prime}~\epsilon^{kk^\prime} \nonumber
\\
&&~+~ Y^\prime_{34}~\big(\overline{\Delta_L} \big)_{ijk}~(\Sigma_R)^i_{j^\prime}~H^*_{k^\prime}~\epsilon^{jj^\prime}~\epsilon^{kk^\prime} ~+~ Y_{45}~\big(\overline{\Delta_L} \big)_{ijk}~\big(\Phi_R\big)^{ijk\ell}~H^{\ell^\prime}~\epsilon_{\ell \ell^\prime} \nonumber
\\
&&~+~ Y^\prime_{45}~\big(\overline{\Delta_R}\big)_{ijk}~\big(\widetilde \Phi_R \big)_{~i^\prime j^\prime k^\prime \ell}~H^{\ell}~\epsilon^{ii^\prime}~\epsilon^{jj^\prime}~\epsilon^{kk^\prime}~; \nonumber
\\
-{\cal L}_{\rm mass} &=& M_\Sigma~\big(\overline{\Sigma_R}\big)^i_j~\big(\Sigma_L\big)_i^j ~+~M_\Delta~\big(\overline{\Delta_R}\big)_{ijk}~\big(\Delta_L\big)^{ijk} \nonumber
\\
&& +~ \frac{M_\Phi}{2}~\left(\overline{\widetilde\Phi_R}\right)^{ijk\ell}~\big(\Phi_R\big)^{i^\prime j^\prime k^\prime \ell^\prime}~\epsilon_{ii^\prime}~\epsilon_{jj^\prime}~\epsilon_{kk^\prime}~\epsilon_{\ell \ell^\prime}~;
\end{eqnarray}
where $Y_{23} ,Y_{34}, Y_{34}^\prime, Y_{45} \mathrm{~and~} Y_{45}^\prime$ are Yukawa matrices (in general, complex), $M_\Sigma, M_\Delta$ and $M_\Phi$ are mass matrices for the heavy leptons, $\widetilde \chi$ denotes the charge-conjugation of $\chi$, {\it i.e.,} $\widetilde \chi~=~C(\bar \chi)^T$, where $C$ is the charge-conjugation matrix. The dummy $SU(2)$ indices ($i,j,k$, {\it e.t.c.}) are summed over 1 and 2. For convenience, the generation indices are kept implicit here. $\epsilon$ is the anti-symmetric permutation tensor of rank-2 with $\epsilon^{12}_{(12)}=+1$ and $\epsilon^{21}_{(21)}=-1$. $\Sigma$ is a rank-2 tensor, and $\Delta$ and $\Phi$ are, respectively, symmetric rank-3 and rank-4 tensors with the following components:
\begin{itemize}
\item $\Sigma^1_1=-\Sigma^2_2=\Sigma^+/\sqrt{2},~~\Sigma^1_2=\Sigma^{++},~~\Sigma^2_1=\Sigma^0$~;
\item $\Delta^{111}=\Delta^{++},~~\Delta^{112}=\Delta^+/\sqrt{3},~~\Delta^{122}=\Delta^0/\sqrt{3},~~\Delta^{222}=\Delta^-$~;
\item $\Phi^{1111}=\Phi^{++},~~\Phi^{1112}=\Phi^+/\sqrt{4},~~\Phi^{1122}=\Phi^0/\sqrt{6},~~\Phi^{1222}=\Phi^-/\sqrt{4},~~\Phi^{2222}=\Phi^{--} $~.
\end{itemize}
Being vector-like, one can write Dirac mass terms for $\Sigma$ and $\Delta$, whereas the gauge quantum numbers allow Majorana mass terms for $\Phi$ (see Eq.~\ref{eq:yukawa_int}). Therefore, the masses of these heavy leptons are not constricted by the electroweak symmetry breaking scale, and, hence, these masses can be tuned freely to the new physics scale. It is important to mention that the Dirac and Majorana mass matrices in Eq.~\ref{eq:yukawa_int} are, in general, $3\times 3$ complex matrices containing 18 real parameters each.  However, one has the freedom to choose a basis for these exotic leptons in which the mass matrices are diagonal with real positive elements. The Yukawa matrices in Eq.~\ref{eq:yukawa_int} are also $3\times 3$ complex matrices. However, all the 18 real parameters contained in each of those Yukawa matrices are not independent. For example, one can always choose a basis for the SM lepton doublets ($L$) and singlets ($l_R$) in which $Y_l$ is diagonal with real positive elements. In the basis where the SM charge lepton mass matrix and the exotic leptons mass matrices are diagonal, the Yukawa matrices contain few (in particular, 9-phases\footnote{For example, six phases of $Y_{23}$ may be absorbed in the definition of $\Sigma_{L}$-triplets and $L$-doublets. Phase redefinition of $\Delta_L$($\Delta_R$)-quadruplets can absorb three more phases of $Y_{34}^\prime$($Y_{34}$). In presence of the Majorana mass terms for the $\Phi_R$, the phase redefinition of the quintuplets are not possible without introducing additional phases in the Majorana mass matrix $M_\Phi$.}) unphysical phases which can be absorbed in the definition of the lepton multiplets.

After the EWSB, the Yukawa Lagrangian in Eq.~\ref{eq:yukawa_int} leads to mixing among the leptons (the SM leptons as well as the exotic ones). The parts of Lagrangian involving mass and mixing terms for the neutral (${\cal L}_0$), singly-charged (${\cal L}_1$) and doubly-charged (${\cal L}_2$) leptons are given by:
\begin{equation}
  -\mathcal{L}_0 =\overline{\widetilde\psi^{N}_L} M_N \psi^N_L,~~~ -\mathcal{L}_1 =\ \overline{\psi^E_R} M_E \psi^E_L~~{\rm and}~~-\mathcal{L}_2 =\ \overline{\psi^K_R} M_K \psi^K_L~,
  \label{lag_mass}
\end{equation}
where tilde stands for charge conjugation, and
\begin{eqnarray*}
  \psi^N_L ~=~ \left({\begin{array}{c}
      \nu_L \\
      \tilde{\Sigma}^0_R \\
      \Sigma^0_L \\
      \tilde{\Delta}^0_R \\
      \Delta^0_L \\
      \tilde{\Phi}^0_R \\
      \end{array} }\right)~,~~~
  \psi^E_{L(R)}~=~ \left({\begin{array}{c}
      \ell_{L(R)} \\
      \tilde{\Sigma}^+_{R(L)}\\
      \tilde{\Delta}^+_{R(L)}\\
      \Delta^-_{L(R)}\\
      \widetilde\Phi^{+}_{R}(\Phi^{-}_{R})
      \end{array} }\right)~,~~~
  \psi^K_{L(R)} ~=~\left({\begin{array}{c}
      \tilde{\Sigma}^{++}_{R(L)}\\
      \tilde{\Delta}^{++}_{R(L)}\\
      \tilde{\Phi}^{++}_{R}({\Phi}^{--}_{R}) \\
      \end{array} }\right)~,
\end{eqnarray*}
where the generation indices are kept implicit for brevity. The mass matrices $M_N,~M_E$ and $M_K$ are  $18 \times 18$, $15 \times 15$ and $9 \times 9$ matrices, respectively, and are presented in Appendix~\ref{app:mass_matrices} in $3\times 3$ block-matrix form. $M_N$, being a symmetric matrix, can be diagonalised by a unitary matrix $U_N$: $M_N^{\rm diag}~=~U_N^T M_N U_N~=~\mathrm{diag} (\hat{m}_\nu,-M_\Sigma,M_\Sigma,-M_\Delta,M_\Delta,M_\Phi)$\footnote{The corrections of the order of neutrino masses ($\sim {\cal O}\left(10^{-10}\right)$ GeV) are neglected for the TeV scale exotic leptons.}, whereas the diagonalisation of $M_E(M_{K})$, being arbitrary complex matrix, requires two unitary matrices ($U^L_{E(K)}$ and $U^R_{E(K)}$) for bi-unitary transformations:
\begin{eqnarray}
  M_{E}^{\rm diag}&=&\left(U^R_{E}\right)^\dagger\,\, M_{E}\,\,U^L_{E}~=~\mathrm{diag} (m_\ell,M_\Sigma,M_\Delta,M_\Delta,-M_\Phi),\nonumber\\
M_{K}^{\rm diag}&=&\left(U^R_{K}\right)^\dagger\,\, M_{K}\,\,U^L_{K}~=~\mathrm{diag} (M_\Sigma,M_\Delta,M_\Phi),\nonumber  
\end{eqnarray}
where $\hat m_\nu~=~{\rm diag}\left(m_1,m_2,m_3\right)$ with $m_1,~m_2~{\rm and}~m_3$ being the masses of the SM neutrinos, and $m_l~=~{\rm diag}\left(m_e,m_\mu,m_\tau\right)$ is the SM charged lepton mass matrix. After the diagonalisation of the mass matrices, the physical (denoted by subscript `$m$') neutral, singly-charged and doubly-charged fermions are given by
\begin{eqnarray}
  \psi^N_{mL} = \left({\begin{array}{c}
      \nu_{mL} \\
      \Sigma^1_{mL} \\
      \Sigma^2_{mL} \\
      \Delta^1_{mL} \\
      \Delta^2_{mL} \\
      \tilde{\Phi}^0_{mR} \\
      \end{array} }\right)&=&U_N^\dagger \,\psi^N_{L},~~~
  \psi^E_{mL(R)}= \left({\begin{array}{c}
      \ell_{mL(R)} \\
      \tilde{\Sigma}^+_{mR(L)}\\
      \tilde{\Delta}^+_{mR(L)}\\
      \Delta^-_{mL(R)}\\
      \widetilde\Phi^{+}_{mR}(\Phi^{-}_{mR})
      \end{array} }\right)=\left(U_E^{L(R)}\right)^\dagger \,\psi^E_{L(R)},\nonumber\\
 && \psi^K_{mL(R)} ~=~\left({\begin{array}{c}
      \tilde{\Sigma}^{++}_{mR(L)}\\
      \tilde{\Delta}^{++}_{mR(L)}\\
      \tilde{\Phi}^{++}_{mR}({\Phi}^{--}_{mR}) \\
      \end{array} }\right)=\left(U_K^{L(R)}\right)^\dagger \,\psi^K_{L(R)}.\nonumber
\end{eqnarray}
The mixings among the fermions in different $SU(2)_L$ representations play a crucial role in determining the phenomenology (collider as well as low-energy phenomenology) of this model. One can numerically diagonalise the mass matrices in Eq.~\ref{lag_mass} and obtain the masses and mixings of the physical fermions. However, to have a clear understanding about the dependence of different low-energy observables as well as the collider signatures (which will be discussed in subsequent sections) on different Yukawa couplings and mass matrices, it is instructive to obtain approximate analytic results for the mixing matrices $U_N$, $U_{L(R)}^E$, $U_{L(R)}^K$. In the basis where the SM charged lepton Yukawa matrix $Y_l$ (see Eq.~\ref{eq:yukawa_int}) and bare mass matrices $M_\Sigma$, $M_\Delta$ and $M_\Phi$ are real and diagonal, the approximate analytical expressions for the mixing matrices $U_N$, $U_{L(R)}^E$ and $U_{L(R)}^K$ are presented in appendix \ref{app:mass_matrices}.

The kinetic part of the Lagrangian involving electroweak gauge interactions of the fermionic multiplets is given by
\begin{equation}
  \mathcal{L}_{\text{kin}} \supset \text{Tr}[\overline{\chi} i\slashed{D} \chi]~,
  \label{lag_kin}
\end{equation}
where $D_\mu = \partial_\mu -i g\tau_a W^a_\mu-ig^\prime Y B_\mu$ is the gauge covariant derivative with $W^a_\mu$ and $B_\mu$ being the gauge bosons of $SU(2)_L$ and $U(1)_Y$ respectively. $g$ and $g^\prime$, respectively, are the $SU(2)_L$ and $U(1)_Y$ gauge couplings. $\chi$ symbollically stands for the heavy fermionic multiplets $\Sigma$, $\Delta$ and $\Phi$. $Y$ stands for the hypercharges ($Y=1,1/2$ and $0$ for $\Sigma$, $\Delta$ and $\Phi$, respectively) and $\tau_a=\sigma_a/2$, where $\sigma_a$ are the Pauli matrices, are the generators of fundamental representation of $SU(2)_L$. The Lagrangian in Eq.~\ref{lag_kin} contains the gauge interaction terms of the SM as well as the exotic fermions with the SM electroweak gauge bosons ($W^\pm,~Z$-boson and photon).  On the other hand, apart from leading to the mixings among the fermions in different $SU(2)$ multiplets, the Yukawa Lagrangian in Eq.~\ref{eq:yukawa_int} results into couplings involving the exotic fermions and the SM Higgs boson. The phenomenology (both the low-energy phenomenology as well as the collider signatures) of this model crucially depends of these gauge and Yukawa interactions. Therefore, we have presented the relevant parts of the gauge and Yukawa interactions in appendix \ref{app:gauge_Lag}.

\subsection{Neutrino masses and mixing angles}
\label{sec:nu_mass_mixing}
The block diagonalisation of the mass matrix ($M_N$) for the neutral leptons in Eq.~\ref{lag_mass} (see appendix~\ref{app:mass_matrices}) gives rise to the following $3\times 3$ Majorana mass matrix for light neutrinos at the leading order ${\cal O}\left(\frac{v^6}{\Lambda^5}\right)$:
\begin{eqnarray}
\label{eq:nu_mass}
m_\nu & \approx & -\frac{v^6}{48} Y_{23}^\dagger M_\Sigma^{-1} Y_{34}^{\prime \dagger} M_\Delta^{-1} Y_{45}^\prime M_\Phi^{-1} Y_{45}^{\prime T} M_\Delta^{-1} Y_{34}^{\prime *} M_\Sigma^{-1} Y_{23}^* + \mathcal{O}\left(\frac{v^8}{\Lambda^7}\right) \nonumber
\\
&=& \frac{v^2}{2} Y_{23}^\dagger \mathcal{M}^{-1} Y_{23}^* + \mathcal{O}\left(\frac{v^8}{\Lambda^7}\right)~,
\end{eqnarray}
where $\Lambda$ represents typical heavy lepton mass scale, and $\mathcal{M}$ is a $3\times 3$ complex symmetric matrix defined as $\mathcal{M}=\frac{24}{v^4} \left(M_\Sigma^{-1} Y_{34}^{\prime \dagger} M_\Delta^{-1} Y_{45}^\prime M_\Phi^{-1} Y_{45}^{\prime T} M_\Delta^{-1} Y_{34}^{\prime *} M_\Sigma^{-1}\right)^{-1}$ which, in general, is determined by 12 independent real parameters at high-scale. Note that the light neutrino mass matrix, $m_\nu$, is independent of the Yukawa matrices $Y_{34}$ and $Y_{45}$ upto the leading order. So, one will have nonzero light neutrino mass matrix even for vanishing $Y_{34}$ and $Y_{45}$. Note that $m_\nu$ in Eq.~\ref{eq:nu_mass} is suppressed by higher power of $\Lambda$ (in particular, by $\Lambda^{-5}$ at the leading order) which automatically brings down the new physics scale of $\Lambda$ to TeV for reasonably large Yukawa couplings of $\mathcal{O}(10^{-1}-10^{-2})$. This makes this model tastable at the collider experiments. Eq.~\eqref{eq:nu_mass} has the form of generalised Majorana seesaw formula, and, hence, one may call the underlying neutrino mass generation mechanism a modified/generalised seesaw one. Instead, the neutrino masses, in this model, result from a dimension-9 operator generated at tree-level via the Feynman diagram depicted in Fig.~\ref{fig:main}.  Also, if one follows the convention of labeling a seesaw model with the size of the representation of the fermionic field (more precisely, the fermionic field which yields lepton number violation) it employs, {\it i.e.}, if one labels a seesaw model employing a fermion transforming as $n$-representation of $SU(2)_L$ as type-$n$ or $n$-plet seesaw model, then one may call this model a Type-V or quintuplet seesaw model.

\begin{figure}[h]
\centering 
\includegraphics[width=.45\textwidth]{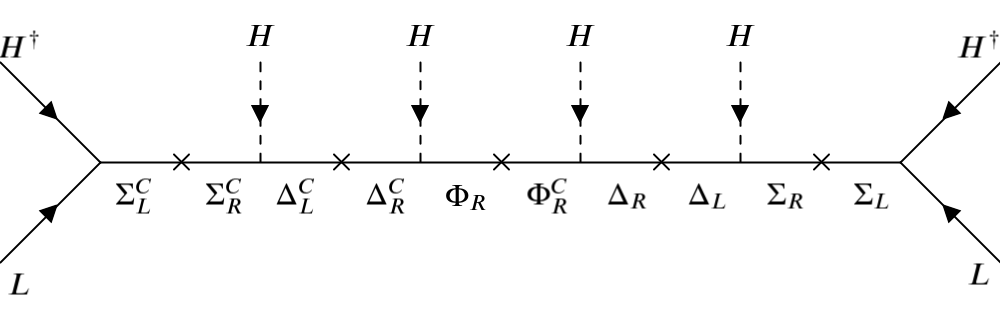}
\caption{\label{fig:main} Tree level diagram generating neutrino masses via low-energy effective operator of mass dimension-9. `$\times$' denotes mass insertion- either Majorana or Dirac type.}
\label{fig:main}
\end{figure}

It is needless to mention that the neutrino mass formula in Eq.~\ref{eq:nu_mass} contains a large number of Yukawa and mass parameters at high-scale. However, the low-energy effective theory (low-energy neutrino phenomenology) is determined by only 9 physical and measurable parameters --- 3 neutrino masses, 3 mixing angles and 3 phases. In order to ensure consistency between the theories, or in other words, to inspect the phenomenological implications of the seesaw theory in the neutrino oscillation experiments, proper parameterisation of Yukawa matrices is required. To find the general texture of the Yukawa matrix $Y_{23}$ which reproduces the measured mixing angles and mass spectrum of light neutrinos, we use the well-known Casas-Ibarra parametrization \cite{Casas_Ibarra,Ibarra_Ross}. Using the said parametrisation, the following most general texture for $Y_{23}$ is procured:
\begin{eqnarray}
\label{eq:parametrisation_0}
\frac{v}{\sqrt{2}} Y_{23}^* &=& U_\mathcal{M} \sqrt{\hat{\mathcal{M}}} R \sqrt{\hat{m}_\nu}V_{\rm PMNS}^\dagger\\
\downarrow && \qquad \qquad \downarrow ~\quad \downarrow \quad \downarrow \nonumber \\ 
\mathrm{\#~of~parameters}: \qquad 15 && \qquad \qquad 6 \qquad 3 \quad 6 \nonumber 
\end{eqnarray}
where $V_{\rm PMNS}$ is the so-called the Pontecorvo-Maki-Nakagawa-Sakata matrix determined by 3 angles and 3 phases --- one Dirac phase and two Majorana phases, $\hat{m}_\nu=\mathrm{diag}(m_1,m_2,m_3)$ is the diagonal light neutrino mass matrix, and $R$ is an arbitrary orthogonal matrix which can be parameterised by three complex angles. $R$ encodes the residual freedom in $Y_{23}$ once the other parameters are fixed, or in other words, the information lost in the decoupling of the heavy fermions is parameterised in $R$. 

\subsection{Mass spectrum of exotic fermions}
To explain three non-zero neutrino masses readily, this model employs three generations of vector-like $SU(2)_L$ triplets ($\Sigma_{L,R}$) and quadruplets ($\Delta_{L,R}$) as well as three copies of chiral $SU(2)_L$ quintuplets ($\Phi_R$). The particle spectrum of the model contains several exotic doubly charged, singly charged and neutral fermions arising from the triplets, quadruplets and quintuplets. For example, each triplet, quadruplet and quintuplet contain a exotic doubly charged fermion, namely, $\Sigma^{++}$, $\Delta^{++}$ and $\Phi^{++}$, respectively. While each triplet and quintuplet contribute a singly charged fermion ($\Sigma^{+}$ and $\Phi^{+}$, respectively) in the exotic fermion spectrum, two singly charged fermions ($\Delta^+$ and $\Delta^-$) result from a single generation of quadruplet. The number of neutral fermions resulting from each generations of triplets, quadruplets and quintuplets can be counted as two ($\Sigma^{1,2}$), two ($\Delta^{1,2}$) and one ($\Phi^0$), respectively. At the tree-level, the masses of these exotic fermions are given by bare mass matrices, $M_\Sigma,M_\Delta$ and $M_\Phi$, appearing in the Lagrangian in  Eq.~\eqref{eq:yukawa_int}. The corrections to the heavy lepton masses due to the mixing among different $SU(2)_L$ multiplets being very small, the components of a fermionic multiplet can be assumed to be exactly degenerate at tree level. The radiative corrections, dominantly induced by the electroweak gauge bosons, lift the degeneracy and induce a mass-splitting amongst the components of a fermionic multiplet. The radiative mass-splitting between the components $\chi^Q$ and $\chi^{Q^\prime}$ of a given multiplet $\chi$ ($=\Sigma,\Delta,\Phi$) with (tree-level) mass $M_\chi$ and hypercharge $Y_\chi$ is given by \cite{cirelli}
\begin{equation*}
M_{\chi^Q} - M_{\chi^{Q^\prime}} = \frac{\alpha_2 M_\chi}{4\pi} \left[(Q^2-Q^{\prime 2}) s_w^2 f\left(r_Z\right) + (Q-Q^\prime)(Q+Q^\prime-2Y_\chi)\left\{ f\left(r_W\right)-f\left(r_Z\right) \right\} \right]~,
\end{equation*}
where $s_w=\sin \theta_w$, $\theta_w$ being the weak-mixing angle; $\alpha_2=\alpha_{m_Z}/s_w^2$, $\alpha_{m_Z}$ being the electromagnetic fine structure constant; $Q$ and $Q^\prime$ denote the electric charges, $r_W=\frac{m_W}{M_\chi}$, $r_Z=\frac{m_Z}{M_\chi}$, and the function $f(r)$ has the following from
\begin{equation*}
f(r) =\frac{r}{2} \left[2r^3 \ln r -2r+\sqrt{r^2-4}~(r^2+2) \ln \left\{\frac{r^2-2-r\sqrt{r^2-4}}{2} \right\} \right]~.
\end{equation*}
For the mass range of our interest ($M_\chi >> m_{W,Z}$), these splittings are almost constant, and are completely determined by the electric- and hyper-charges of the particles involved. These mass-splittings are summarized in the following: $M_{\Sigma^{++}} - M_{\Sigma^+} \sim [830-860] \rm ~MeV, M_{\Sigma^{+}} - M_{\Sigma^0} \sim [480-515] \rm ~MeV,$
$M_{\Delta^{++}} - M_{\Delta^+} \sim [670-690] \rm ~MeV, M_{\Delta^{+}} - M_{\Delta^{1,2}} \sim [330-345] \rm ~MeV,$
$M_{\Phi^{++}} - M_{\Phi^+} \sim [510-520] \rm ~MeV, M_{\Phi^{+}} - M_{\Phi^0} \sim 170 \rm ~MeV$ and $M_{\Delta^-} - M_{\Delta^0} \sim [5-15] \rm ~MeV.$
Such small (of the order of a few hundreds of MeVs) mass-splittings can be safely inconsiderated in the context of most of the LHC studies. However, as the lightest neutrino approaches towards zero, Yukawa coupling induced 2-body decays into a SM fermion and a boson get highly suppressed for some of exotic fermions, and, hence, the mass-splitting induced decays into pion/leptons and a lighter exotic fermion start to play crucial role in determining the decay lengths of the heavy fermions. We will discuss these issues in details in section~\ref{sec:decay}. 

To summarize, the extended fermion sector of this model contains, in total, 9 doubly charged (3 from each multiplet), 12 singly charged (3 from triplet and quintuplet each, and 6 from quadruplet) and 15 neutral fermions (6 from triplet and quadruplet each, and 3 from quintuplet). All the fermions belonging to the same generation in a multiplet are degenerate at tree level. Though the radiative corrections lift the degeneracy inducing small mass-splittings among the differently charged fermions of a multiplet, in the context of the collider analysis, all the same generation fermions in a multiplet can be assumed to be degenerate. 

\subsection{A simplified model for collider phenomenology}
\label{subsec:simplified_model}
The explanation for the three tiny non-zero neutrino masses and mixings requires atleast three generations of additional fermions in the framework of different fermionic seesaw models like the Type-I/III seesaw as well as the model introduced in the previous section. Each generation of new fermions brings additional free parameters like, the bare fermionic mass parameters, Yukawa couplings, {\it etc.} in the model. Note that the collider studies of a scenario with a large number of free parameters is challenging. Moreover, the discovery signatures at the colliders rarely depend on all the parameters. For example, in the presence of three generations of fermionic multiplets, it will be the lightest one that will be more easily discovered, and, thus, the mass and Yukawa couplings of the lightest one will be the only relevant parameters in the context of collider phenomenology. As a result, the discovery signatures of fermionic seesaw models are, in general, studied in the framework of a simplified version of the model. For example, in the context of Type-III seesaw model, both the phenomenological \cite{biggio2011} as well as experimental \cite{cms_multilepton_137,lhc_1,lhc_2,lhc_3,lhc_4} searches are performed in the framework of a simplified model with only one generation of triplet leptons. The model introduced in the previous section includes three generations of fermionic triplets, quadruplets and quintuplets, and, hence, a number of bare mass and Yukawa matrices which are, in general, arbitrary complex matrices. The simplified version of our model assumes following simplified structures\footnote{Note that these assumptions are purely phenomenological without invoking additional symmetries which might results into such structures. However, the (in)sensitivity of the discovery signatures on such assumptions will be discussed in the collider phenomenology section.} for the mass and Yukawa matrices.

Note that we can always work in a basis where the fermion mass matrices $M_\Sigma,M_\Delta~{\rm and}~M_\Phi$ are real and diagonal. We further assumed degenerate copies of triplets, quadruplets and quintuplets, {\it i.e.,} $M_\Sigma~=~m_\Sigma\,{\rm \bf I}_{3\times 3},M_\Delta~=~m_\Delta\,{\rm \bf I}_{3\times 3}~{\rm and}~M_\Phi~=~m_\Phi\,{\rm \bf I}_{3\times 3}$, where $m_\Sigma,~m_\Delta$ and $m_\Phi$ are the common masses for the triplet, quadruplet and quintuplet generations. The Yukawa matrices $Y_{34}^\prime$ and $Y_{45}^\prime$, which appear in Eq.~(\ref{eq:parametrisation_0}), are also assumed to be diagonal and proportional to identity: $Y_{34}^\prime~=~y_{34}^\prime\,{\rm \bf I}_{3\times 3}$ and $Y_{45}^\prime~=~y_{45}^\prime\,{\rm \bf I}_{3\times 3}$. With these assumptions and $R~=~{\rm \bf I}_{3\times 3}$, Eq.~(\ref{eq:parametrisation_0}) reduces to %
\begin{equation}
\label{eq:parametrisation}
Y_{23}=\frac{\sqrt{48}}{v^3} \frac{m_\Sigma m_\Delta m_\Phi^{1/2}}{y_{34}^\prime y_{45}^\prime} \sqrt{\hat{m}_\nu} V_{\rm PMNS}^T~.
\end{equation}
For a given set of $m_\Sigma,m_\Delta,m_\Phi,y_{34}^\prime$ and $y_{45}^\prime$ and experimentally measured values for $\hat{m}_\nu$ and $V_{\rm PMNS}$, $Y_{23}$ can be completely determined using the above equation. We use this simplified parametrisation for further analysis. Note that the simple choice $R={\rm \bf 1}_{3\times 3}$ is commensurate with the fact that there exists a basis in which $Y_{23}$ and $\mathcal{M}$ are simultaneously diagonal but not $Y_\ell$. However, we accentuate that a different choice of $R$,  specifically a non-real choice of $R$, may lead to different collider limits. We use the following parametrisation of $V_{\rm PMNS}$:
\begin{equation*}
V_{\rm PMNS} = \left( \begin{array}{ccc}
c_{12} c_{13} & s_{12} c_{13} & s_{13} e^{-i\delta} \\
-s_{12} c_{23} -c_{12} s_{13} s_{23} e^{i\delta} & c_{12} c_{23} -s_{12} s_{13} s_{23} e^{i\delta} & c_{13} s_{23} \\
s_{12} s_{23} -c_{12} s_{13} c_{23} e^{i\delta} & -c_{12} s_{23} -s_{12} s_{13} c_{23} e^{i\delta} & c_{13} c_{23}
\end{array} \right) \times \mathrm{diag}(e^{-i \phi/2},e^{-i \phi^\prime/2},1) ~,
\end{equation*}
where $c_{ij}=\cos\theta_{ij}$ and $s_{ij}=\sin\theta_{ij}$; $\theta_{12},\theta_{13}$ and $\theta_{23}$ are the light neutrino mixing angles, $\delta$ is the Dirac CP phase, and $\phi$ and $\phi^\prime$ are the Majorana phases. For simplicity, we take the Majorana phases to be zero throughout this work. The following best fit values and $3\sigma$ range for the neutrino oscillation parameters \cite{nu_osci_param} are used:
\begin{itemize}
\item For NH:
$\theta_{12}=33.82^\circ~[31.61^\circ \to 36.27^\circ]$,~
$\theta_{13}=8.60^\circ~[8.22^\circ \to 8.98^\circ]$,~
$\theta_{23}=48.6^\circ~[41.1^\circ \to 51.3^\circ]$,~
$\Delta m_{21}^2 \times 10^5 ~\rm eV^{-2}=7.39~[6.79 \to 8.01]$,~
$\Delta m_{31}^2 \times 10^3 ~\rm eV^{-2}=2.528~[2.436 \to 2.618]$~and~
$\delta=221^\circ~[144^\circ \to 357^\circ]$~,
\item For IH:
$\theta_{12}=33.82^\circ~[31.61^\circ \to 36.27^\circ]$,~
$\theta_{13}=8.64^\circ~[8.26\circ \to 9.02^\circ]$,~
$\theta_{23}=48.8^\circ~[41.4^\circ \to 51.3^\circ]$,~
$\Delta m_{21}^2 \times 10^5 ~\rm eV^{-2}=7.39~[6.79 \to 8.01]$,~
$\Delta m_{32}^2 \times 10^3 ~\rm eV^{-2}=-2.510~[-2.601 \to -2.419]$~and~
$\delta=282^\circ~[205^\circ \to 348^\circ]$~,
\end{itemize}
where $\Delta m_{ij}^2=m_i^2-m_j^2$.

\section{Bounds from the lepton flavour violating processes}
\label{sec:lfv}
The presence of additional fermionic multiplets with Yukawa couplings with the SM lepton doublets leads to lepton flavour violation in this model. In this section, we will discuss the constraints on the model parameters, namely the masses of the triplets, quadruplets and quintuplets ($m_\Sigma,~m_\Delta$ and $m_\Phi$, respectively) and the Yukawa couplings $y_{34}^\prime$ and $y_{45}^\prime$ in the framework of the simplified scenario, from  the lepton flavour violating observables. We begin the discussion with the analytical results for the decay rates of lepton flavour violating magnetic transitions $\mu \to e\gamma$ and $\tau \to \ell \gamma$ ($\ell=e,\mu$). The amplitude\footnote{Note that we have assumed the outgoing photon to be on-shell while writing this amplitude. For on-shell photon, $\epsilon_\mu^*(q)q^\mu=0$ and $q^2=0$.} of such radiative decay can be written as \cite{Abada_et_al,Lavoura}
\begin{equation*}
T(\ell_\alpha \to \ell_\beta \gamma)=e\epsilon^*_\mu \bar u_\beta(p-q) \left[i\sigma^{\mu \nu} q_\nu \left( \sigma_L^{\beta \alpha} P_L + \sigma_R^{\beta \alpha} P_R \right) \right]u_\alpha(p)~,
\end{equation*}
where $\alpha$ and $\beta$ are charged lepton flavour indices; $p,p-q$ and $q$ are, respectively, the momentum of incoming lepton, outgoing lepton and outgoing photon; $\epsilon_\mu^*$ is the polarisation of outgoing photon, and $\sigma_L$ and $\sigma_R$ are form factors with inverse mass dimension. The decay rate for $\ell_\alpha \to \ell_\beta \gamma$ is given by \cite{Lavoura}
\begin{equation*}
\Gamma(\ell_\alpha \to \ell_\beta \gamma)=\frac{e^2(m_\alpha^2-m_\beta^2)^3\left(\left|\sigma_L^{\beta \alpha}\right|^2 + \left|\sigma_R^{\beta \alpha}\right|^2 \right)}{16\pi m_\alpha^3}~,
\end{equation*}
where $m_\alpha$ and $m_\beta$, respectively, denote the masses of $\ell_\alpha$ and $\ell_\beta$. We will work in the limit of massless final state, i.e., $m_\beta \to 0$.
\begin{figure}[htb!]
\centering 
\includegraphics[width=.45\textwidth]{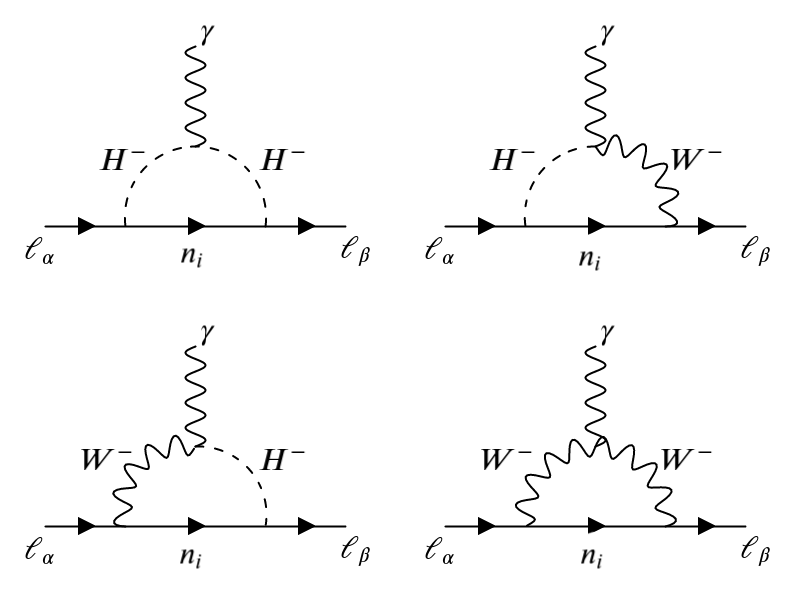}
\includegraphics[width=.45\textwidth]{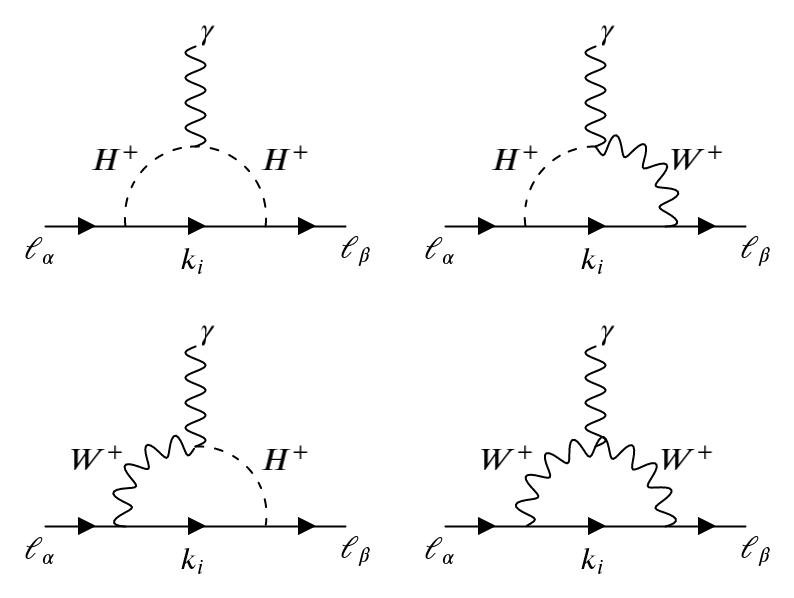}
\includegraphics[width=.45\textwidth]{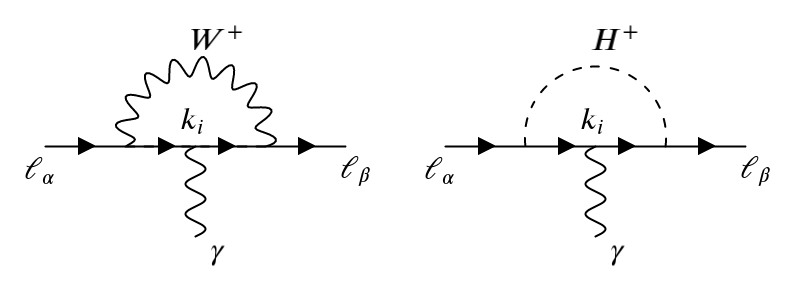}
\includegraphics[width=.45\textwidth]{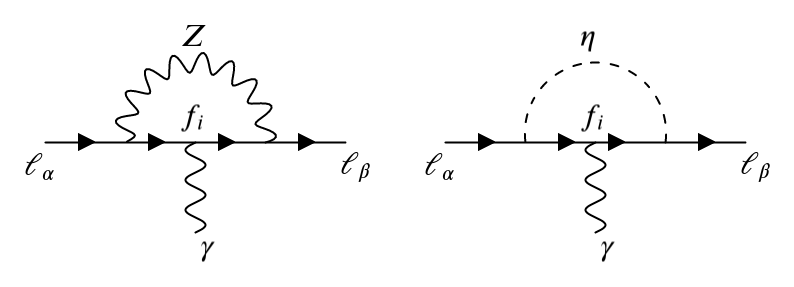}\\
\includegraphics[width=.225\textwidth]{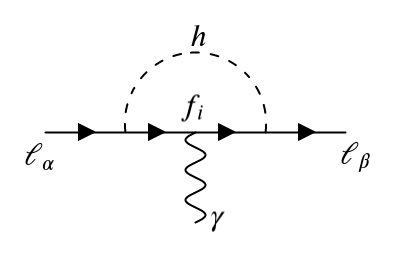}
\caption{Vertex-type diagrams. $n_i$,$f_i$ and $k_i$ symbollically stand for any of the neutral, singly-charged and doubly-charged fermions, respectively.}
\label{vtype}
\end{figure}
\begin{figure}[htb!]
\centering 
\includegraphics[width=.45\textwidth]{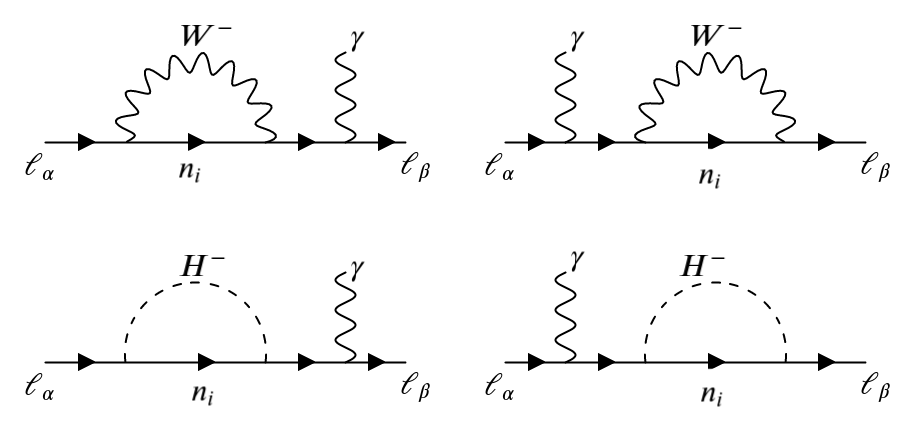}
\includegraphics[width=.45\textwidth]{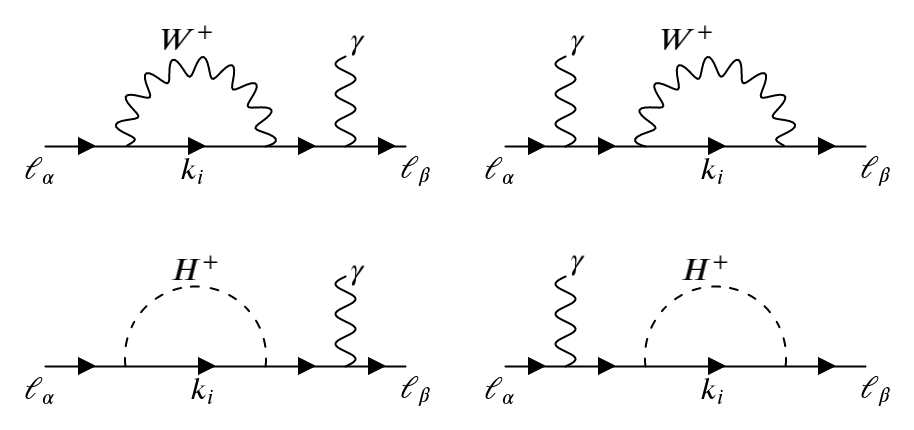}
\includegraphics[width=.45\textwidth]{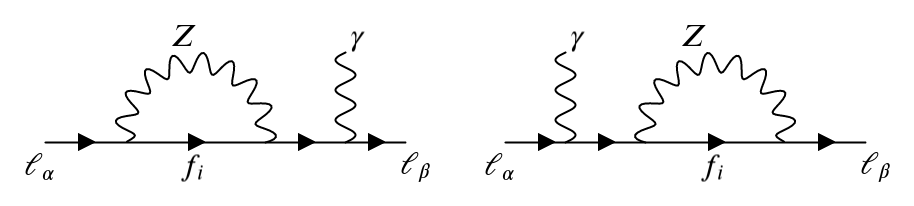}
\includegraphics[width=.45\textwidth]{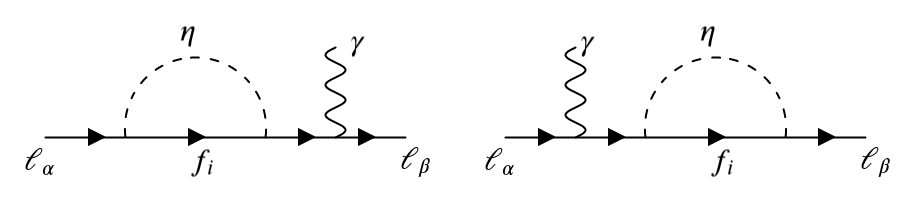}
\includegraphics[width=.45\textwidth]{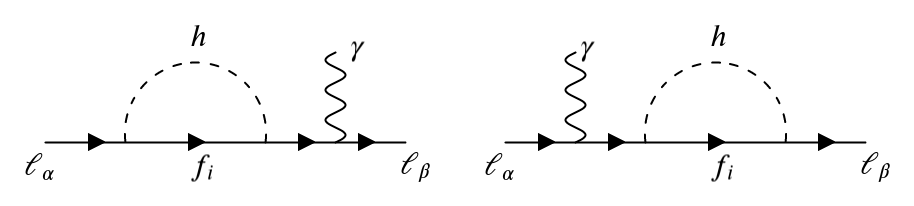}
\caption{Self-energy diagrams. $n_i$,$f_i$ and $k_i$ symbollically stand for any of the neutral, singly-charged and doubly-charged fermions, respectively.}
\label{setype}
\end{figure}
There are as many as 27\footnote{13 of them are of vertex-type diagrams (see Fig.~\ref{vtype}) and the rest are self-energy diagrams (see Fig.~\ref{setype}).} diagrams depicted in Figs.~\ref{vtype} and \ref{setype} which contributes to lepton flavour violating radiative decay $\ell_\alpha \to \ell_\beta \gamma$. These diagrams can be grouped into few sets of diagrams depending on the particles running in the loop. For example, the diagrams in which a neutral lepton, $\nu_i$, (doubly charged lepton, $(\Sigma^{++})^C$,)  a $W^-$-boson and/or a `would-be charged Goldstone boson' $H^-$  are circulating can be grouped together, and we call the corresponding amplitude as $T^{W^-,H^-}_{\nu_i}$ ($T^{W^-,H^-}_{(\Sigma^{++})^C}$). Similarly, the amplitudes of the loops involving a singly charged lepton ($l$ or ${(\Sigma^{+})^C}$),  a $Z/H$-boson and/or a `would-be neutral Goldstone boson' $\eta$ are denoted by $T^{Z,\eta}_{\ell},~T^{Z,\eta}_{(\Sigma^+)^C}$ and $T^{H}_{(\Sigma^+)^C}$. All the loop calculations are performed in the Feynman-'t Hooft gauge. The non-vanishing amplitudes corresponding to the different sets of diagrams are presented in the following upto $\mathcal{O}(\frac{Y^2v^2}{\Lambda^2})$: 
\begin{eqnarray*}
T^{W^-,H^-}_{\nu} &=&i \mathcal{A} \sum_{i=1}^3 \left\{({\rm \bf I}+\lambda)V_{\rm PMNS}\right\}_{\beta i} \bigg[\left(V_{\rm PMNS}^\dagger \right)_{i\alpha} F_1(w_{\nu_i}) - \left\{V_{\rm PMNS}^\dagger ({\rm \bf I}+\lambda) \right\}_{i\alpha} F_2(w_{\nu_i}) \bigg]~,
\\
T^{W^-,H^-}_{(\Sigma^{++})^C} &=&i \mathcal{A} v^2 \sum_{i=1}^3 \left(Y_{23}^T M_\Sigma^{-1} \right)_{\beta i} \left(M_\Sigma^{-1} Y_{23}^* \right)_{i\alpha} F_3(w_{\Sigma_i}), \nonumber
\\
T^{Z,\eta}_{\ell} &=&4i \mathcal{A}\lambda_{\beta \alpha} \sum_{i=1}^3 \bigg[F_4(z_{\ell_i}) +F_{5}(z_{\ell_i}) \sin^2\theta_w \bigg]~,\\
T^{Z,\eta}_{(\Sigma^+)^C} &=&i \mathcal{A} v^2 \sum_{i=1}^3 \left(Y_{23}^T M_\Sigma^{-1} \right)_{\beta i} \left(M_\Sigma^{-1} Y_{23}^* \right)_{i\alpha} F_6(z_{\Sigma_i})~,
\\
T^{H}_{(\Sigma^+)^C} &=&i \mathcal{A}v^2 \sum_{i=1}^3 \left(Y_{23}^T M_\Sigma^{-1} \right)_{\beta i} \left(M_\Sigma^{-1} Y_{23}^* \right)_{i\alpha}  F_7(h_{\Sigma_i})~,
\end{eqnarray*} 
where $w(z/h)_{X_i}=\frac{m_{X_i}^2}{m_{W(Z/h)}^2}$, 
$\lambda=\frac{v^2}{8}Y_{23}^T M_\Sigma^{-2}Y_{23}^*$,
$\theta_w$ is the weak mixing angle, 
$$\mathcal{A}=\frac{G_F}{\sqrt{2}} \frac{e}{4\pi^2} m_\alpha \epsilon^*_\mu \bar u_\beta(p-q) \sigma^{\mu \nu} q_\nu P_R u_\alpha(p),$$ 
and $F_1,F_2,F_3,F_4,F_5,F_6$ and $F_7$ are the loop functions:
\begin{align*}
F_1(r)&=\frac{-1+4r-3r^2+2r^2\log{r}}{4(-1+r)^3}~,
\\
F_2(r)&=\frac{-7+33r-57r^2+31r^3+6r^2(1-3r) \log{r}}{12(-1+r)^4}~,
\\
F_3(r)&=\frac{46-79r+42r^2-13r^3+4r^4+18(-2+r)^2r\log{r}}{24(-1+r)^4}~,
\\
F_4(r)&=\frac{4-9r+5r^3+6r(1-2r) \log{r}}{12(-1+r)^4}~,
\\
F_5(r)&=\frac{2+3r-6r^2+r^3+6r \log{r}}{6(-1+r)^4}~,
\\
F_6(r)&=\frac{-14+50r-39r^2+2r^3+r^4-6r(2-7r+2r^2)\log{r}}{96(-1+r)^4}~,
\\
F_7(r)&=\frac{6-28r+45r^2-24r^3+r^4+6r^2(-1+2r)\log{r}}{96(-1+r)^4}.
\end{align*}
After adding up all the contributions, we obtain the following form factors: 
\begin{align*}
&\sigma_L^{\beta \alpha}=0~~~{\rm and}\nonumber\\
&\sigma_R^{\beta \alpha} =\frac{G_F}{\sqrt{2}} \frac{1}{4\pi^2} m_\alpha \sum_{i=1}^3 \bigg[\left\{({\rm \bf I}+\lambda)V_{\rm PMNS}\right\}_{\beta i} \left\{ \left(V_{\rm PMNS}^\dagger \right)_{i\alpha} F_1(w_{\nu_i}) - \left\{V_{\rm PMNS}^\dagger ({\rm \bf I}+\lambda) \right\}_{i\alpha} F_2(w_{\nu_i}) \right\} + \nonumber
\\
& v^2 \left(Y_{23}^T M_\Sigma^{-1} \right)_{\beta i} \left(M_\Sigma^{-1} Y_{23}^* \right)_{i\alpha} \Big\{ F_3(w_{\Sigma_i}) + F_6(z_{\Sigma_i}) +F_7(h_{\Sigma_i}) \Big\} + 4 \lambda_{\beta \alpha} \Big\{F_4(z_{\ell_i}) + F_{5}(z_{\ell_i}) \sin^2\theta_w \Big\} \bigg]~.
\end{align*}
Note that the $SU(2)_L$ and $U(1)_Y$ charges of the exotic fermion multiplets do not allow Yukawa interactions involving the SM singlet leptons and the exotic multiplets. Moreover, to simplify the calculations, we have neglected the mass of the final state lepton, {\it i.e.,} $m_\beta=0$. These two facts together result into $\sigma_L^{\beta \alpha}=0$. Finally, the lepton flavour violating branching ratio $Br(\ell_\alpha \to \ell_\beta \gamma)$ reads as
\begin{align}
\label{eq:lfv_br}
{\rm Br}(\ell_\alpha \to \ell_\beta \gamma) = \frac{48\pi^3 \alpha}{G_F^2 m_\alpha^2} \left|\sigma_R^{\beta \alpha}\right|^2 \times {\rm Br}(\ell_\alpha \to \ell_\beta \bar{\nu_\beta} \nu_\alpha)~,
\end{align}
where $Br(\ell_\alpha \to \ell_\beta \bar{\nu_\beta} \nu_\alpha)$ is 1 for $\mu \to e \gamma$ and $0.1784\pm 0.0005$ for $\tau \to \ell \gamma$ ($\ell=e,\mu$), and $\alpha$ is the electromagnetic fine-structure constant. In the limit $w(z/h)_{\Sigma_i} \to \infty$ {\it i.e.,} $M_\Sigma >> m_{W,Z,h}$ and $w(z/h)_{\nu_i(\ell_i)} \to 0$, the branching ratio reduces to
\begin{align}
\label{eq:lfv_br1}
{\rm Br}(\ell_\alpha \to \ell_\beta \gamma) = \frac{3\alpha}{2\pi} \left| \frac{51+16\sin^2\theta_w}{12} \lambda_{\beta \alpha} \right|^2 \times {\rm Br}(\ell_\alpha \to \ell_\beta \bar{\nu_\beta} \nu_\alpha)~.
\end{align}
The following constraints on the $\lambda_{\beta \alpha}$'s are obtained using the current experimental bounds (see table \ref{table:exp_limit_LFV}) on $\ell_\alpha \to \ell_\beta \gamma$ \cite{muegamma,tauellgamma}:
\begin{eqnarray}
\label{eq:limit_lambda}
\left|\lambda_{e\mu(e\tau)[\mu \tau]}\right| = \left|\left(\frac{v^2}{8}Y_{23}^T M_\Sigma^{-2}Y_{23}^*\right)_{e\mu(e\tau)[\mu \tau]}\right| \leq 2.3\times 10^{-6}~(1.5\times 10^{-3})~[1.8\times 10^{-3}]~.
\end{eqnarray}
Note that these limits are more constraining (by almost a factor of $\sim 2.5$) than those in type-III seesaw model. In the context of LFV, there is mainly two differences between the present model and the type-III seesaw one. First,  in the present model, there is a LFV contribution from diagrams with a  doubly charged lepton, a $W$-boson and/or a `would-be Goldstone boson' circulating in the loop. Second, the LFV contribution from diagrams with a heavy neutral lepton, a $W$-boson and/or a `would-be Goldstone boson' circulating on the loop is absent in the present model upto $\mathcal{O}(Y^2v^2/\Lambda^2)$\footnote{The reason behind that the corresponding vertex vanishes upto $\mathcal{O}(Y^2v^2/\Lambda^2)$ has been discussed in the next to next section.}, which is not the case with the type-III seesaw model. These two reasons together account for the difference among the bounds for the two models.

We use the parametrisation in Eq.~\eqref{eq:parametrisation} to comprehend how the bounds in Eq.~\eqref{eq:limit_lambda} translate to parameters of the simplified scenario described in subsection \ref{subsec:simplified_model}. In this scenario, $\lambda_{\beta \alpha}$ is completely determined by $m_\Delta, m_\Phi, y_{34}^\prime, y_{45}^\prime$ and the neutrino oscillation parameters:
\begin{eqnarray}
\label{eq:lambda_param}
\left|\lambda_{\beta \alpha}\right| = \frac{6m_\Delta^2 m_\Phi}{v^3 |y^\prime_{34} y^\prime_{45}|^2} \left|\frac{1}{v}\left( V_{\rm PMNS} \hat{m}_\nu V_{\rm PMNS}^\dagger \right)_{\beta \alpha} \right|~.
\end{eqnarray}
Using Eq.~\eqref{eq:lambda_param}, the bound on $|\lambda_{e \mu}|$\footnote{One can also translate the bounds on $|\lambda_{e \tau}|$ and $|\lambda_{\mu \tau}|$, but those bounds are more liberal in comparison to that on $|\lambda_{e \mu}|$.} in Eq.~\eqref{eq:limit_lambda} can be translated on $\frac{m_\Delta^2 m_\Phi}{v^3 |y^\prime_{34} y^\prime_{45}|^2}$ as in the following:
\begin{eqnarray}
\label{eq:limit_ana}
\frac{m_\Delta^2 m_\Phi}{v^3 |y^\prime_{34} y^\prime_{45}|^2} \leq 1.2\times 10^7~(6.4\times 10^7)~[1.8\times 10^7~(7.9\times 10^7)]~.
\end{eqnarray}
These bounds correspond to the lightest neutrino mass $m_1[m_3]=0.0 (0.1)$ eV for NH[IH] light neutrino mass spectrum. Note that $\mu \to e \gamma$ and $\tau \to \ell \gamma$ ($\ell=e,\mu$) are not the only LFV observables that have been searched experimentally. There are several other LFV observables which could render stronger constraints on the parameters of the simplified model. The experimental searches for different LFV processes and their impact on our model will be discussed in the following.  

\subsection{Experimental bounds on LFV processes:}
\label{sec:Expt_LFV}
Even after long-standing efforts from various experimental collaborations, lepton flavour violation in the charged fermion sector is yet to be observed. Being not observed, the experimental upper limits on various lepton flavour violating (LFV) decays put stringent constraints on the Yukawa couplings as well as the new physics scale. Among all the experiments performed so far to find LFV radiative decay $\ell_\alpha \to \ell_\beta \gamma$, the best existing bound is $4.2\times 10^{-13}$ on the branching ratio of $\mu \to e\gamma$ from MEG Collaboration \cite{muegamma}, whereas the most stringent bound on $\ell_\alpha \to \ell_\beta \ell_\gamma \ell_\delta$ is $1.0\times 10^{-12}$ on the branching ratio of $\mu^+ \to e^+e^+e^-$ from SINDRUM collaboration \cite{mu3e}. In case of $\mu \to e$ conversion in a target nucleus, the most stringent bound on the branching ratio is $7\times 10^{-13}$ on Gold from SINDRUM-II collaboration \cite{mueau}. The COMET \cite{mueal} collaboration at J-PARC and the Mu2e \cite{mueal2} collaboration at FNAL both have projected a future sensitivity of $1.0\times 10^{-16}$ on the branching ratio of $\mu \to e$ conversion on Al. All these experimental bounds are summarised in table~\ref{table:exp_limit_LFV}.

\begin{table}[htb!]
\centering
\begin{tabular}{|c|c|c|c|}
\hline
\bf LFV process & \bf Expt. upper limit & \bf Model prediction & \bf Model prediction \\
& & \bf for BP1 & \bf for BP2 \\
\hline
$\mu^+ \to e^+ \gamma$ & $4.2\times 10^{-13}$ \cite{muegamma} & $3.8 \times 10^{-13} (3.7 \times 10^{-13})$ & $5.7 \times 10^{-17} (5.6 \times 10^{-17})$
\\
$\tau^\pm \to e^\pm \gamma$ & $3.3\times 10^{-8}$ \cite{tauellgamma} & $3.4 \times 10^{-15} (3.3 \times 10^{-15})$ & $5.2 \times 10^{-19} (5.0 \times 10^{-19})$
\\
$\tau^\pm \to \mu^\pm \gamma$ & $4.4\times 10^{-8}$ \cite{tauellgamma} & $5.1 \times 10^{-13} (5.0 \times 10^{-13})$ & $7.8 \times 10^{-17} (7.6 \times 10^{-17})$
\\
$\mu^+ \to e^+e^+e^-$ & $1.0\times 10^{-12}$ \cite{mu3e} & $3.1 \times 10^{-12}$ & $4.7 \times 10^{-16}$ 
\\
$\tau^- \to e^-e^+e^-$ & $2.7\times 10^{-8}$ \cite{tau3ell} & $2.7 \times 10^{-14}$ & $4.2 \times 10^{-18}$
\\
$\tau^- \to \mu^-\mu^+\mu^-$ & $2.1\times 10^{-8}$ \cite{tau3ell} & $4.1 \times 10^{-12}$ & $6.3 \times 10^{-16}$
\\
$\tau^- \to e^-\mu^+\mu^-$ & $2.7\times 10^{-8}$ \cite{tau3ell} & $1.7 \times 10^{-14}$ & $2.7 \times 10^{-18}$
\\
$\tau^- \to \mu^-e^+e^-$ & $1.8\times 10^{-8}$ \cite{tau3ell} & $2.6 \times 10^{-12}$ & $4.0 \times 10^{-16}$
\\
$\tau^- \to e^+\mu^-\mu^-$ & $1.7\times 10^{-8}$ \cite{tau3ell} & $2.4 \times 10^{-23}$ & $5.8 \times 10^{-31}$
\\
$\tau^- \to \mu^+e^-e^-$ & $1.5\times 10^{-8}$ \cite{tau3ell} & $1.6 \times 10^{-25}$ & $3.8 \times 10^{-33}$
\\
$\mu \to e$ on Pb & $4.6\times 10^{-11}$ \cite{muepb} & $7.9 \times 10^{-11}$ & $1.2 \times 10^{-14}$
\\
$\mu \to e$ on Ti & $1.7\times 10^{-12}$ \cite{mueti} & $5.9 \times 10^{-11}$ & $9.1 \times 10^{-15}$
\\
$\mu \to e$ on Au & $7.0\times 10^{-13}$ \cite{mueau} & $8.1 \times 10^{-11}$ & $1.2 \times 10^{-14}$
\\
$\mu \to e$ on Al & $1.0\times 10^{-16}$ \cite{mueal,mueal2} & $2.7 \times 10^{-11}$ & $4.1 \times 10^{-15}$
\\
\hline
\end{tabular}
\caption{Experimental upper limit and model prediction for two benchmark points on the branching ratio of various LFV processes. The last line in the table shows future sensitivity on $\mu \to e$ conversion on Al. Bracketed values correspond to the branching ratio calculated using the analytical expressions for $\ell_\alpha \to \ell_\beta \gamma$.}
\label{table:exp_limit_LFV}
\end{table}

The model predictions for the LFV observables are presented for two benchmark points (BPs) defined as follows:
\begin{itemize}
\item \textbf{BP1}:$\frac{m_\Delta^2 m_\Phi}{v^3 |y^\prime_{34} y^\prime_{45}|^2}=1.1\times 10^7$ \footnote{A choice like $m_\Delta, m_\Phi \sim 1$ TeV and $y^\prime_{34}, y^\prime_{45} \sim 0.05$ will give rise to BP1, whereas $m_\Delta, m_\Phi \sim 1$ TeV and $y^\prime_{34}, y^\prime_{45} \sim 0.15$ will give rise to BP2.} and $m_1=10^{-5}$ eV~,
\item \textbf{BP2}:$\frac{m_\Delta^2 m_\Phi}{v^3 |y^\prime_{34} y^\prime_{45}|^2}=1.3\times 10^5$ and $m_1=10^{-5}$ eV~,
\end{itemize}
where we have assumed NH light neutrino mass spectrum. To numerically evaluate the LFV observables, we have implemented\footnote{Note that we have already derived the analytical results for the LFV radiative decays, $\ell_\alpha \to \ell_\beta \gamma$, in the previous section. We used these analytical results to validate our model implementation. The bracketed numbers in the first three rows of table~\ref{table:exp_limit_LFV} are the analytical predictions for $\ell_\alpha \to \ell_\beta \gamma$ which are in excellent agreement with numerical results from SPheno.} the model in SARAH \cite{sarah,sarah2}, generated modules for SPheno, and used SPheno \cite{spheno,spheno2} for the numerical evaluation. The numerical values of the LFV branching ratios for the BPs are presented in Table~\ref{table:exp_limit_LFV}. Clearly, BP1 gives rise to $\mu \to 3e$ branching ratio and $\mu \to e$ conversion rates which are larger than the experimental bounds, and hence is ruled out. Whereas, BP2 is allowed by all the LFV processes at present\footnote{Non observations of LFV decays from the COMET and the Mu2e collaboration will rule out the BP2.}. We have scanned over the parameter space in the simplified scenario and procured the following upper limits on $\frac{m_\Delta^2 m_\Phi}{v^3 |y^\prime_{34} y^\prime_{45}|^2}$ from $\mu \to e\gamma$, $\mu \to 3e$, $\mu \to e$ conversion in Al and $\mu \to e$ conversion in Au for two values of the lightest neutrino mass, $m_1 (m_3)=0.0$ and $0.1$ eV in NH (IH) light neutrino mass spectrum\footnote{Analytically procured upper limit on $\frac{m_\Delta^2 m_\Phi}{v^4 |y^\prime_{34} y^\prime_{45}|^2}$ from $\mu \to e\gamma$ in Eq.~\eqref{eq:limit_ana} are in quite excellent agreement with those obtained numerically (SPheno) in second column in Table \ref{table:limit_spheno}.}.
\begin{table}[htb!]
\centering
\scalebox{0.84}{
\begin{tabular}{|c|c|c|c|c|}
\hline
$m_1~(m_3)$ in eV &  $\mu \to e\gamma$ & $\mu \to 3e$ & $\mu \to e$ in Al & $\mu \to e$ in Au \\
\hline
0.0 & $1.1\times 10^7 (1.7\times 10^7)$ & $6.2\times 10^6 (9.3\times 10^6)$ & $2.1\times 10^4 (3.2\times 10^4)$ & $1.1\times 10^6 (1.5\times 10^6)$\\
0.1 & $6.2\times 10^7 (7.6\times 10^7)$ & $3.4\times 10^7 (4.2\times 10^7)$ & $1.1\times 10^5 (1.4\times 10^5)$ & $5.4\times 10^6 (6.6\times 10^6)$
\\
\hline
\end{tabular}
}
\caption{Upper limit on $\frac{m_\Delta^2 m_\Phi}{v^3 |y^\prime_{34} y^\prime_{45}|^2}$ from various LFV processes for the lightest neutrino mass, $m_1 (m_3)=0.0$ and $0.1$ eV in NH (IH) light neutrino mass spectrum.}
\label{table:limit_spheno}
\end{table}

\section{Collider Phenomenology}
One of the prime advantages of obtaining neutrino masses from a tree-level dimension-9 operator is a world of new physics at the TeV scale which is potentially testable, and hence falsifiable at the LHC. In the framework of this model (see section~\ref{sec:model}), the explanation of the neutrino oscillation data ({\emph i.e.,} the tiny neutrino masses and mixings) requires several exotic fermionic multiplets at TeV scale, and hence gives rise to interesting collider signatures which will be discussed in the following.

\subsection{Production of exotic fermions at the LHC}
\label{sec:production}
Owing to their non-trivial representation under the $SU(2)_L$ gauge group, the exotic fermions can be pair\footnote{The mixings between the SM and exotic fermions give rise to single production of exotic fermions in association with a SM fermion via the interactions listed in equations (\ref{eq:dec_CC_3})-(\ref{eq:dec_NC_5}). However, such single productions are highly suppressed due to the mixing.} produced at the LHC via the gauge interactions listed in equations (\ref{eq:gauge_prod_3})-(\ref{eq:gauge_prod_5}). The pair productions are quark-antiquark initiated processes with a $\gamma/Z$ or $W^\pm$ exchange in the $s$-channel, namely, the Drell-Yan processes:
\begin{subequations}
\begin{align*}
q~\overline{q^\prime} &\to \gamma/Z \to \chi^{++}\chi^{--} / \chi^+ \chi^- / \chi^0 \chi^0~, \hspace{1cm} q~\overline{q^\prime} \to W^\pm \to \chi^{\pm \pm} \chi^\mp / \chi^\pm \chi^0~,
\end{align*}
\end{subequations}
where $\chi$ symbollically stands for the heavy fermionic multiplets $\Sigma$, $\Delta$ and $\Phi$ with superscripts denoting the electric charge.
Some of these exotics, being electrically charged, also couples to photons, and, hence, they can also be pair produced at collider via $t/u$-channel photon-photon fusion process (see the Feynman diagram in the left panel of figure \ref{fig:photon_pair}). Although the parton density of photon inside a proton is very small compared to the other partons, namely, gluons and quarks, the contribution of the photon initiated pair production of charged exotics becomes comparable (or even more significant) to that of the Drell-Yan pair productions for large masses of the doubly charged exotics. For large masses of the exotics, the photon initiated pair production becomes significant for the following two reasons \cite{babu_photon,kirti_photon,agar_photon,avnish_photon}. First, the Drell-Yan processes, being $s$-channel ones, falls rapidly with the parton center of mass energy, $\sqrt{\hat{s}}$, whereas the photon initiated pair productions, being $t/u$-channel processes, don't fall so rapidly. Second, the coupling of photons with a pair of fermions is proportional to the electric charge of the fermions. Hence, at the partonic level, the pair production cross section of doubly charged fermions is enhanced by a factor of $2^4$ compared to that of the singly charged ones.

\begin{figure}[h]
\includegraphics[scale=0.6]{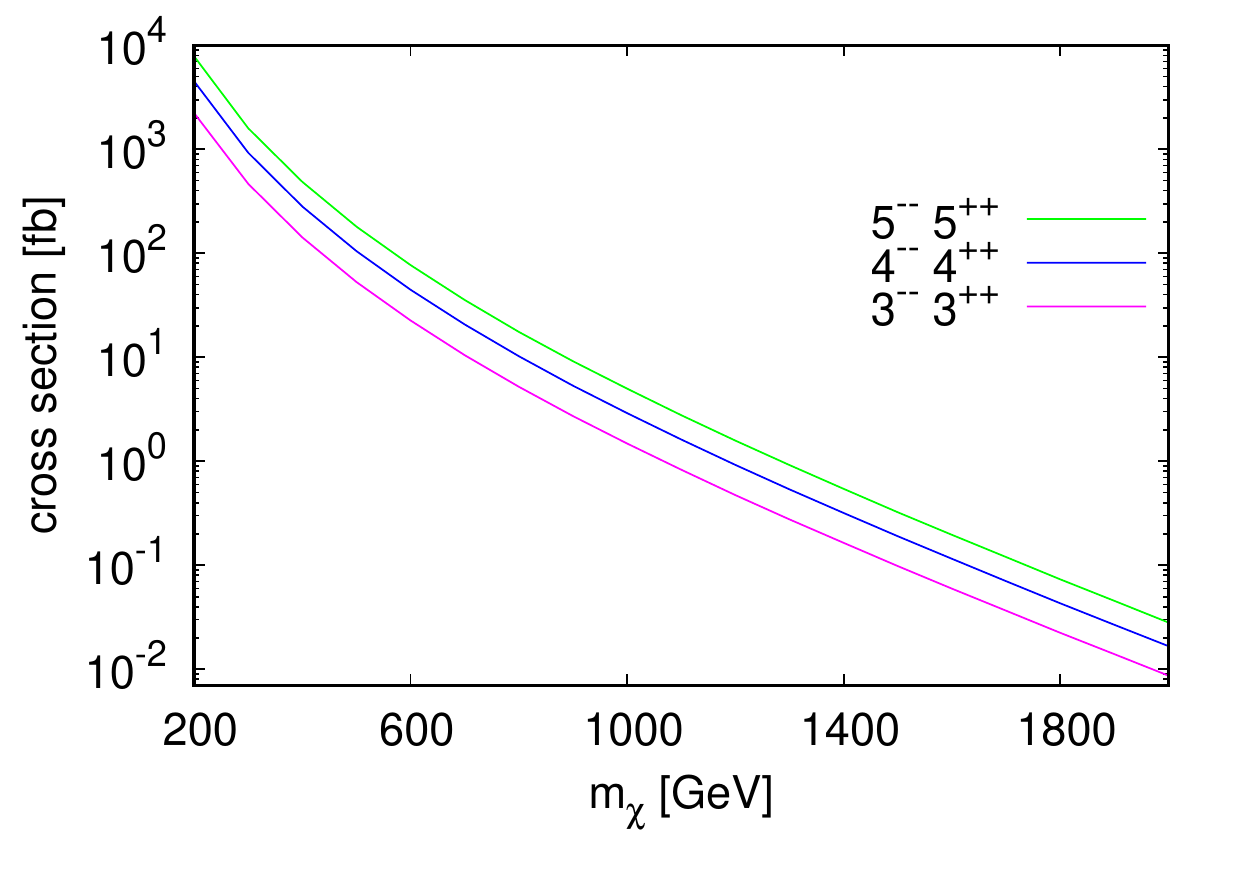}
\includegraphics[scale=0.6]{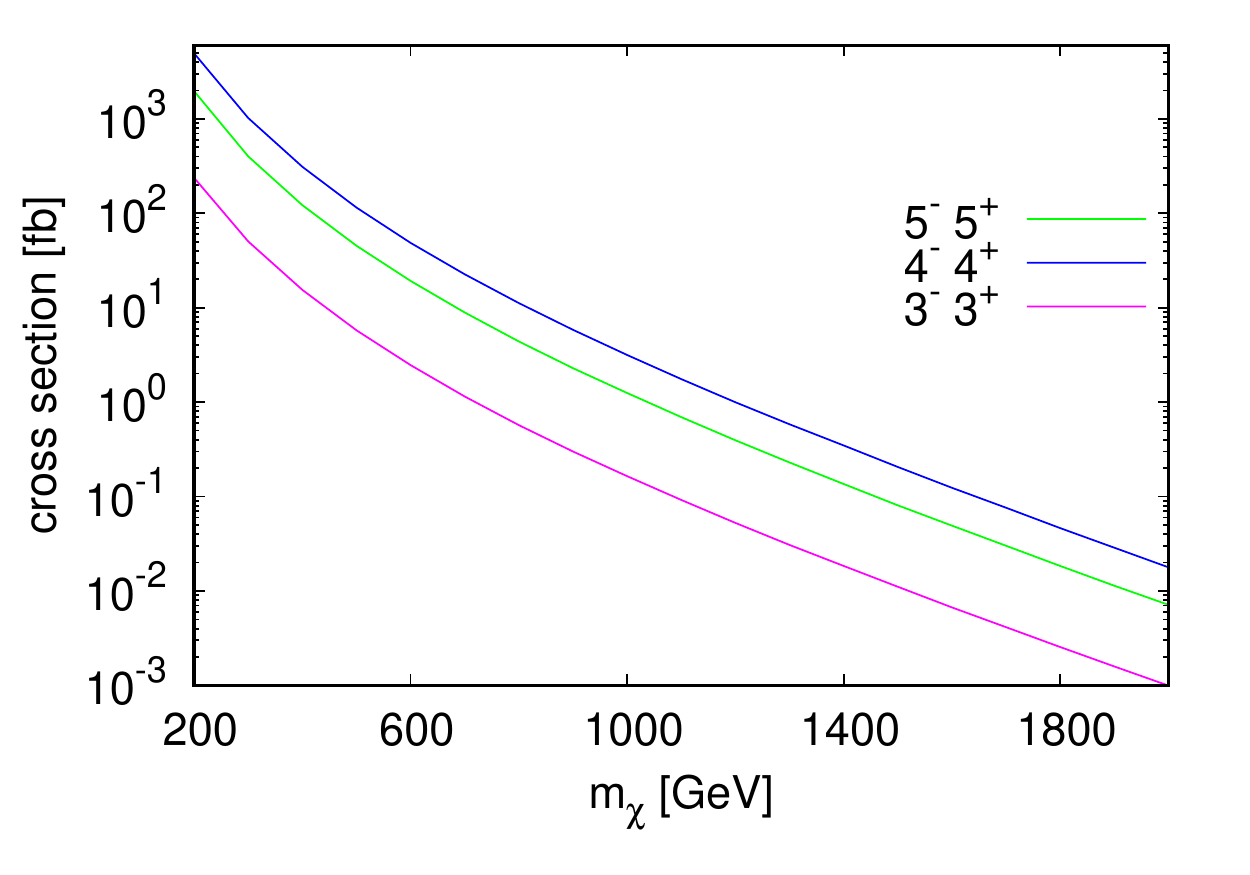}\\
\includegraphics[scale=0.6]{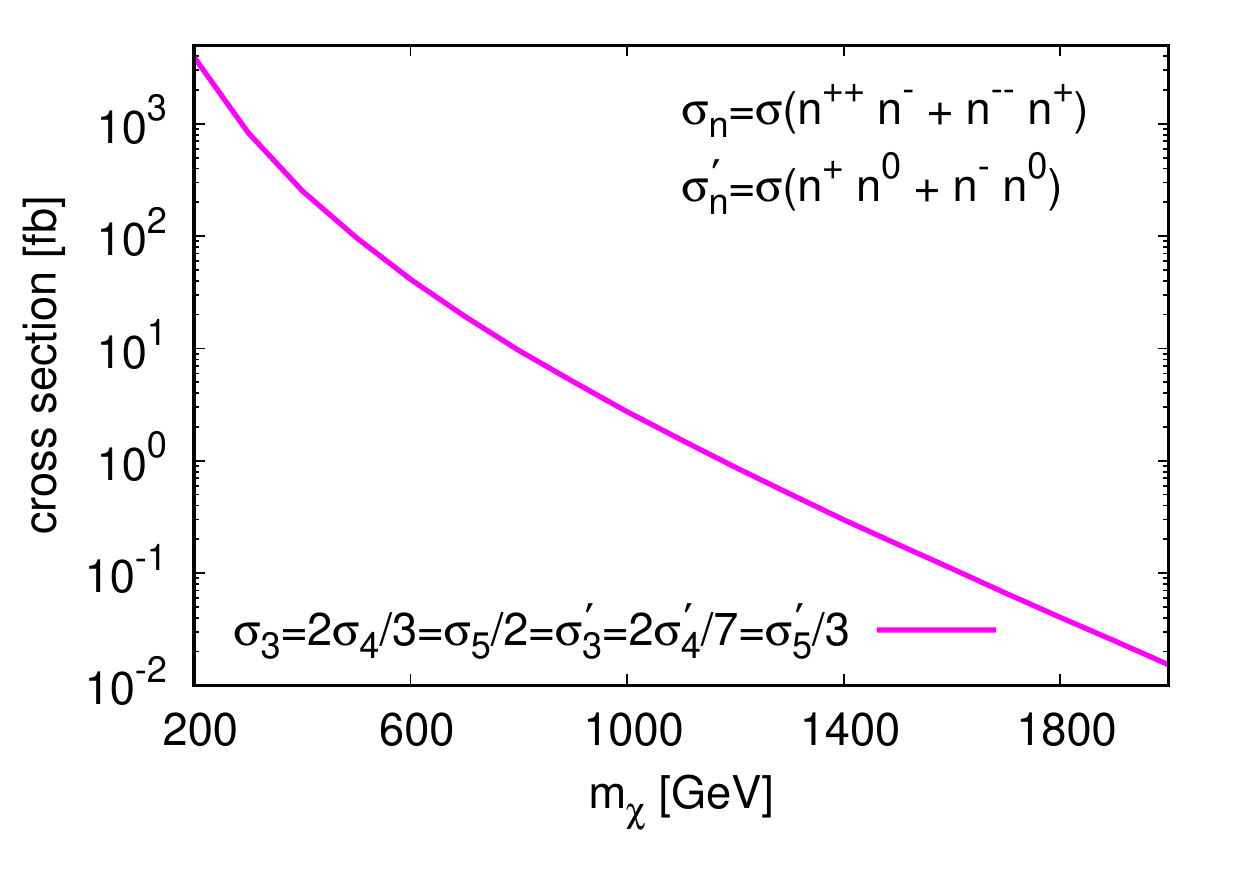}
\includegraphics[scale=0.6]{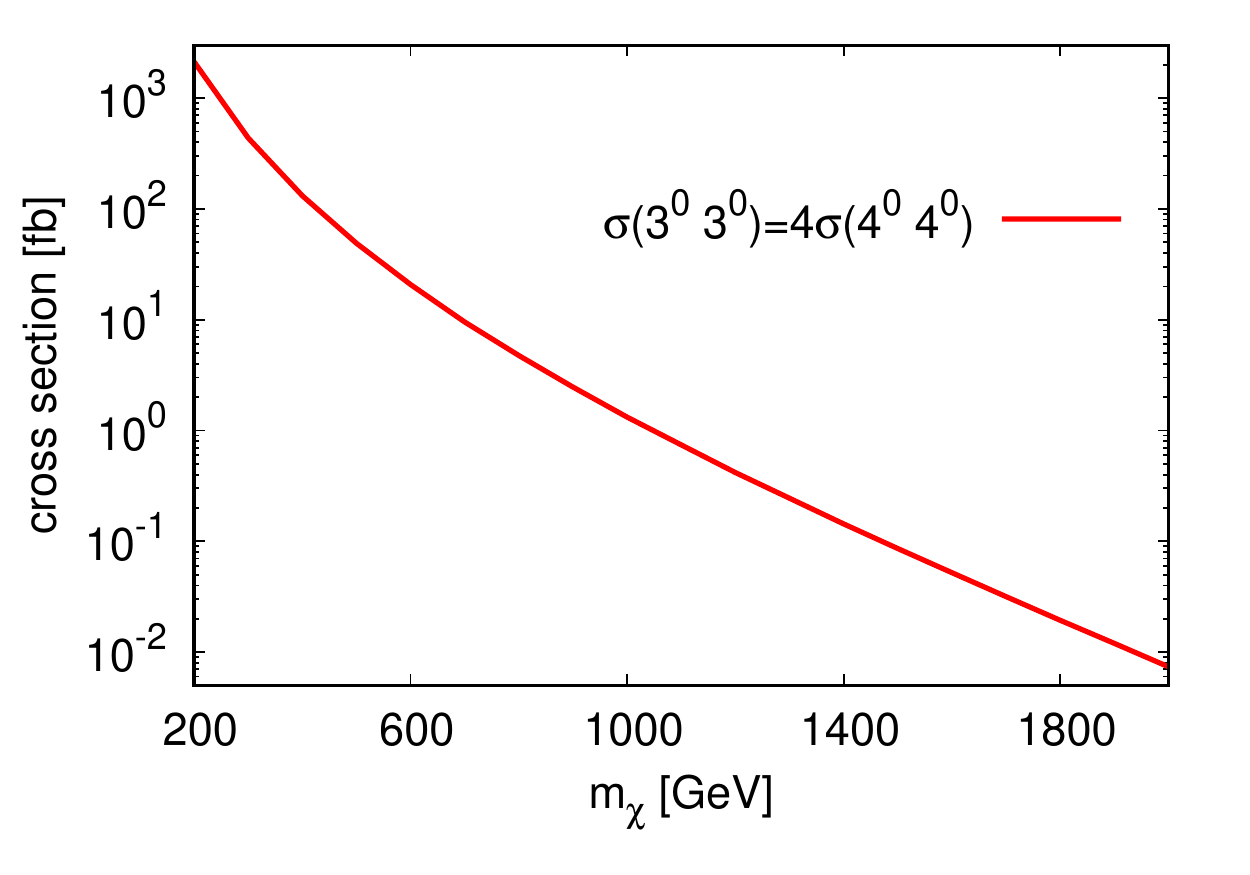}
\caption{\label{fig:photon_pair} Cross sections for Drell-Yan pair (top and bottom right panel) and associated (bottom left panel) productions of exotic fermions at 13 TeV LHC.}
\label{fig:DY_prod}
\end{figure}

For the numerical evaluation of the production cross sections of the exotics at the LHC with $\sqrt s~=~13$ TeV, MadGraph5aMC$@$NLO \cite{mg5,mg52} has been used with the {\bf NNPDF23LO} \cite{nnpdf,nnpdf2} parton distribution function. Figure \ref{fig:DY_prod} shows cross sections for different Drell-Yan pair and associated productions of exotic fermions \footnote{The notations used to denote the exotic fermions is as follows: While $\chi^Q$ denotes a particular component of the multiplet $\chi$ with electric charge $Q$, $N^Q$ includes all the components with electric charge $Q$ in a $N$-plet ($N=3,4,5$). For example, $\Delta^0$ denotes either of $\Delta^1$ and $\Delta^2$, whereas $4^0$ includes both of them.} as a function of the mass of the corresponding exotic. 
It can be seen from figure \ref{fig:DY_prod} that all the Drell-Yan production processes (both pair production and associated production) of the exotics have a sizeable cross section at the 13 TeV LHC.
In the case of pair production of doubly charged fermions, those from the quintuplet have the largest cross section, whereas in the case of pair production for singly charged fermions, those from the quadruplet have the largest cross section. Though the coupling strength of all the doubly (singly) charged fermions with the photon is equal, they couples differently with the $Z$-boson \footnote{The coupling of a fermion with electric charge $Q$ in a $N$-plet with the $Z$-boson is proportional to $T_3-Q\sin^2\theta_w$, where $T_3$ is the diagonal generator of $N$-representation of $SU(2)_L$, and hence different for identically charged exotics coming from different $SU(2)_L$ multiplets.}. This makes the cross sections different for one multiplet than others. The bottom left panel in figure \ref{fig:DY_prod} shows associated production cross sections. Production of doubly charged fermion in association with singly charged fermion from the quintuplet has the largest cross section, whereas, for the production of singly charged fermion in association with neutral fermion, the largest cross section corresponds to the quadruplet. This can be understood from their coupling to the $W$-boson. The $W$-boson couplings involving a doubly- and a singly- charged exotics are, respectively, $\sqrt{2},\sqrt{3}$ and $2$ for triplet, quadruplet and quintuplet; whereas those involving a neutral and a singly-charged exotics are $\sqrt{2},\sqrt{7}$ and $\sqrt{6}$, respectively, in units of $g/\sqrt{2}$.
Using these couplings, one can get the following relations among different associated production cross sections:
$\sigma_3(3^{++}3^-+3^{--}3^+)=2\sigma_4(4^{++}4^-+4^{--}4^+)/3=\sigma_5(5^{++}5^-+5^{--}5^+)/2=\sigma_3^\prime(3^+3^0+3^-3^0)=2\sigma_4^\prime(4^+4^0+4^-4^0)/7=\sigma_5^\prime(5^+5^0+5^-5^0)/3~$. While figure~\ref{fig:DY_prod} (bottom left panel) shows only $\sigma_3(3^{++}3^-+3^{--}3^+)$ as a function of triplet mass, one needs to multiply appropriate factors (discussed above) to obtain other associated production cross-sections. The bottom right panel in figure \ref{fig:DY_prod} shows pair production cross section for neutral fermions. Pair productions for neutral fermions are determined by their coupling strengths with $Z$-boson which are $-1$,$-1/2$ and $0$ in units of $g/\cos_{\theta_w}$, respectively, for triplet, quadrupled and quintuplet. This is why the pair production cross section for neutral fermions from quadruplet is 4 times smaller than that from triplet. Note that the neutral components of the quintuplet ($\Phi^0$) do not couple directly with the $Z$-boson or photon. Hence, they can not be produced profusely in pairs.
\begin{figure}[h]
\centering
\begin{tikzpicture}[scale=0.4]
\begin{feynman}
\vertex(a);
\vertex[below=of a] (b);
\vertex[above left=of a] (i1){\(\gamma\)};
\vertex[below left=of b] (i2){\(\gamma\)};
\vertex[above right=of a] (f1){\(f^{++}/f^{+}\)};
\vertex[below right=of b] (f2){\(f^{--}/f^{-}\)};
\diagram* {
(i1)-- [boson] (a),
(i2)-- [boson] (b),
(a)-- [fermion] (f1),
(b)-- [fermion] (f2),
(a)-- [fermion,edge label'=\(f^{--}/f^-\)] (b),
};
\end{feynman}
\end{tikzpicture}
\includegraphics[scale=0.6]{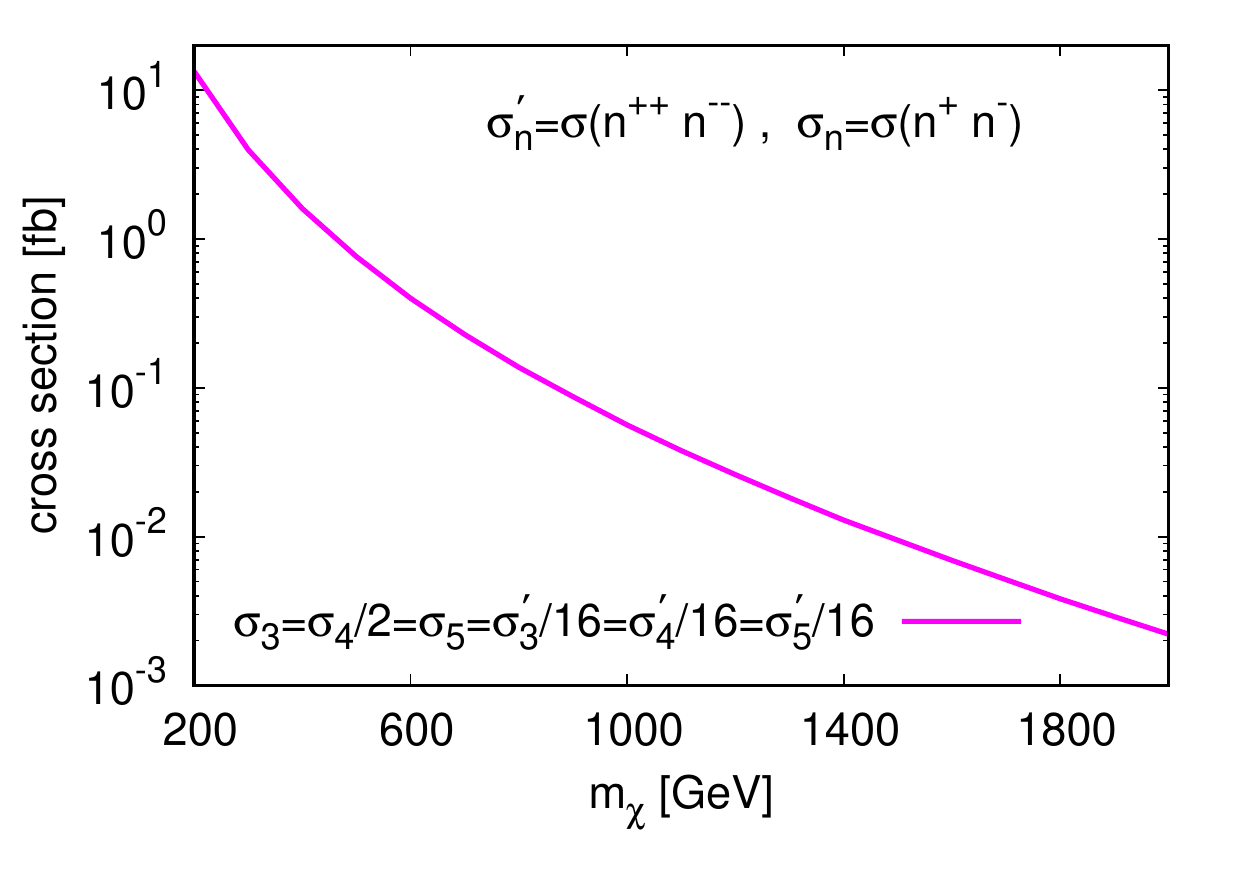}
\caption{\label{fig:photon_pair} Feynman diagram for photon initiated pair production of exotic charged fermions. To the right, cross section for photon initiated pair production of exotic charged fermions at 13 TeV LHC.}
\end{figure}
Figure \ref{fig:photon_pair} shows the cross sections for photon initiated pair productions of exotic charged fermions as a function of the mass of the corresponding exotic. Photon initiated pair production cross section of doubly charged fermions is $2^4=16$ times larger than that of the singly charged ones. Noting that the quadruplet contains two independent singly charged fermions, $\Delta^+$ and $\Delta^-$, the photo-fusion cross-sections of charged exotics of different $SU(2)_L$ multiplets satisfy following relations: $\sigma_3(3^+3^-)~=~\sigma_4(4^+4^-)/2~=~\sigma_5(5^+5^-)~=~\sigma_3^\prime(3^{++}3^{--})/16~=~\sigma_4^\prime(4^{++}4^{--})/16~=~\sigma_5^\prime(5^{++}5^{--})/16$. Figure \ref{fig:photon_pair} and \ref{fig:DY_prod} show that in comparison to Drell-Yan, photo-production contributes significantly (particularly, for the doubly charged fermions) for larger exotic masses, and hence can not be neglected.

In principle, the cross sections can be a discriminator between exotics resulting from different multiplets in this model provided that the masses of the exotics are known from kinematic observables. Likewise, a given multiplet in this model can be distinguished from a similar multiplet in another seesaw-inspired model. For example, the production cross-sections of triplet fermions in this model are different from the production cross sections of triplet fermions in the type-III seesaw model. Note that in type-III seesaw, the neutral component of triplet can not pair-produced at the LHC as it does not couple directly with the Z-boson or photon which is not the case for this model. Similarly, the neutral component of quintuplet in this model can not be pair produced copiously at the LHC which is not the case for the model studied in Ref.~\cite{kumericki} which employs a quintuplet with non-zero hypercharge. However, it is important to note that the LHC, being a hadron collider, cross-section measurements are always associated with large theoretical as well as experimental uncertainties.

\subsection{Decays of exotic fermions}
\label{sec:decay}
After discussing the productions at the LHC, this section will be devoted to study the decays of the exotic fermions. The mixings between the components of different multiplets allow the exotics of heavier multiplets to decay into the lighter ones in association with a SM boson ($W^{\pm},~Z$ or the Higgs boson). The other possible decays being kinematically forbidden, the components of lightest multiplets decay into the SM particles. The mixings among the SM leptons and exotic fermions being small, the decay widths of the exotics belonging to the lightest multiplet are usually suppressed. However, it will be the group of lightest exotics that will be dominantly produced at the LHC and will determine the collider signature of the scenario. Therefore, in this section, we are interested to study the decays of the exotic fermions belonging to the lightest multiplet.

The decays of the exotic fermions belonging to the lightest multiplet can be classified into two categories: {\it Category I:} the decays of the heavier exotics to the lighter ones; {\it Category II:} two-body decays  into a SM lepton and a boson ($W^\pm/Z/h$-boson). \\

\noindent {\it Category I:} The mass-splittings among the differently charged fermions of a given multiplet (except the mass-splitting between $\Delta^-$ and $\Delta^0$), being larger than the charged pion mass ($m_\pi$), open up decay channels like $\chi^Q \to \chi^{Q-1} \pi^+$ (and equivalently $\chi^{-Q} \to \chi^{-Q+1} \pi^-$) with $Q=2,1$. The decay rate for such single charged pion final state is given by \cite{cirelli,ibe,kumericki,mcdonald}
\begin{equation*}
\label{eq:dec_pion}
\Gamma (\chi^Q \to \chi^{Q-1} \pi^+) =\left(g^{W\chi}\right)^2 \frac{2}{\pi} G_F^2 |V_{ud}|^2 f_\pi^2 \Delta M^3 \sqrt{1-\frac{m_\pi^2}{\Delta M^2}}~,
\end{equation*}
where $\Delta M = m_{\chi^Q}-m_{\chi^{Q-1}}$, $|V_{ud}|=0.97420\pm 0.00021$ \cite{Vud}, $G_F$ is the Fermi coupling constant, $f_\pi \simeq 130.2 \pm 0.8$ MeV \cite{fpi} is the pion decay constant and $g^{W\chi}$ is the coupling with the W-boson in units of $g$. The couplings $g^{W\chi}$ can be extracted from eq. (\ref{eq:gauge_prod_3})-(\ref{eq:gauge_prod_5}) and they are
$\left|g^{W\Sigma}\right|=1,\frac{1}{\sqrt{2}} {\rm ~for~} Q=2,1;~$
$\left|g^{W\Delta}\right|=\sqrt{\frac{3}{2}},1,\frac{\sqrt{3}}{2} {\rm ~for~} Q=2,1,0~$ and $\left|g^{W\Phi}\right|=\sqrt{2},\sqrt{3} {\rm ~for~} Q=2,1, {\rm ~respectively~}.$
Since the mass-splittings between doubly charged exotics ($\chi^{++}$) and singly charged exotics ($\chi^+$) is bigger than the mass of charged kaon ($m_K$), the decays $\chi^{++} \to \chi^+ K^+$ are also kinematically allowed. One will have similar expression for the decay width of such single charged kaon decay mode as in the above equation with $f_\pi$, $m_\pi$ and $V_{ud}$ replaced by $f_K$, $m_K$ and $V_{us}$, respectively. But this decay mode is suppressed compared to the pion decay mode as $|V_{us}|=0.2250\pm 0.0027$ \cite{Vus} is much smaller than $|V_{ud}|$. Also, there are other kinematically allowed three-body decays like $\chi^Q \to \chi^{Q-1} \ell^+ \nu$ with $\ell=e,\mu$ and $\chi^Q \to \chi^{Q-1} \pi^+ \pi^0$. The decay width for the three-body decay to leptonic final state is given by \cite{kumericki,mcdonald}
\begin{equation*}
\Gamma (\chi^Q \to \chi^{Q-1} \ell^+ \nu) =\left(g^{W\chi}\right)^2 \frac{2}{15\pi^3} G_F^2 \Delta M^5 \sqrt{1-\frac{m_\ell^2}{\Delta M^2}} ~P\left(\frac{m_\ell}{\Delta M} \right)~,
\end{equation*}
where $m_\ell$ denotes mass of the charged lepton $\ell (=e,\mu)$, and the function $P(r)$ has the following form
\begin{equation*}
P(r)=1-\frac{9}{2}r^2-4r^4+\frac{15r^4}{2\sqrt{1-r^2}} {\rm tanh}^{-1} \sqrt{1-r^2}~.
\end{equation*}
The splitting $m_{\Sigma^{++}}-m_{\Sigma^+}$, being greater than the mass of the $\rho(770)$, opens up the $\rho$ resonance enhanced decay channel to two pion final states:
\begin{equation*}
\Gamma (\Sigma^{++} \to \Sigma^+ \pi^+ \pi^0) \simeq \Gamma (\Sigma^{++} \to \Sigma^+ \rho) = \frac{12}{\pi} G_F^2 |V_{ud}|^2 f_\rho^2 \Delta M^3 \sqrt{1-\frac{m_\rho^2}{\Delta M^2}} \left(1-\frac{4m_\pi^2}{\Delta M^2} \right)^{-1}~,
\end{equation*}
where $f_\rho \simeq 166$ MeV \cite{frho} is the $\rho$-meson decay constant, $m_\rho$ is the $\rho$-meson mass and $\Delta M \simeq 850$ MeV. Compared to the two body decay $\Sigma^{++} \to \Sigma^+ \pi^+$, the decay $\Sigma^{++} \to \Sigma^+ \pi^+ \pi^0$ shows an enhancement almost by a factor of $4$.
For some other cases, though the decays to two pions final state $\chi^Q \to \chi^{Q-1} \pi^+ \pi^0$ are kinematically allowed, the mass-splitting being smaller than the mass of the $\rho(770)$ resonance, those decays are suppressed by the three-body phase space.
The mass splitting between $\Delta^-$ and $\Delta^0$ being smaller than $m_\pi$, $m_K$ and $m_\mu$, the only kinematically accessible decay for $\Delta^-$ is $\Delta^- \to \Delta^0 e^- \nu$ but with a negligible decay width. Note that these kind of decays (transition between two differently charged heavy states belonging to the same multiplet) do not depend on any free parameter in the model, and the decay rates are completely determined by their mass-splitting and coupling with the W-boson. While these decay widths are always suppressed by the small mass-splitting and/or the three-body phase space, they become important when the SM two-body final state decay modes (which will be discussed in the next part) are also suppressed. All the approximate partial decay widths (in units of GeV) are listed below in table \ref{table:dec_heavier_lighter}.
\begin{table}[htb!]
\centering
\begin{tabular}{|c|c|c|c|}
\hline
$\chi$ & $\Gamma(\chi^{++} \to \chi^+ \pi^+)$ [GeV] & $\Gamma(\chi^{++} \to \chi^+ e^+ \nu)$ [GeV] & $\Gamma(\chi^{++} \to \chi^+ \mu^+ \nu)$ [GeV] \\
\hline
$\Sigma$ & $8.4\times 10^{-13}$ & $2.6\times 10^{-13}$ & $2.4\times 10^{-13}$ \\
$\Delta$ & $6.4\times 10^{-13}$ & $1.3\times 10^{-13}$ & $1.1\times 10^{-13}$ \\
$\Phi$ & $3.7\times 10^{-13}$ & $4.2\times 10^{-14}$ & $3.5\times 10^{-14}$ \\
\hline
\hline
$\chi$ & $\Gamma(\chi^+ \to \chi^0 \pi^+)$ [GeV] & $\Gamma(\chi^+ \to \chi^0 e^+ \nu)$ [GeV] & $\Gamma(\chi^+ \to \chi^0 \mu^+ \nu)$ [GeV] \\
\hline
$\Sigma$ & $8.3\times 10^{-14}$ & $9.1\times 10^{-15}$ & $7.4\times 10^{-15}$ \\
$\Delta$ & $5\times 10^{-14}$ & $2.7\times 10^{-15}$ & $1.7\times 10^{-15}$ \\
$\Phi$ & $1.2\times 10^{-14}$ & $2.5\times 10^{-16}$ & $3.4\times 10^{-17}$ \\
\hline
\end{tabular}
$$\Gamma (\Sigma^{++} \to \Sigma^+ \pi^+ \pi^0) \approx 3.3 \times 10^{-12} \rm{GeV}$$
\caption{\label{table:dec_heavier_lighter} Partial decay widths [in GeV] of the heavier charged exotics to the lighter ones and pion/lepton final state.}
\end{table}

\noindent {\it Category II}: Let us turn our discussion to the decays into the SM two-body final states. Except the decays  $\Sigma^+ \to \ell^+ h$ and $\Sigma^{1,2} \to \nu h$, all the decays are induced by the charged- and neutral-current interactions (listed in equations \eqref{eq:dec_CC_3}--\eqref{eq:dec_NC_5}) involving a heavy exotic fermion, a SM fermion and an EW gauge boson ($W$ or $Z$-boson). Note that the interactions in equations \eqref{eq:dec_CC_3}--\eqref{eq:dec_NC_5} are consequences of mixing among the SM and exotic fermions. Therefore, mixings\footnote{The (approximate) analytical results for the mixing matrices in the neutral and singly charged fermion sectors are presented in appendix~\ref{app:mass_matrices}.} in the fermionic sector play a important role in determining the branching ratios as well as the total decay widths. 

Before proceeding further, we define the following functions and variables for convenience: 
\begin{subequations}
\begin{align*}
f_1(r)&= (1-r^2)^2 (1+2r^2)~,~ f_2(r)=(1-r^2)^2~,~ w_\chi=\frac{m_W}{m_\chi},~z_\chi=\frac{m_Z}{m_\chi} \mathrm{~and~} h_\chi=\frac{m_h}{m_\chi}~.
\end{align*}
\end{subequations}
Let us first consider the pointlike decays of doubly charged fermions, $\chi^{++}$. Doubly charged fermions undergo 2-body decay into a SM charged lepton, $\ell^+$ ($\ell=e,\mu,\tau$) and $W^+$-boson. The partial 2-body decay widths\footnote{We only give the non-zero leading order approximate expression for the partial decay widths throughout this work. Although the approximate expressions for partial decay widths are obtained, we numerically evaluate the branching ratios using SPheno for the plots.} for the doubly-charged fermions resulting from different multiplets are given by
\begin{subequations}
\begin{align*}
\Gamma(\Sigma^{++}_i \to \ell^+_j W^+) &\approx \frac{g^2}{32\pi} \frac{M_\Sigma^3}{M_W^2} \left[ \left| \left(U^R_{\tilde{\Sigma}^+_L \ell}\right)_{ij} \right|^2 + \left| \left(U^L_{\tilde{\Sigma}^+_R \ell}\right)_{ij} \right|^2 \right] f_1(w_\Sigma)~,
\\
\Gamma(\Delta^{++}_i \to \ell^+_j W^+) &\approx \frac{g^2}{32\pi} \frac{M_\Delta^3}{M_W^2} \frac{3}{2} \left[ \left| \left( U^R_{\tilde{\Delta}^+_L \ell}\right)_{ij} \right|^2 + \left| \left( U^L_{\tilde{\Delta}^+_R \ell} \right)_{ij} \right|^2 \right] f_1(w_\Delta)~,
\\
\Gamma(\Phi^{++}_i \to \ell^+_j W^+) &\approx \frac{g^2}{32\pi} \frac{M_\Phi^3}{M_W^2} 2 \left[ \left| \left( U^R_{\Phi^-_R \ell} \right)_{ij} \right|^2 + \left| \left( U^L_{\tilde{\Phi}^+_R \ell} \right)_{ij} \right|^2 \right] f_1(w_\Phi)~,
\end{align*}
\end{subequations}
where 
$i~{\rm and}~j$ are the generation indices, and the relevant elements of the mixing matrices $U^L$ and $U^{R}$ are given in Appendix \ref{app:mass_matrices}. The model includes three generations of exotics ({\it i.e.,} $i=1,2,3$) with degenerate masses. It is not possible to distinguish among the different copies of a given multiplet at the LHC unless a big conspiracy makes these copies very different from one another. Therefore, in the context of the LHC, it is more realistic to consider the average branching ratios, ${\rm BR}_{\rm avg}\left(\sum_i\chi_i\to XY\right)$, of three copies instead of branching ratios, ${\rm BR}\left(\chi_i\to XY\right)$, for individual copies, where $XY$ is a generic decay mode of $\chi_i$'s, and ${\rm BR}_{\rm avg}\left(\sum_i\chi_i\to XY\right)$ is defined as
\begin{equation}
  {\rm BR}_{\rm avg}\left(\sum_i\chi_i\to XY\right)~=~\frac{1}{3}\,\sum_{i=1}^3\,{\rm BR}\left(\chi_i\to XY\right)~=~\frac{1}{3}\,\sum_{i=1}^3 \frac{\Gamma\left(\chi_i\to XY\right)}{\Gamma_{\rm TOT}\left(\chi_i\right)},
\end{equation}
where $\Gamma_{\rm TOT}\left(\chi_i\right)~=~\sum_{X,Y}\Gamma\left(\chi_i\to XY\right)$ is the total decay width of $\chi_i$. The particular structure ({\it i.e.,} $U^L_{\tilde{\Sigma}^+_R \ell}\left(U^L_{\tilde{\Sigma}^+_R \ell}\right)^{\dagger} \propto \hat m_\nu$) of individual $3\times 3$ components of the mixing matrices in the simplified scenario ensures lepton flavour universality of the average branching ratios {\it i.e.,} 
\begin{eqnarray}
BR_{\rm avg}\left(\sum_i\chi_i^{++} \to e^+ W^+\right)=BR_{\rm avg}\left(\sum_i\chi_i^{++} \to \mu^+ W^+\right)&=BR_{\rm avg}\left(\sum_i\chi_i^{++} \to \tau^+ W^+\right)\nonumber\\ &\approx 33.33\%~,\nonumber
\end{eqnarray}
as long as $e^+W^+,~\mu^+W^+$ and $\tau^+W^+$ are the dominant decay modes for $\chi_i^{++}$ {\it i.e.,} $\Gamma_{\rm TOT}\left(\chi_i^{++}\right)\approx \sum_{j}\Gamma\left(\chi_i^{++}\to l_j^+ W^+\right)$. In the simplified scenario described in section \ref{subsec:simplified_model}, the total (two-body) decay width of $\Sigma_i^{++}$ can be approximated as
\begin{equation}
\label{eq:tot_dec_width_sigma}
\Gamma_{\rm TOT}\left(\Sigma_i^{++}\right)\approx \sum_{j}\Gamma\left(\Sigma^{++}_i\to l_j^+ W^+\right) \approx \frac{3g^6}{128 \pi}\frac{m_\Sigma^3 m_\Delta^2 m_\Phi}{y_{34}^{\prime 2}~y_{45}^{\prime 2}m_W^6}\,m_i~,
\end{equation}
where $m_i$ is the mass of $i^{\rm th}$ light neutrino. Using the above equation and the values quoted in Table~\ref{table:dec_heavier_lighter}, one can estimate the scale of $m_1(m_3)$ at which the lepton flavour universality of the average branching ratios of $\Sigma^{++}$'s breaks down for a given set of $m_\Sigma, m_\Delta, m_\Phi, y^\prime_{34}$ and $y^\prime_{45}$. For $m_\Sigma,~m_\Delta,~m_\Phi \sim 1$ TeV and $y_{34}^\prime,~y_{45}^\prime\sim 0.1$\footnote{We use this benchmark point ($m_\Sigma,~m_\Delta,~m_\Phi \sim 1$ TeV and $y_{34}^\prime,~y_{45}^\prime\sim 0.1$) throughout this section. For a different choice of parameters, the very nature of the average branching ratios plots will remain same.}, $\Gamma_{\rm TOT} \sim \mathcal{O}(10^7)\,m_i$. Therefore, $\Sigma_i^{++} \to l^+_jW^+$ decay modes dominate over the decay modes into pion(s)/leptons and $\Sigma_i^{+}$ (see Table~\ref{table:dec_heavier_lighter}) for $m_i> 10^{-9}$ eV. Since the lightest neutrino mass, $m_1(m_3)$, is a free parameter for NH(IH), $\Sigma_{1(3)}^{++}$ dominantly decays to pion(s)/leptons and $\Sigma_i^{+}$ for $m_i< 10^{-10}$ eV breaking down the lepton flavour universality of the average leptonic branching ratios of $\Sigma^{++}$'s.
\begin{figure}[htb!]
\includegraphics[scale=0.6]{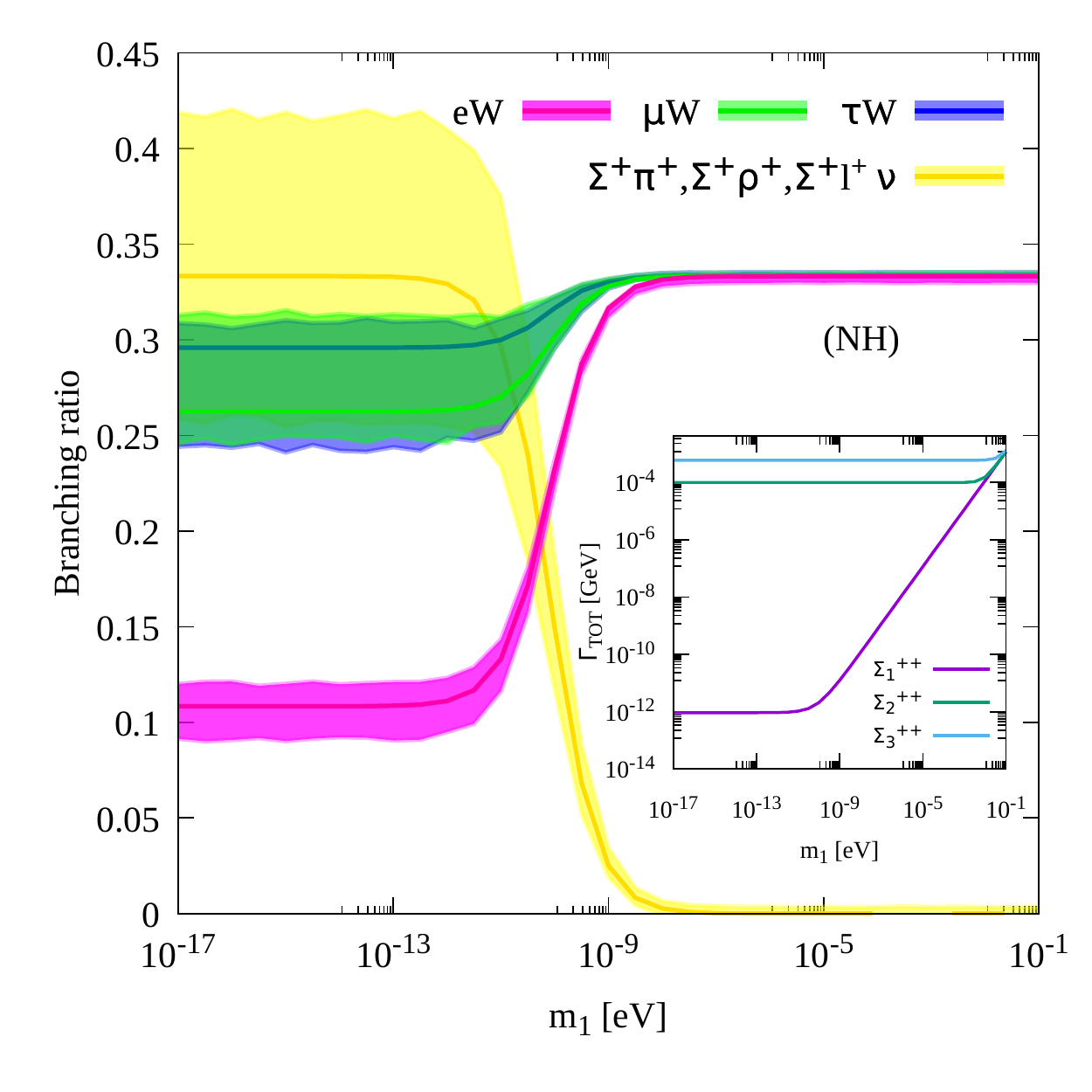}
\includegraphics[scale=0.6]{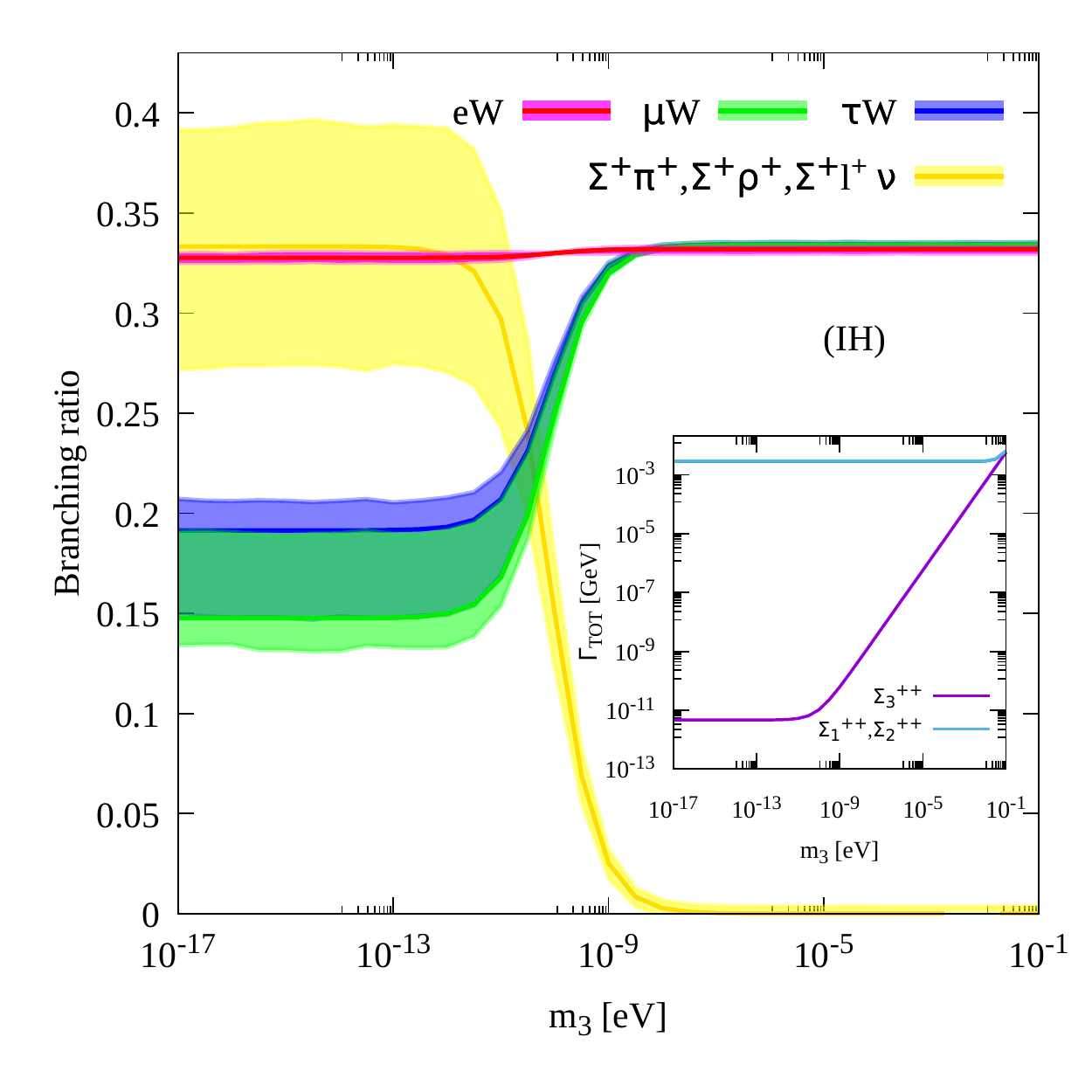}
\caption{\label{fig:BR_doubly} Average branching ratios for the decay modes of $\Sigma^{++}$ as a function of $m_1(m_3)$ for NH(IH). The total decay width ($\Gamma_{\rm TOT}$) of $\Sigma^{++}_i$'s are presented as a function of $m_1(m_3)$ for NH(IH) in the inset. See text for details.}
\end{figure}
Figure \ref{fig:BR_doubly} shows average branching ratio for the different decay modes of $\Sigma^{++}$'s as a function of the lightest neutrino mass, $m_1 (m_3)$, for NH (IH) light neutrino mass spectrum. Uncertainty in the measured neutrino oscillation parameters has been reflected in the branching ratio plots as bands (the solid line within the bands correspond to the best fit values of the neutrino oscillation parameters).
In the inset of Fig.~\ref{fig:BR_doubly}, the total decay width ($\Gamma_{\rm TOT}$) \footnote{The total decay widths ($\Gamma_{\rm TOT}$), shown in the inset of Fig.~\ref{fig:BR_doubly}, include contribution from both decay categories -- transition between two heavy states and SM two-body final state decays.} of $\Sigma^{++}_i$'s are presented as a function of $m_1(m_3)$ for NH(IH). One will have similar figures as Figure \ref{fig:BR_doubly} for $\Delta^{++}$'s and $\Phi^{++}$'s except for the fact that the lepton flavour universality of the average branching ratios of $\Delta^{++}$'s and $\Phi^{++}$'s break down below $m_1(m_3) \sim 10^{-7}$ and $10^{-4}$ eV, respectively. We avert to show the average branching ratio plots for $\Delta^{++}$'s and $\Phi^{++}$'s for brevity.

\begin{figure}[htb!]
\includegraphics[scale=0.6]{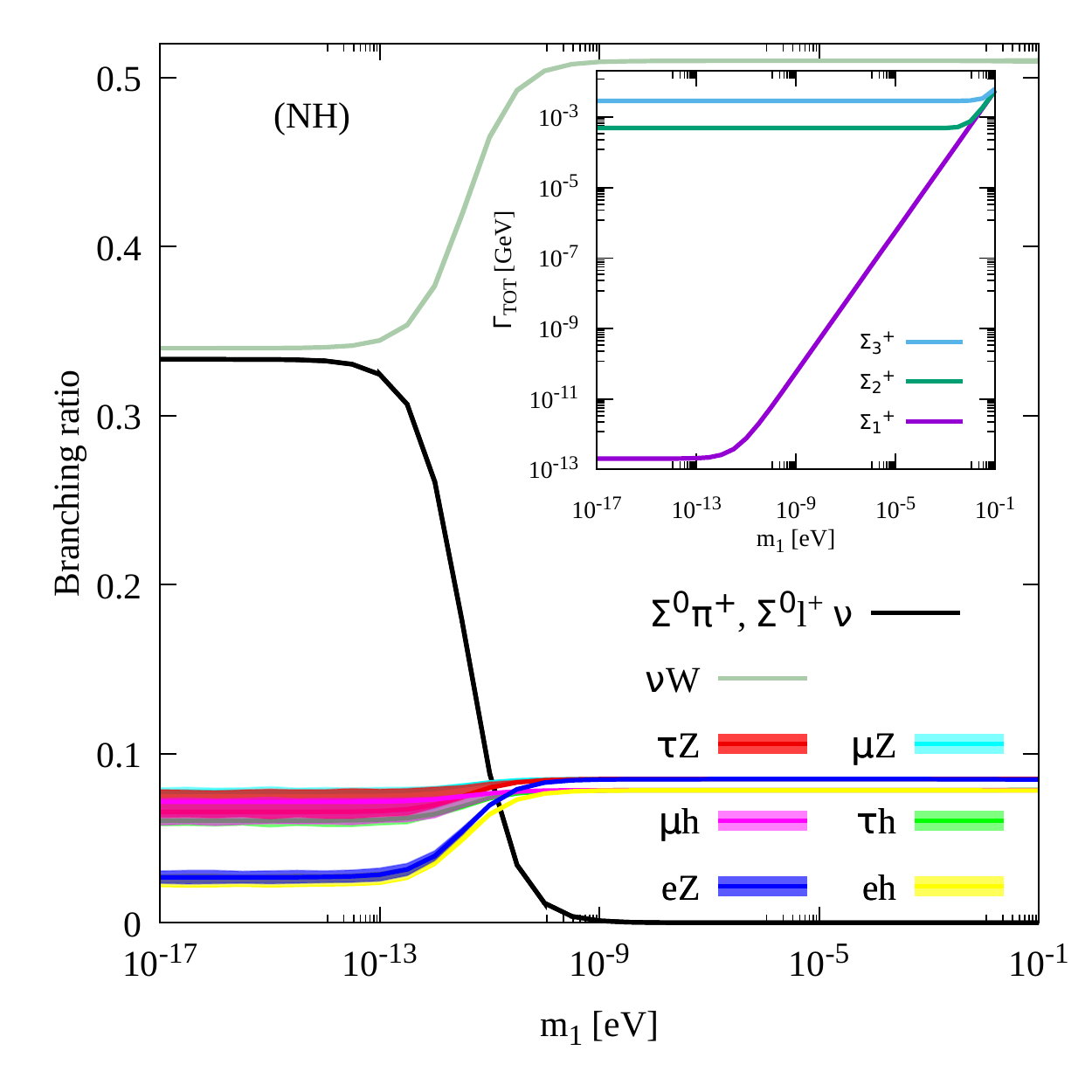}
\includegraphics[scale=0.6]{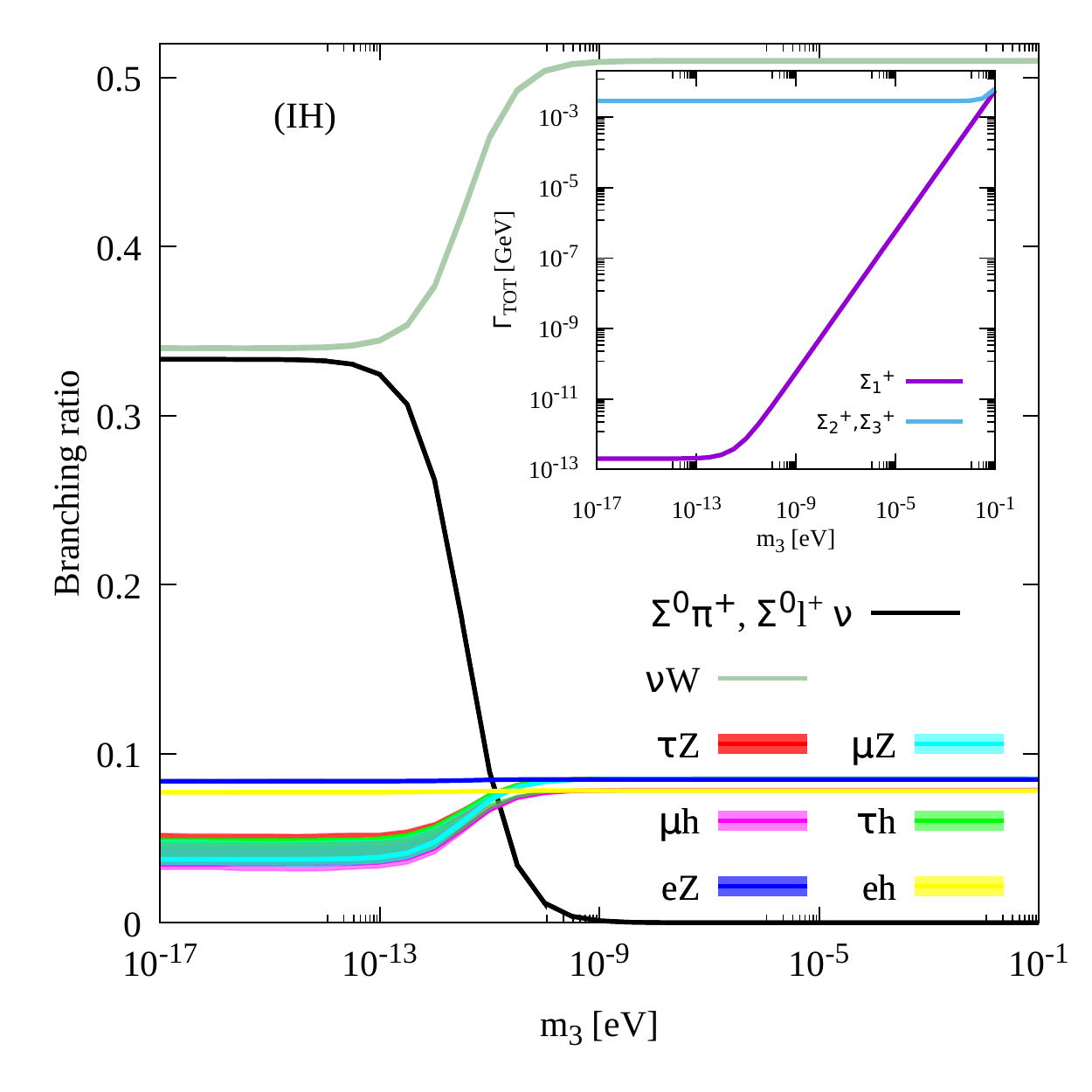}
\caption{\label{fig:BR_singly_3} Average branching ratios for the decay modes of $\Sigma^{+}$ as a function of $m_1(m_3)$ for NH(IH). The total decay width ($\Gamma_{\rm TOT}$) of $\Sigma^{+}_i$'s are presented as a function of $m_1(m_3)$ for NH(IH) in the inset.}
\end{figure}

The singly-charged exotic fermions can decay into  $\nu_j W^\pm$, $\ell^\pm Z$ or $\ell^\pm h$. The partial decay widths \footnote{In the simplified scenario, the total (two-body) decay width of differently charged components of triplet are equal and is given by Eq.~\eqref{eq:tot_dec_width_sigma}.} for the singly-charged triplet fermions ($\Sigma^+_i$) are as follows:
\begin{subequations}
\begin{align*}
\Gamma(\Sigma^+_i \to \sum_j \nu_j W^+) &\approx \frac{g^2}{32\pi} \frac{m_\Sigma^3}{m_W^2} \sum_j \left[ \left| \left( U^0_{\Sigma^0_L \nu_L} \right)_{ij} \right|^2 + \left| \left( U^0_{\tilde{\Sigma}^0_R \nu_L}+\frac{1}{\sqrt{2}} U^L_{\tilde{\Sigma}^+_R \ell} \right)_{ij} \right|^2 \right] f_1(w_\Sigma)~,
\\
\Gamma(\Sigma^+_i \to \ell^+_j Z) &\approx \frac{g^2}{32\pi} \frac{m_\Sigma^3}{m_W^2} \frac{1}{4} \left| \left( U^L_{\tilde{\Sigma}^+_R \ell} \right)_{ij} \right|^2 f_1(z_\Sigma)~,
\\
\Gamma(\Sigma^+_i \to \ell^+_j h)&\approx \frac{m_\Sigma}{32\pi} \frac{1}{2} \left[ \left| \left( U^{L*}_{\tilde{\Sigma}^+_R \ell} Y_\ell \right)_{ij} \right|^2 + \frac{1}{2} \left| \left( Y_{23} \right)_{ij} \right|^2 \right] f_2(h_\Sigma)~.
\end{align*}
\end{subequations}
Figure \ref{fig:BR_singly_3} shows average branching ratio for the different decay modes of $\Sigma^{+}$'s as a function of the lightest neutrino mass, $m_1 (m_3)$, for NH (IH) light neutrino mass spectrum. As long as $\Sigma^+$'s decay dominantly into the SM two-body final states, the average leptonic branching ratios of $\Sigma^{+}$'s show lepton flavour universality with $\sum_j {\rm BR_{avg}}(\Sigma^+_i \to \nu_j W^+) \approx  50\%$, ${\rm BR_{avg}}(\Sigma^+_i \to l^+ Z)\approx \frac{25}{3}\%$ and ${\rm BR_{avg}}(\Sigma^+_i \to l^+ h) \approx \frac{25}{3}\%$ where $l \ni e,~\mu$ and $\tau$. The lepton flavour universality of the average leptonic branching ratios of $\Sigma^{+}$'s breaks down below $m_1(m_3) \sim 10^{-10}$ eV.
The partial decay widths for the singly-charged quadruplet fermions ($\Delta^+$ and $\Delta^-$)\footnote{Note that quadruplet contains two independent singly charged particles namely, $\Delta^+$ and $\Delta^-$.} and singly-charged quintuplet fermion ($\Phi^+$) are given in the following:
\begin{subequations}
\begin{align*}
\Gamma(\Delta^+_i \to \sum_j \nu_j W^+) &\approx \frac{g^2}{32\pi} \frac{m_\Delta^3}{m_W^2} 2 \sum_j \left[ \left| \left( U^0_{\Delta^0_L \nu_L} \right)_{ij} \right|^2 + \left| \left( U^0_{\tilde{\Delta}^0_R \nu_L}+\frac{1}{2} U^L_{\tilde{\Delta}^+_R \ell} \right)_{ij} \right|^2 \right] f_1(w_\Delta)~,
\\
\Gamma(\Delta^+_i \to \ell^+_j Z) &\approx \frac{g^2}{32\pi} \frac{m_\Delta^3}{m_W^2} \frac{1}{4}  \left| \left( U^R_{\tilde{\Delta}^+_L \ell} \right)_{ij} \right|^2 f_1(z_\Delta)~,
\\
\Gamma(\Delta^+_i \to \ell^+_j h)&\approx \frac{m_\Delta}{32\pi} \left[ \left| \left( \frac{y_{34}}{\sqrt{3}}U^{R*}_{\tilde{\Sigma}^+_L \ell} + \frac{1}{\sqrt{2}}U^{L*}_{\tilde{\Delta}^+_R \ell} Y_\ell \right)_{ij} \right|^2 + \frac{1}{3} \left| \left( y_{34}^\prime U^L_{\tilde{\Sigma}^+_R \ell} \right)_{ij} \right|^2 \right] f_2(h_\Delta)~,
\end{align*}
\end{subequations}
\begin{subequations}
\begin{align*}
\Gamma(\Delta^-_i \to \sum_j \nu_j W^-) &\approx \frac{g^2}{32\pi} \frac{m_\Delta^3}{m_W^2} \frac{3}{2} \sum_j \left[ \left| \left( U^0_{\Delta^0_L \nu_L}+\frac{1}{\sqrt{3}} U^L_{\Delta^-_L \ell} \right)_{ij} \right|^2 + \left| \left( U^0_{\tilde{\Delta}^0_R \nu_L} \right)_{ij} \right|^2 \right] f_1(w_\Delta)~,
\\
\Gamma(\Delta^-_i \to \ell^-_j Z) &\approx \frac{g^2}{32\pi} \frac{m_\Delta^3}{m_W^2} \left| \left( U^L_{\Delta^-_L \ell} \right)_{ij} \right|^2 f_1(z_\Delta)~,
\\
\Gamma(\Delta^-_i \to \ell^-_j h) &\approx \frac{m_\Delta}{32\pi} \frac{1}{8} \left[ \left| \left( m_{45}^\prime U^L_{\tilde{\Phi}^+_R \ell} \right)_{ij} \right|^2 + \left| \left( -2U^L_{\Delta^-_L \ell}Y_\ell +m_{45} U^R_{\Phi^-_R \ell} \right)_{ij} \right|^2 \right] f_2(h_\Delta)~.
\end{align*}
\end{subequations}
\begin{subequations}
\begin{align*}
\Gamma(\Phi^+_i \to \sum_j \nu_j W^+) &\approx \frac{g^2}{32\pi} \frac{m_\Phi^3}{m_W^2} \sum_j \left[ \left| \left( \sqrt{3} U^0_{\tilde{\Phi}^0_R \nu_L} \right)_{ij} \right|^2 + \left| \left( \sqrt{3} U^0_{\tilde{\Phi}^0_R \nu_L}+\frac{1}{\sqrt{2}} U^L_{\tilde{\Phi}^+_R \ell} \right)_{ij} \right|^2 \right]  f_1(w_\Phi)~,
\\
\Gamma(\Phi^+_i \to \ell^+_j Z) &\approx \frac{g^2}{32\pi} \frac{m_\Phi^3}{m_W^2} \frac{1}{4} \left| \left( U^L_{\tilde{\Phi}^+_R \ell} \right)_{ij} \right|^2 f_1(z_\Phi)~,
\\
\Gamma(\Phi^+_i \to \ell^+_j h) &\approx \frac{m_\Phi}{32\pi} \frac{3}{8} \left[ \left| \left( -\sqrt{\frac{4}{3}}U^{L*}_{\tilde{\Phi}^+_R \ell}Y_\ell +m_{45}^\dagger U^R_{\tilde{\Delta}^+_L \ell} \right)_{ij} \right|^2 + \left| \left( m_{45}^\dagger U^L_{\tilde{\Delta}^+_R \ell} \right)_{ij} \right|^2 \right] f_2(h_\Phi)~.
\end{align*}
\end{subequations}
As long as $\Delta^+,~\Delta^-$'s and $\Phi^+$'s decay dominantly into the SM two-body final states, their average branching ratios are given by:
$\sum_j {\rm BR_{avg}}(\Delta^\pm_i \to \nu_j W^\pm) \approx  67\%$,
${\rm BR_{avg}}(\Delta^\pm_i \to l^\pm Z) \approx 0.6\%$,
${\rm BR_{avg}}(\Delta^\pm_i \to l^\pm h) \approx 10\%$,
$\sum_j {\rm BR_{avg}}(\Phi^+_i \to \nu_j W^+) \approx 50\%$,
${\rm BR_{avg}}(\Phi^+_i \to l^+ Z)\approx 0.0\%$ and 
${\rm BR_{avg}}(\Phi^+_i \to l^+ h)\approx \frac{50}{3}\%$ where $l\ni e,~\mu,~{\rm and}~\tau$.
One will have similar figures as Figure \ref{fig:BR_singly_3} for $\Delta^{+},~\Delta^{-}$'s and $\Phi^{+}$'s except for the fact that the lepton flavour universality of the average branching ratios of $\Delta^{\pm}$'s and $\Phi^{+}$'s break down below $m_1(m_3) \sim 10^{-8}$ and $10^{-6}$ eV respectively. For brevity, we do not show the average branching plots for $\Delta^{+},~\Delta^{-}$'s and $\Phi^{+}$'s.

Now, we consider the decays of neutral fermions. The possible decay modes for the neutral exotics are $\ell^\pm W^\mp$, $\nu_j Z$ and $\nu_j h$. Here we list the partial decay widths of neutral triplets ($\Sigma^{1,2}_i$) and quadruplets ($\Delta^{1,2}_i$) in the following:
\begin{subequations}
\begin{align*}
&\Gamma(\Sigma^1_i\to \ell^\pm_j W^\mp) \approx \Gamma(\Sigma^2_i \to \ell^\pm_j W^\mp) \approx \frac{g^2}{32\pi} \frac{m_\Sigma^3}{m_W^2} \frac{1}{2} \left[ \left| \left( U^L_{\tilde{\Sigma}^+_R \ell} + \frac{1}{\sqrt{2}}U^0_{\tilde{\Sigma}^0_R \nu_L} \right)_{ij} \right|^2 + \left| \left( U^R_{\tilde{\Sigma}^+_L \ell} \right)_{ij} \right|^2 \right] f_1(w_\Sigma)~,
\\
&\Gamma(\Sigma^1_i\to \sum_j \nu_j Z) \approx \Gamma(\Sigma^2_i \to \sum_j \nu_j Z) \approx \frac{g^2}{32\pi} \frac{m_\Sigma^3}{m_W^2} \frac{1}{4} \sum_j \left| \left( U^0_{\tilde{\Sigma}^0_R \nu_L} \right)_{ij} \right|^2 f_1(z_\Sigma)~,
\\
&\Gamma(\Sigma^1_i\to \sum_j \nu_j h) \approx \Gamma(\Sigma^2_i \to \sum_j \nu_j h) \approx \frac{m_\Sigma}{32\pi} \frac{1}{2} \sum_j \left| \left( y_{23} \right)_{ij} \right|^2 f_2(h_\Sigma)~.
\\
&\Gamma(\Delta^1_i \to \ell^\pm_j W^\mp) = \Gamma(\Delta^2_i \to \ell^\pm_j W^\mp) \approx \frac{g^2}{32\pi} \frac{m_\Delta^3}{m_W^2} \left[ \left| \left( U^L_{\tilde{\Delta}^+_R \ell} +\frac{1}{2}U^0_{\tilde{\Delta}^0_R \nu_L} \right)_{ij} \right|^2 + \left| \left( U^R_{\tilde{\Delta}^+_L \ell} \right)_{ij} \right|^2 \right] f_1(w_\Delta)~,
\\
&\Gamma(\Delta^1_i \to \sum_j \nu_j Z) = \Gamma(\Delta^2_i \to \sum_j \nu_j Z) \approx \frac{g^2}{32\pi} \frac{m_\Delta^3}{m_W^2} \frac{1}{4} \sum_j \left| \left(U^0_{\Delta^0_L \nu_L} \right)_{ij} \right|^2 f_1(z_\Delta)~,
\\
&\Gamma(\Delta^1_i \to \sum_j \nu_j h)\approx \Gamma(\Delta^2_i \to \sum_j \nu_j h)=\frac{m_\Delta}{32\pi} \frac{1}{6} \sum_j \left| \left( y_{34}^{\prime *} U^0_{\tilde{\Sigma}^0_R \nu_L} \right)_{ij} \right|^2 f_2(h_\Delta)~.
\end{align*}
\end{subequations} 
$\Sigma^{1,2}$s have equal average branching ratios to the $Z$- and $h$-boson decay modes --- $50\%$ each, whereas the W-boson decay mode has negligible branching ratio\footnote{The reason for negligible branching ratio of the W-boson decay mode is as follows. The corresponding vertex factor gets dominant contributions via mixing from two terms --- the usual SM CC interaction, $\frac{g}{2} \overline{\nu}\gamma^\mu P_L \ell W^+_\mu$, and the BSM CC interaction, $g\overline{\Sigma^+}\gamma^\mu \left(P_R \Sigma^0_c +P_L \Sigma^0 \right) W^+_\mu$. These two contributions exactly cancel each other at leading order ($\mathcal{O}(v/M_\Sigma)$). This can be understood simply by looking at the hypercharge assignments. $\Sigma$, $W_\mu$, $H$ and $\ell$ have hypercharge of $1,0,1/2$ and $-1/2$, respectively. Thus, to connect those fields in a gauge invariant way, one needs at least three units of $H$ (equivalently $H$ induced masses)  making said vertex of $\mathcal{O}(v^3/M^3_\Sigma)$. This explains the negligible branching ratio for the W-boson decay mode. Note that this is not the case for type-III seesaw model. Because, in type-III seesaw model, the triplet field has a hypercharge of $0$, and thus one needs just one unit of $H$ to connect the those fields in a gauge invariant way making the corresponding vertex of $\mathcal{O}(v/M_\Sigma)$.}. Neutral quadruplets, $\Delta^{1,2}$'s, have the following average branching ratios for all values of $m_1(m_3)$:
$\sum_i {\rm BR_{avg}}(\Delta^{1,2} \to \nu_i Z) \approx 4\%,~ \sum_i {\rm  BR_{avg}}(\Delta^{1,2} \to \nu_i h) \approx 73\%$ and 
${\rm BR_{avg}}(\Delta^{1,2} \to l^\pm W^\mp) \approx 8\%$. Finally, the partial decay widths of neutral quintuplet fermions are procured below:
\begin{subequations}
\begin{align*}
\Gamma(\Phi^0_i \to \ell^\pm_j W^\mp) &\approx \frac{g^2}{32\pi} \frac{m_\Phi^3}{m_W^2} \left[ \left| \left( \sqrt{3}U^L_{\tilde{\Phi}^+_R \ell} +\frac{1}{\sqrt{2}}U^0_{\tilde{\Phi}^0_R \nu_L} \right)_{ij} \right|^2 + \left| \left( \sqrt{3}U^R_{\Phi^-_R \ell} \right)_{ij} \right|^2 \right] f_1(w_\Phi)~,
\\
\Gamma(\Phi^0_i \to \sum_j \nu_j Z) &\approx \frac{g^2}{32\pi} \frac{m_\Phi^3}{m_W^2} \frac{1}{2} \sum_j \left| \left( U^0_{\tilde{\Phi}^0_R \nu_L} \right)_{ij} \right|^2 f_1(z_\Phi)~,
\\
\Gamma(\Phi^0_i \to \sum_j \nu_j h)&\approx \frac{m_\Phi}{32\pi} \frac{1}{2} \sum_j \left| \left( y_{45}^{\prime T} U^0_{\tilde{\Delta}^0_R \nu_L} \right)_{ij} \right|^2 f_2(h_\Phi)~.
\end{align*}
\end{subequations}
$\Phi^0$'s have the following average branching ratios for all values of $m_1(m_3)$:
$\sum_i {\rm BR_{avg}}(\Phi^0 \to \nu_i Z) \approx 0.4\%,~ \sum_i BR(\Phi^0 \to \nu_i h) \approx 70\%$,
${\rm BR_{avg}}(\Phi^0 \to l^\pm W^\mp)\approx 10\%$.

Note that as the lightest neutrino mass ($m_1$ or $m_3$) approaches towards zero, SM two-body decay modes for the heavy charged fermions become insignificant and the heavy state transitions become the dominant ones. Unlike the SM two-body  decay modes which depend on the free parameters of the model, and hence on neutrino mass spectrum, the decay rates for the heavy state transitions are completely determined by their coupling with the $W$-boson and the mass-splitting between the differently charged components belonging to a particular multiplet. These mass-splittings are almost independent of the heavy lepton bare masses for the mass range of our interest ($m_{\Sigma,\Delta,\Phi} >> m_{W,Z,h}$) making the total decay rates almost constant as $m_1(m_3) \to 0$. Unlike the heavy charged fermions, the heavy neutral fermions, being the lightest component of a multiplet, can not decay into  pions/leptons in association with another component of the same multiplet. This is why, as $m_1(m_3) \to 0$, the decay length of the first (third) generation of heavy neutral fermions can be arbitrarily large making them long-lived having disappearing track signatures
or displaced vertex at detector. This has been discussed in Section \ref{sec:displaced}.

 \subsection{Collider Searches}
\label{sec:collider_signals}

After discussing the production and decays of the exotic fermions in the last two sections, we are now equipped enough to discuss the signatures of these exotics at the LHC with 13 TeV center of mass energy. The signatures of the exotic fermions at the LHC can be broadly categorized into two classes depending on their total decay widths. While larger value for the lightest neutrino mass ($m_1(m_3)$ for NH(IH)) ensures prompt decays of all exotics into SM leptons and bosons and hence, give rise to multilepton final states at the LHC, interesting signatures like displaced vertex, vanishing charge tracks {\it, etc.,} arise from some of the exotics which could be long-lived for smaller $m_1(m_3)$ for NH(IH). First, we will discuss the prompt decay signatures in detail in the following.

The pair production of the exotic fermions and their subsequent prompt decays at the LHC lead to a variety of final state signatures including a pair of SM leptons ($\ell,\nu$) in association with different combinations of SM bosons ($W,Z,h$) pair. All possible final state signatures arising from the pair production and subsequent decays of exotic fermions at the LHC have enlisted in table \ref{table:signatures}. The list of possible final states in table~\ref{table:signatures} includes smoking gun signatures like di-Higgs, mono-Higgs, di-gauge boson, {\it etc.} in association with multiple leptons. Note that some of the final states\footnote{The kinematic reconstruction of the invariant masses are possible for the final states which do not include missing neutrinos. For example, the pair production of the doubly-charged fermions gives rise to a pair of $W^\pm$-boson in association with a pair of charged SM lepton in the final state (see table~\ref{table:signatures} first column). The subsequent decay of one of the $W^\pm$-bosons into visible final states ({\it i.e.,} $W^{\pm}\to$ quark-antiquark, $q\bar q^\prime$ pairs) gives rise to smoking-gun signatures like an invariant mass peak in the $W$-jet charged-lepton invariant mass distribution, and hence allow us to reconstruct the masses of the doubly charged fermions. Note that such reconstructions usually require large statistics.} also allow kinematic reconstructions of the masses of the doubly-charged, singly-charged and neutral exotic fermions in the theory. While these final states are very common to a large class of modified type-III seesaw models or seesaw-inspired models \cite{aguila,aguila2,li}, the rate of production of a given final state as well as the characteristic phase-space distributions are in general different for different models. Therefore, each model requires a dedicated collider study for its own sake. Note that there has been a number of LHC searches based on different final state topologies dedicated to type-III seesaw only \cite{cms_multilepton_137,lhc_1,lhc_2,lhc_3,lhc_4}.

\begin{table}[htb!]
\huge
\centering
\scalebox{0.45}{
\begin{tabular}{|c||c|ccc|ccc|} 
\hline
& $\bm{\chi^{++} \to \ell^+ W^+}$ & \multicolumn{3}{c|}{$\bm{\chi^{+} \to \nu W^+}$, $\bm{\ell^+ Z}$, $\bm{\ell^+ h}$} & \multicolumn{3}{c|}{$\bm{\chi^0 \to \ell^\pm W^\mp}$, $\bm{\nu Z}$, $\bm{\nu h}$} \\
\hline
\hline
$\bm{\chi^{--} \to \ell^- W^-}$ & $\ell^- \ell^+ W^- W^+$ & $\ell^- \nu W^- W^+$ & $\ell^- \ell^+ W^- Z$ & $\ell^- \ell^+ W^- h$ & -- & -- & -- \\
\hline
$\bm{\chi^{-} \to \nu W^-}$ & $\nu \ell^+ W^- W^+ $ & $\nu \nu W^- W^+ $ & $\nu \ell^+ W^- Z $ & $\nu \ell^+ W^- h $ & $\nu \ell^\pm W^- W^\mp$ & $\nu \nu W^- Z $ & $\nu \nu W^- h $\\
$\bm{\chi^{-}  \to \ell^- Z}$ & $\ell^- \ell^+ Z W^+$ & $\ell^- \nu Z W^+$ & $\ell^- \ell^+ Z Z$ & $\ell^- \ell^+ Z h$ & $\ell^- \ell^\pm Z W^\mp$ & $\ell^- \nu Z Z$ & $\ell^- \nu Z h$ \\
$\bm{\chi^{-} \to \ell^- h}$ & $\ell^- \ell^+ h W^+$ & $\ell^- \nu h W^+ $ & $\ell^- \ell^+ h Z $ & $\ell^- \ell^+ h h$ & $\ell^- \ell^\pm h W^\mp$ & $\ell^- \nu h Z $ & $\ell^- \nu h h$ \\
\hline
$\bm{\chi^0 \to \ell^\pm W^\mp}$ & -- & $\ell^\pm \nu W^\mp W^+$ & $\ell^\pm \ell^+ W^\mp Z$ & $\ell^\pm \ell^+ W^\mp h$ & $\ell^\pm \ell^{\pm(\mp)} W^\mp W^{\mp(\pm)}$ & $\ell^\pm \nu W^\mp Z$ & $\ell^\pm \nu W^\mp h$
\\
$\bm{\chi^0 \to \nu Z}$& -- & $\nu \nu Z W^+$ & $\nu \ell^+ Z Z$ & $\nu \ell^+ Z h$ & $\nu \ell^\pm Z W^\mp$ & $\nu \nu Z Z$ & $\nu \nu Z h$ \\
$\bm{\chi^0 \to \nu h}$& --& $\nu \nu h W^+$ & $\nu \ell^+ h Z$ & $\nu \ell^+ h h$ & $\nu \ell^\pm h W^\mp$ &$\nu \nu h Z$ & $\nu \nu h h$ \\
\hline
\end{tabular}
}
\caption{Decays of exotic fermions to SM particles along with the possible final state signatures of pair/associated productions.} 
\label{table:signatures}
\end{table}

Although in our model, there are various final state signatures (listed in table~\ref{table:signatures}) which are interesting in their own rights, here, we emphasize only on multilepton (three or more leptons) final states. Without focusing on a particular decay channel of a particular exotic, we studied multilepton final states resulting from all possible productions and decays of the exotics present in the model. Note that all the final states in table~\ref{table:signatures} contribute to the multilepton signatures with different lepton multiplicities\footnote{Lepton multiplicities may vary between 0--10. Depending on the decay modes, pair productions of singly-charged exotics can give rise to 0 as well as 10 lepton multiplicity final states. Zero lepton multiplicity final state arises when $\chi^\pm \to \nu W^\pm$ followed by hadronic decays of $W^\pm$-boson. Whereas, $\chi^\pm \to l^\pm h$ followed by $h\to ZZ^* \to 4l$ results into 10 lepton final state. However, the rates of such high lepton multiplicity final states are usually suppressed by various branching ratios. Therefore, in this work, we consider all possible final states with three or more leptons.}. While, the pair productions and subsequent decays of singly-charged exotics ($\chi^\pm$) into a $l^\pm Z$ pair followed by the leptonic decays of the $Z$-boson give rise to 6-lepton final state ($pp \to \chi^- \chi^+ \to \ell^- \ell^+ Z Z \rm{~with~} Z\to \ell^- \ell^+$), 5-lepton final state results from the production channels: 
\begin{align*}
pp & \to \chi^- \chi^{++} \to \ell^- \ell^+ Z W^+ \rm{~with~} Z\to \ell^- \ell^+~ \rm{and~} W \to \ell \nu~,
\\
pp & \to \chi^0 \chi^+ \to \nu \ell^+ Z Z \rm{~with~} Z\to \ell^- \ell^+~,
\\
pp & \to \chi^0 \chi^+ \to \ell^\pm \ell^+ W^\mp Z \rm{~with~} Z\to \ell^- \ell^+~ \rm{and~} W \to \ell \nu.
\end{align*}
Almost all possible combinations of exotic fermion pair productions give rise to 3 and 4 lepton final states (see table~\ref{table:signatures}). Note that the SM contributions to three or more lepton final states are highly suppressed at the LHC. The SM processes, giving rise to three or more leptons, can be classified into two classes -- reducible and irreducible backgrounds. The reducible backgrounds are from the SM processes like $Z$+jets, $t\overline{t}$+jets, {\it etc.}, with additional leptons resulting from jet misidentifications or from the decay of heavy quark mesons. The irreducible ones are from diboson ($WZ,ZZ,Z\gamma$) and triboson ($ZZZ,WWZ$,etc.) production and processes like $t\overline{t}W$, $t\overline{t}Z$ and Higgs boson production, {\it etc.} Because of suppressed SM backgrounds and excellent efficiencies of the LHC detectors to identify electrons and muons, multilepton signatures are considered as one of the cleanest channels to probe new physics scenarios.

The multilepton final states have already been searched by the CMS collaboration \cite{cms_multilepton_137} at the LHC with $\sqrt s=13$ TeV and 137.1 fb$^{-1}$ integrated luminosity data. The CMS analysis in Ref.~\cite{cms_multilepton_137} is dedicated to the search of heavy triplet leptons in the context of type-III seesaw mechanism. To constrain\footnote{While the productions and subsequent decays of the exotic fermions in our model as well as type-III seesaw scenario give rise to multilepton final states, the CMS multilepton search limits on the masses of heavy triplet leptons in type-III seesaw model are not trivially applicable to limit the masses of triplet, quadruplet and quintuplet fermions in the present scenario.} the parameter space of the simplified type-V scenario, we performed a multilepton search similar to the search in Ref.~\cite{cms_multilepton_137}. The technical details of our implementation of multilepton search strategies of Ref.~\cite{cms_multilepton_137} and the resulting bounds on the masses of triplet, quadruplet, and quintuplet fermions in type-V seesaw model will be discussed in the next section.

\subsubsection{Bounds from the CMS multilepton search at the LHC}
\label{sec:multilepton}
The CMS collaboration has recently published a multilepton search with an integrated luminosity of $137.1$ fb$^{-1}$ of pp collisions at $\sqrt{s}=13$ TeV \cite{cms_multilepton_137}. The CMS multilepton search was targetted to probe pair productions of type-III seesaw heavy fermions. In view of the observations being consistent with the expectations from standard model processes, the CMS analysis in Ref.~\cite{cms_multilepton_137} excluded triplet fermions masses below 880 GeV at $95\%$ confidence level (CL) in the flavour democratic scenario (identical branching fractions across all lepton flavors) of type-III seesaw. While both the present model and the type-III seesaw have quite a similar multilepton final state signatures, the CMS bound on the masses of the triplet fermions in the type-III seesaw model can not be directly applicable to the multiplets of the present model withal\footnote{Even the productions and decays of the triplet exotics in this model are very different from the production and decays of type-III seesaw triplets. For example, the triplet of the present model has non-zero hypercharge ($Y=1$), and hence contains a doubly charged fermion in addition to the singly charged and neutral ones. Moreover, the neutral triplet fermion of the present model has a negligible branching ratio to the $W$-boson decay mode which is the dominant decay mode for the neutral heavy fermions in type-III seesaw model.}. Although the 95\% CL CMS observed upper limits (see Fig.~11 of Ref.~\cite{cms_multilepton_137}) on the total pair production cross-sections of triplet fermions in type-III seesaw, being highly dependent on the production and decays of the triplet fermions in type-III seesaw, can not be applicable for other models\footnote{Note that a realistic type-III seesaw scenario requires two or more generations of, in general, non-degenerate triplets to explain the neutrino oscillation data. Moreover, the decays of different triplet generations are not necessarily flavor democratic. The 95\% CL CMS observed upper limits in Ref.~\cite{cms_multilepton_137} on the total pair production cross-sections of triplet fermions are not even applicable for such a realistic type-III seesaw model \cite{asha_kirti}.}, one can always obtain 95\% CL upper limits on the total pair production cross-sections of exotic fermions in a different model which also gives rise to multilepton final states via using the CMS observed number of events\footnote{The CMS observed data for this particular analysis is presented in Fig.~3 and 4 of Ref.~\cite{cms_multilepton_137}, and also in Ref.~\cite{cms137hepdata}.} in different statistically independent signal bins\footnote{The classification of multilepton events, based on lepton multiplicity as well as invariant masses of different opposite sign same flavors (OSSF) lepton pairs, into different statistically independent signal bins is presented in table~1 of Ref.~\cite{cms_multilepton_137}.}. In the following, we embark on a mission to derive 95\% CL upper limits on the total production cross-sections of triplet, quadruplet and quintuplet fermions in type-V seesaw from the CMS multilepton data in Ref.~\cite{cms_multilepton_137}. Before going into the final results, a brief description about the event simulation, reconstruction of various objects (jets, leptons. {\it etc.}), event selection criteria, classification of different statistically independent signal regions (SRs) are presented in the following.

The signal events are simulated using MadGraph \cite{mg5} and showered with PYTHIA \cite{pythia}, and then DELPHES \cite{delphes} is used for detector simulation. Since the relevant backgrounds in our analysis are exactly the same as that of \cite{cms_multilepton_137}, we do not simulate the SM background processes, instead we use distributions of expected SM backgrounds and observed events given in figure 3 and 4 of \cite{cms_multilepton_137}, and also in \cite{cms137hepdata}. To reconstruct different physics objects (jets, electrons, muons, missing transverse energy, {\it etc.}), we have closely followed the object reconstruction criteria used by CMS collaboration in Ref.~\cite{cms_multilepton_137}.

\paragraph{Object reconstruction and selection:}Jets are reconstructed using the anti-kT algorithm \cite{anti-kt} with a distance parameter $\Delta R = 0.4$. Reconstructed jets with $p_T^j > 30$ GeV and $|\eta^j|<2.1$ are considered for further analysis. Electron(muon) candidates are required to have $p_T^l > 10$ GeV and within $|\eta^l|<2.5(2.4)$. In addition, following set of lepton isolation and displacement requirements are enforced to suppress the reducible backgrounds. The relative isolation ($I$), defined as the scalar $p_T$ sum, normalized to the lepton $p_T$, of photons and hadrons within a cone of $\Delta R$ around the lepton, for the electron candidates is required to be smaller than $I=0.0478+0.506/p_T^e$ ($I=0.0658+0.963/p_T^e$) \footnote{We thank Adish Vartak and Maximilian Heindl from CMS PAS EXO-19-002 group for providing the exact dependence of relative isolation on $p_T^e$.} within barrel(endcap) {\it i.e.,} $|\eta|<1.479$($|\eta|>1.479$) for $\Delta R~=~0.3$. For muons, we demand a maximum 15\% relative isolation with $\Delta R~=~0.4$. Following requirements on the transverse and longitudinal impact parameters ($d_z$ and $d_{xy}$) with respect to primary vertex has also been implemented: for electron candidate within barrel(endcap), we demand $d_z<0.10$ and $d_{xy}<0.05$ ($d_z<0.20$ and $d_{xy}<0.10$), whereas muon candidates require $d_z<0.10$ and $d_{xy}<0.05$. Lepton isolation trims hadronic activity inside the isolation cone. Lepton isolation along with impact parameter requirements suppresses the backgrounds from the SM processes like $Z$+jets and $t\overline{t}$+jets where a jet is misidentified as lepton or additional leptons originate from B-meson decays. Missing transverse momentum vector $\vec p_T^{\rm mis}$ (with magnitude $p_T^{\rm mis}$) is reconstructed using all remaining visible entities, viz. jets, leptons, photons and all calorimeter clusters not associated to such objects.

Events with at least one electron with $p_T>30(35)$ GeV for $35.9(41.5+59.7)$ fb$^{-1}$ integrated luminosity\footnote{Note that the trigger thresholds for data collection at the 13 TeV LHC were different in the year 2016 (with 35.9 fb$^{-1}$ integrated luminosity data), 2017 (with 41.5 fb$^{-1}$ integrated luminosity data) and 2018 (with 59.7 fb$^{-1}$ integrated luminosity data).} or at least one muon with $p_T>29(26)$ GeV for $41.5(35.9+59.7)$ fb$^{-1}$ integrated luminosity are considered for further analysis. For events with more than four leptons, only the leading-$p_T$ leptons are considered. Events containing a lepton pair with $\Delta R<0.4$ are rejected. Also, events containing a same-flavor lepton pair with dilepton invariant mass below 12 GeV are rejected. This reduces backgrounds from low-mass resonances and final-state radiations. Classification of different signal regions (SRs) based on the number of leptons, characteristic kinematic variables, {\it etc.,} and SR specific selection cuts will be discussed in the following.

\paragraph{Signal regions and event selection criteria:}
Final states with three or more leptons ($e,\mu$) are considered only. The events with four or more leptons containing no opposite-sign same-flavor (OSSF) lepton pair, an OSSF lepton pair and two OSSF lepton pairs are labelled by {\it 4LOSSF0, 4LOSSF1} and {\it 4LOSSF2}, respectively. Likewise, the events with exactly three leptons containing no OSSF lepton pair are labelled by {\it 3LOSSF0}. The events with exactly three leptons containing an OSSF lepton pair are further categorised into three categories based on OSSF dilepton invariant mass ($M_{\rm OSSF}$), and they are labelled by {\it 3L below-Z, 3L on-Z} and {\it 3L above-Z} when $M_{\rm OSSF}$\footnote{In cases of ambiguity (more than one OSSF lepton pair), the OSSF lepton pair with the mass closest to $M_Z$ is considered in the calculation of $M_{\rm OSSF}$.} is below, within and above the $Z$-boson mass window ($M_Z \pm 15$), respectively. Here are SR specific event selection criteria in the following.
\begin{itemize}
\item {\it 3LOSSF on-Z} events with trilepton invariant mass within the Z boson mass window  are vetoed. This censors the background events from $Z \to \ell^- \ell^+ \gamma$ with $\gamma \to \ell^- \ell^+$ where one of the leptons (from $\gamma$) gets lost. {\it 3LOSSF on-Z} events with missing transverse momentum ($p_T^{\rm miss}$) <100 GeV are vetoed.  
\item {\it 4LOSSF2} events with $p_T^{\rm miss}<100$ GeV are vetoed if both OSSF lepton pairs are on-Z. 
\end{itemize}
The CMS multilepton search uses $L_T+p_T^{\rm miss}$ as the primary kinematic discriminant for all signal regions except 3L on-Z, where $L_T$ is the scalar $p_T$ sum of all charged leptons. In 3L on-Z signal region, the transverse mass\footnote{The transverse mass is defined as $M_T=\sqrt{2 p_T^{\rm miss} p_T^\ell [1-\cos(\Delta\phi_{\vec p_T^{\rm miss},\vec p_T^\ell})]}~,$ where $\vec p_T^\ell$ is the transverse momentum vector of the lepton which is not a part of the on-Z pair.} ($M_T$) is used as discriminant. Table \ref{table:signal_regions} shows the binning scheme for different multilepton signal regions. The overflow events are contained in the last bin in each signal region.  As one can see from Table \ref{table:signal_regions} that there are 40 statistically independent signal bins  all together.
\begin{table}[htb!]
\centering
\begin{tabular}{c c c c}
\hline
Label & Kinematic discriminant & Range (GeV) &  Number of bins 
\\
\hline
3L below-$Z$ & $L_T$+$p_T^{\rm miss}$ & $[0,1200]$ & 6
\\
3L on-$Z$~~~~ & $M_T$ & $[0,700]\,$~ & 7
\\
3L above-$Z$ & $L_T$+$p_T^{\rm miss}$ & $[0,1600]$ & 8
\\
3L OSSF0\,~ & $L_T$+$p_T^{\rm miss}$ & $[0,1200]$ & 6
\\
4L OSSF0\,~ & $L_T$+$p_T^{\rm miss}$ & $[0,600]\,$~ & 2
\\
4L OSSF1\,~ & $L_T$+$p_T^{\rm miss}$ & $[0,1000]$ & 5
\\
4L OSSF2\,~ & $L_T$+$p_T^{\rm miss}$ & $[0,1200]$ & 6 \\
\hline
\end{tabular}
\caption{Multilepton signal regions. See text for details.}
\label{table:signal_regions}
\end{table}

\begin{figure}[htb!]
\includegraphics[scale=0.3,angle=-90]{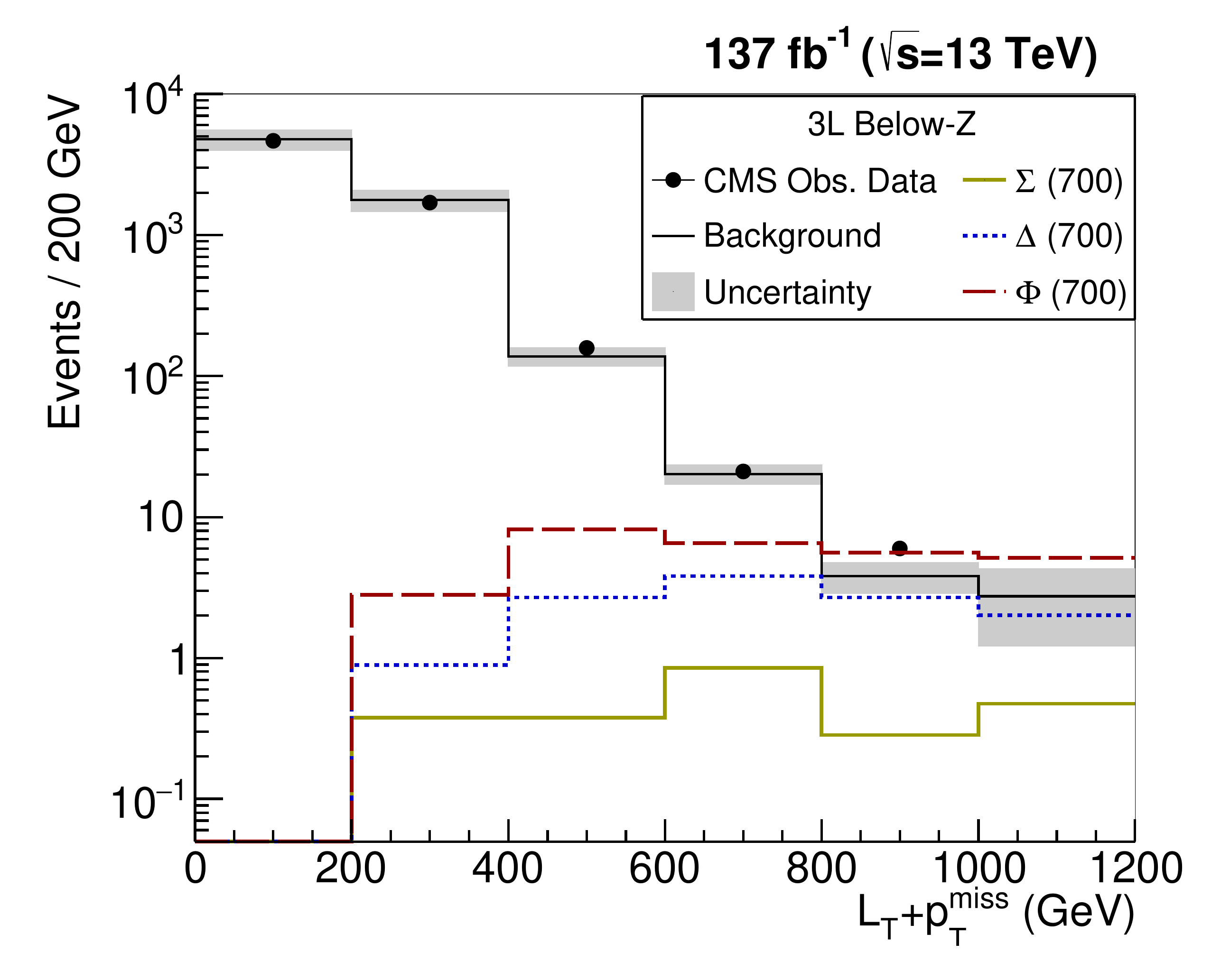}
\includegraphics[scale=0.3,angle=-90]{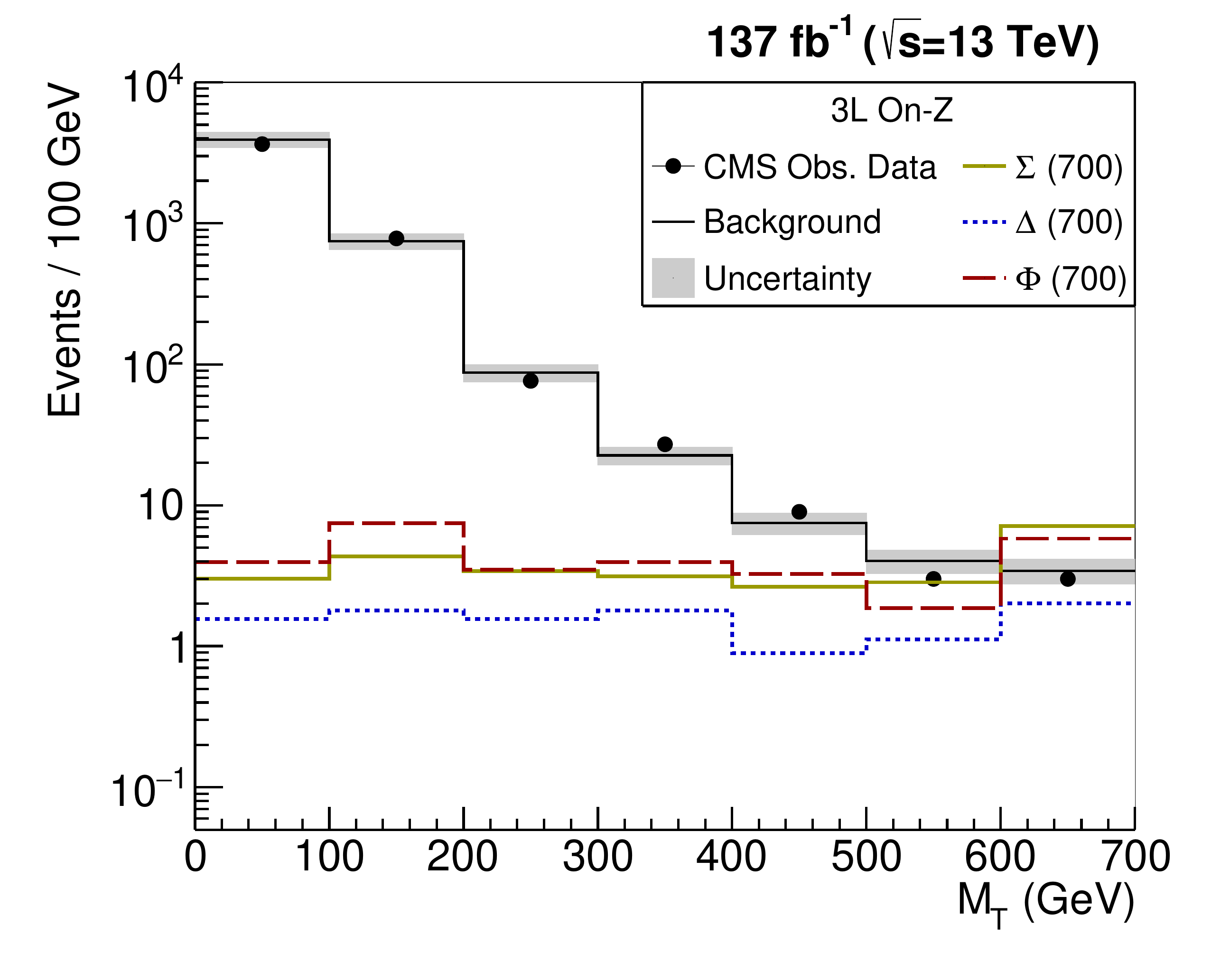}
\includegraphics[scale=0.3,angle=-90]{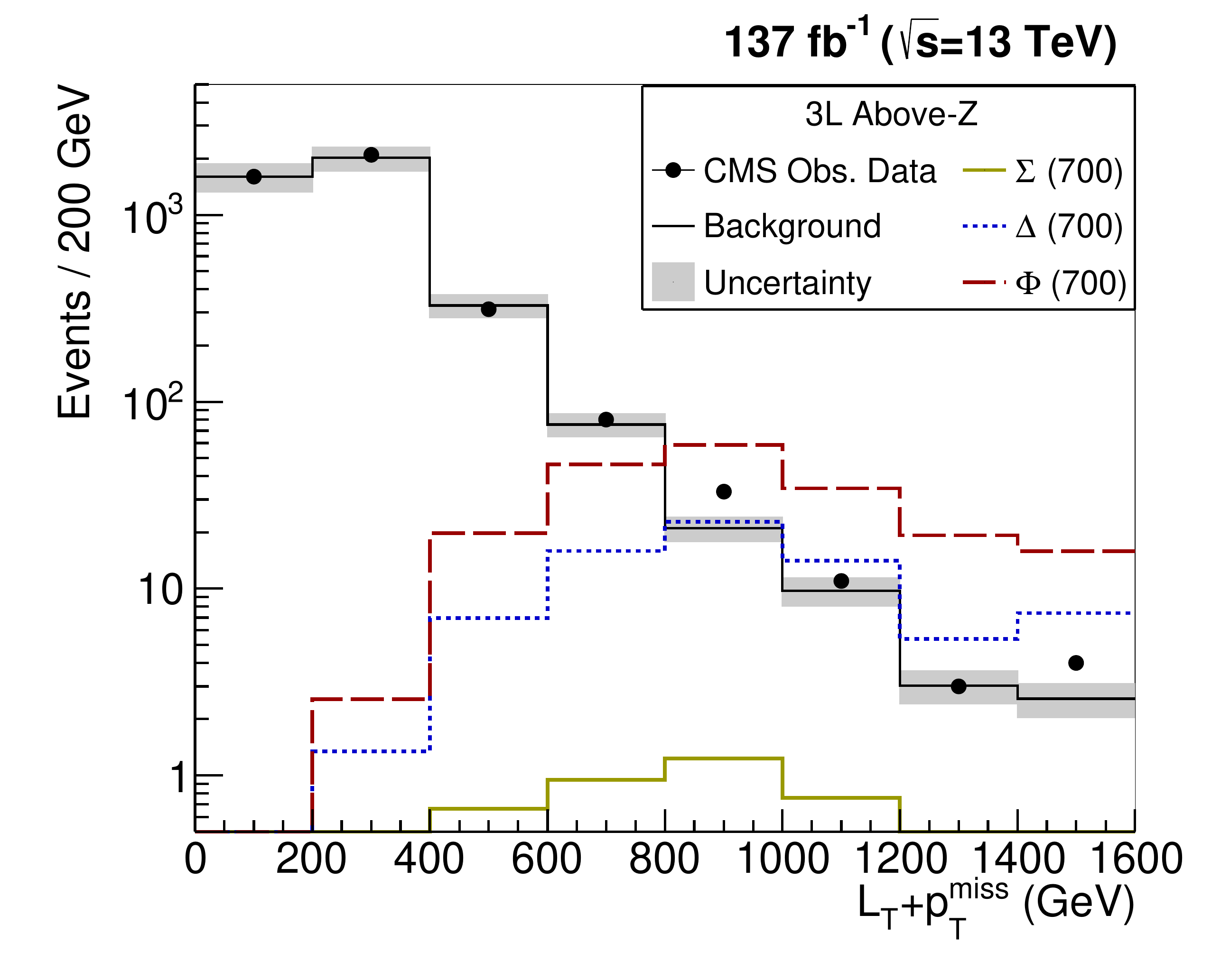}
\includegraphics[scale=0.3,angle=-90]{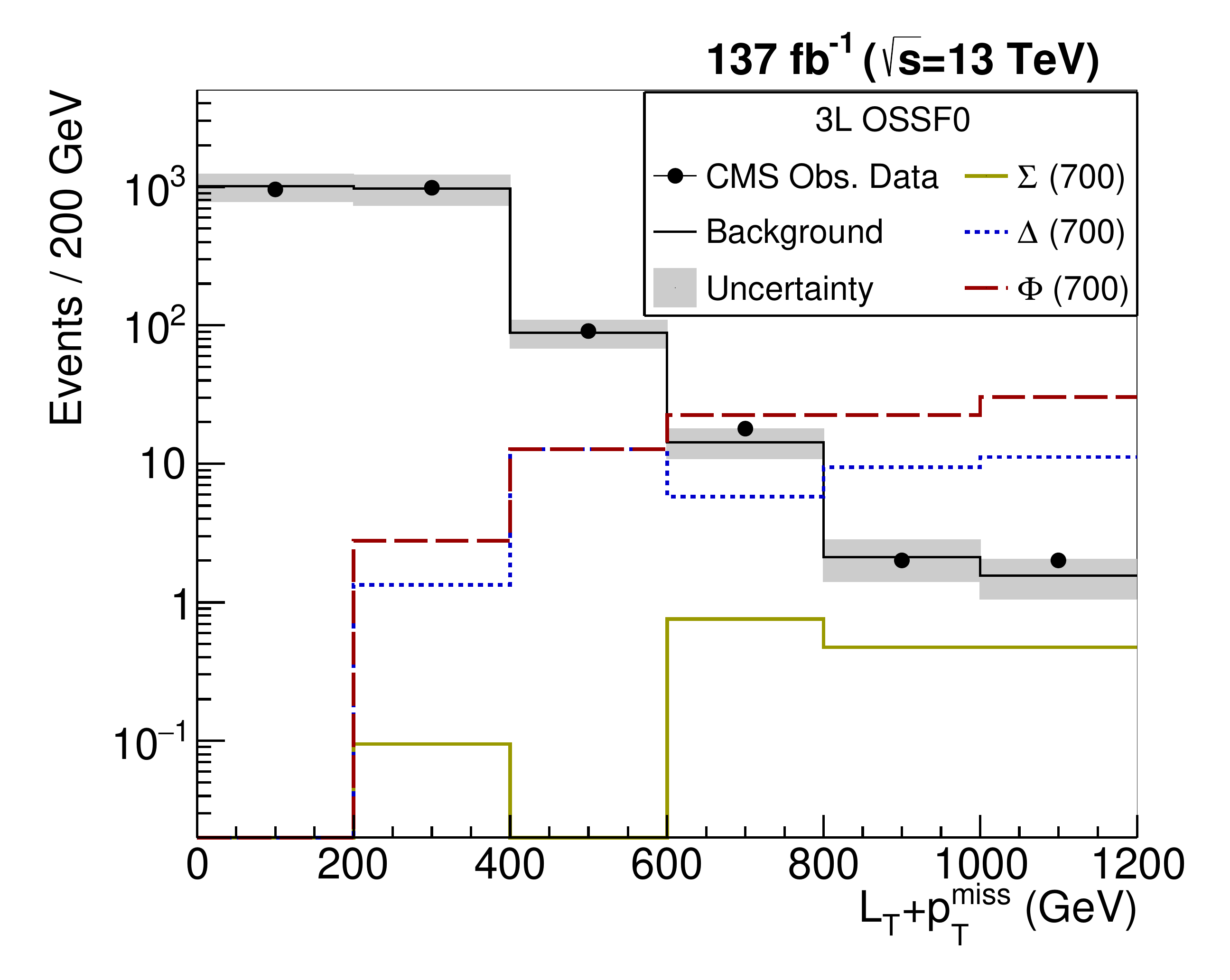}
\caption{For the signal regions 3L below-Z (upper left), on-Z (upper right), above-Z (lower left) and OSSF0 (lower right), the total expected SM backgrounds are shown as a histograms with black line, whereas the total observed events are shown as big black dots. The gray bands represent the total (systematic and statistical) uncertainty of the background estimation. The signal predictions for the three benchmark points are also shown. See text for details.}
\label{fig:signal_regions_3}
\end{figure}

\begin{figure}[htb!]
\includegraphics[scale=0.3,angle=-90]{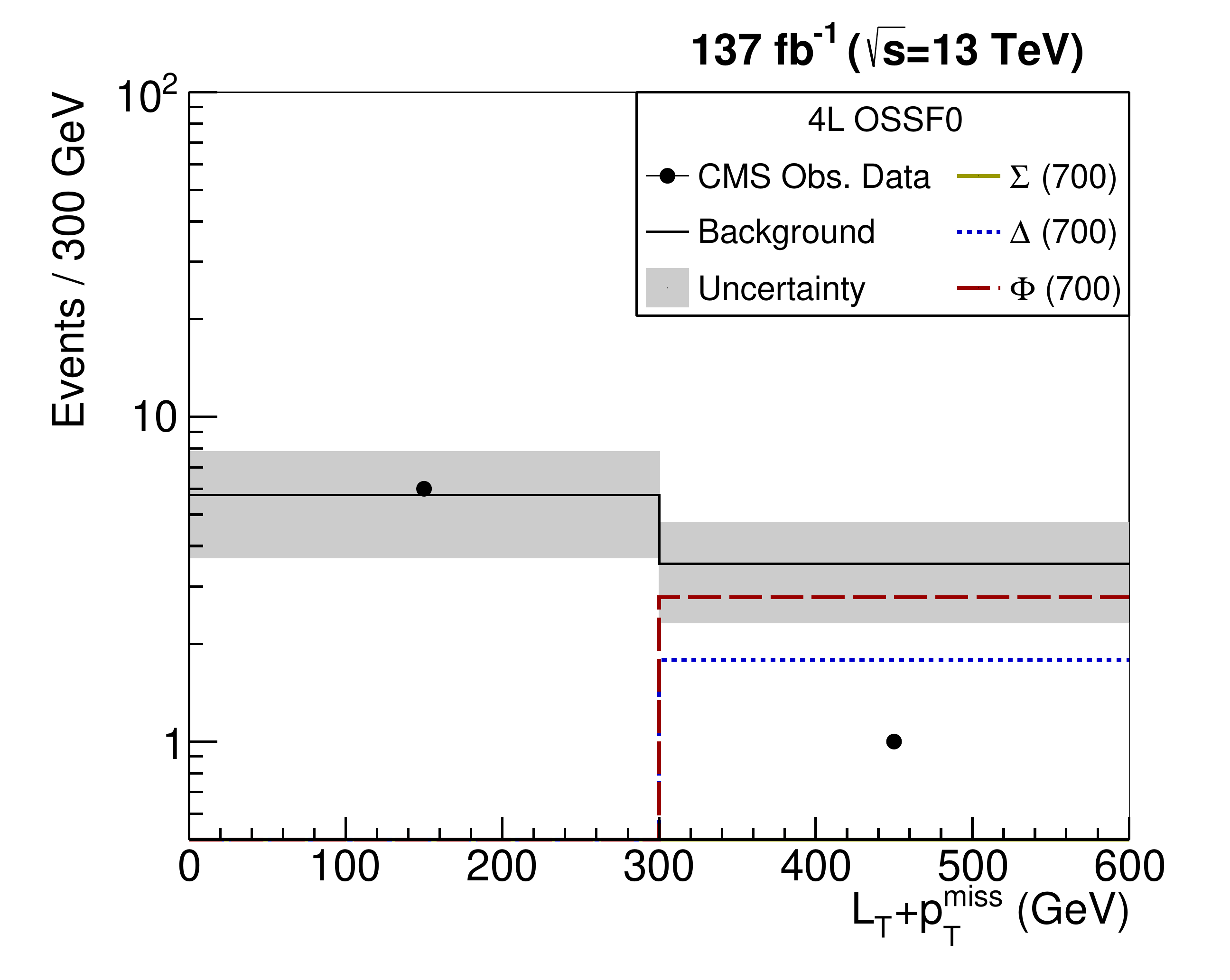}
\includegraphics[scale=0.3,angle=-90]{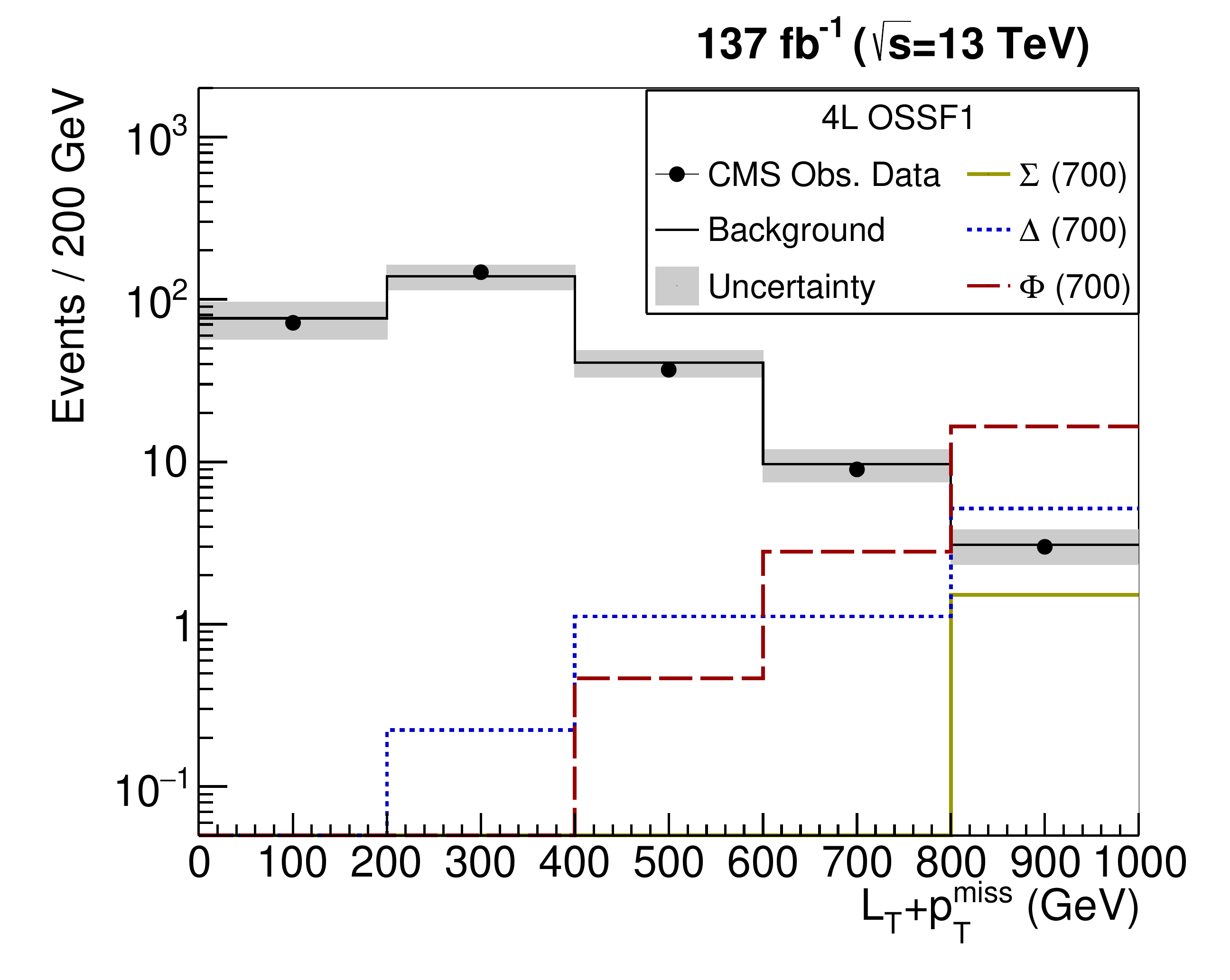}
\begin{center}
\includegraphics[scale=0.3,angle=-90]{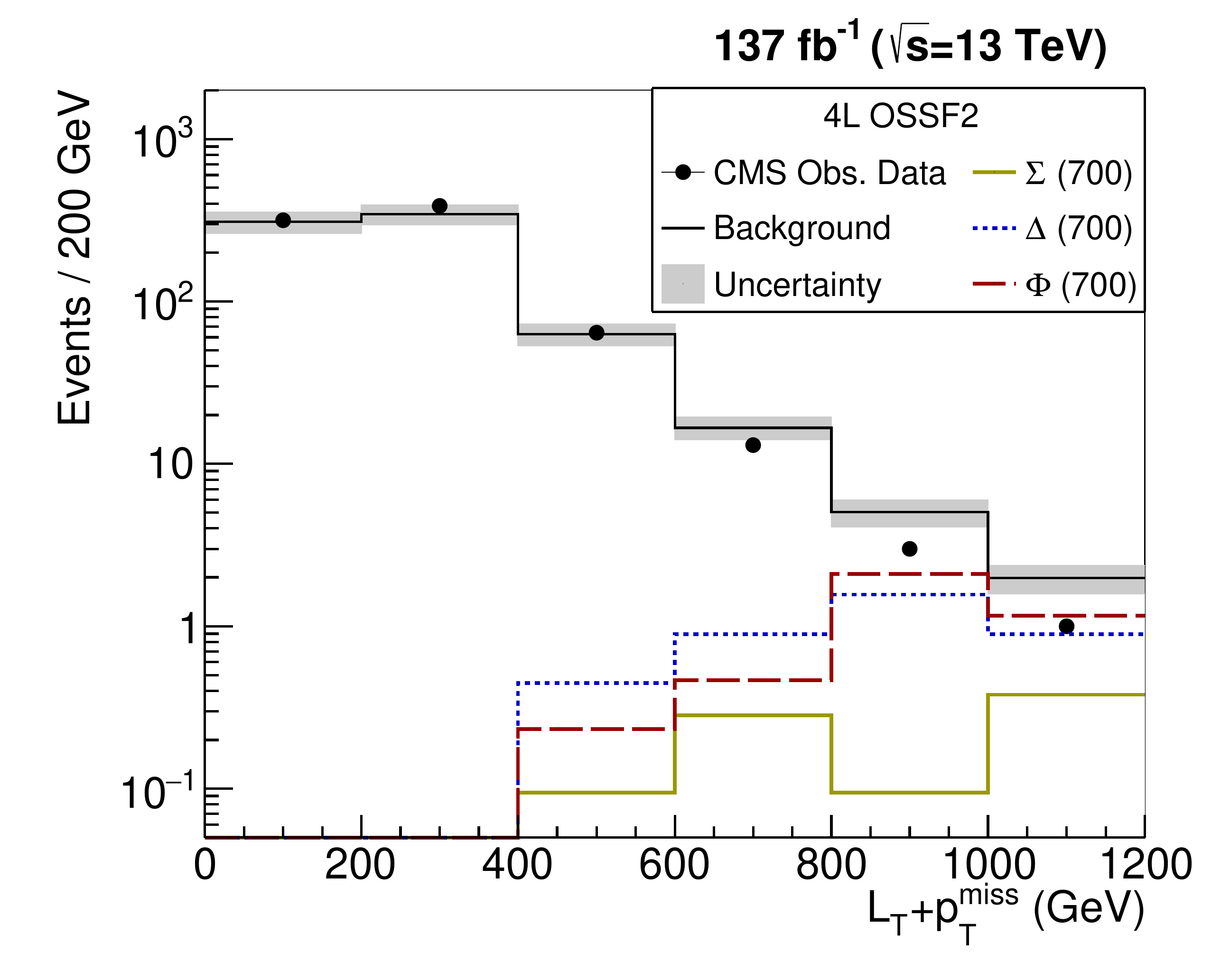}
\end{center}
\caption{Same as figure~\ref{fig:signal_regions_3} for 4-lepton signal regions {\it i.e.,} OSSF0 (upper left), OSSF1 (upper right) and OSSF2 (lower). See text for details.}
\label{fig:signal_regions_4}
\end{figure}

The total number of predicted SM background events\footnote{As acknowledged earlier, since the relevant backgrounds in our analysis are exactly same as that of \cite{cms_multilepton_137}, we do not simulate the SM background processes, instead we use distributions of expected SM backgrounds and observed events given in figure 3 and 4 of \cite{cms_multilepton_137}.} as well as the observed events, corresponding to 137.1 fb$^{-1}$ integrated luminosity data of 13 TeV LHC, in the 40 statistically independent signal bins (see Tabel \ref{table:signal_regions}) are shown in figures \ref{fig:signal_regions_3} and \ref{fig:signal_regions_4}. The expected SM backgrounds and observed events in 3L below-Z, on-Z, above-Z and OSSF0 signal bins are shown in Figure \ref{fig:signal_regions_3}, whereas Figure \ref{fig:signal_regions_4} shows the same in 4L OSSF0, OSSF1 and OSSF2 signal bins. For each signal regions, the total expected SM backgrounds and observed events are shown as histograms with black line and big black dots, respectively, whereas the gray bands represent the total (systematic and statistical) uncertainty of the backgrounds in each bin. In figures \ref{fig:signal_regions_3} and \ref{fig:signal_regions_4}, we have also presented our model predictions corresponding to different signal bins for three benchmark points listed below. 
\begin{itemize}
\item {\bf BP1} [$\Sigma(700)$]: $m_\Sigma=700$ GeV, $m_\Delta=m_\Phi=3$ TeV, $y_{34}^{(\prime)}=y_{45}^{(\prime)}=0.15$~,
\item {\bf BP2} [$\Delta(700)$]: $m_\Delta=700$ GeV, $m_\Sigma=m_\Phi=3$ TeV, $y_{34}^{(\prime)}=y_{45}^{(\prime)}=0.15$~,
\item {\bf BP3} [$\Phi(700)$]: $m_\Phi=700$ GeV, $m_\Sigma=m_\Delta=3$ TeV, $y_{34}^{(\prime)}=y_{45}^{(\prime)}=0.15$~.
\end{itemize}
Notice that the collider signatures of {\bf BP1}, {\bf BP2} and {\bf BP3} are dominated by the production and subsequent decays of light triplets, quadruplets and quintuplets, respectively. The larger(smaller) {\bf BP3}({\bf BP1}) contributions to the signal bins can be attributed to facts that the pair production cross-sections of doubly charged quintuplets(triplets) are the largest(smallest) (see figure~\ref{fig:DY_prod}) as well as the effective branching ratios (see the discussion in section~\ref{sec:decay}) of different combinations of quintuplet(triplet) pairs into  3 or 4-lepton final states are are relatively large(small). 

\paragraph{The $95\%$ CL upper limits on the total production cross sections of triplet, quadruplet and quintuplet fermions in type-V seesaw:}
In the absence of any statistically significant deviations\footnote{CMS collaboration \cite{cms_multilepton_137} found the observations to be globally consistent with the SM predictions within 2.7 standard deviations.} of the data from the SM predictions (see figures \ref{fig:signal_regions_3} and \ref{fig:signal_regions_4}) in all the signal bins, one can obtain bounds on the parameter space of different signal models which is type-V seesaw in our case and type-III seesaw for Ref.~\cite{cms_multilepton_137}. First, we proceed to obtain the 95\% CL upper bounds on the total pair production cross-section of a given multiplet (triplet, quadruplet and quintuplet) assuming the other two multiplets are too heavy to contribute significantly in the signal bins.

\begin{figure}[htb!]
\centering
\includegraphics[scale=0.6]{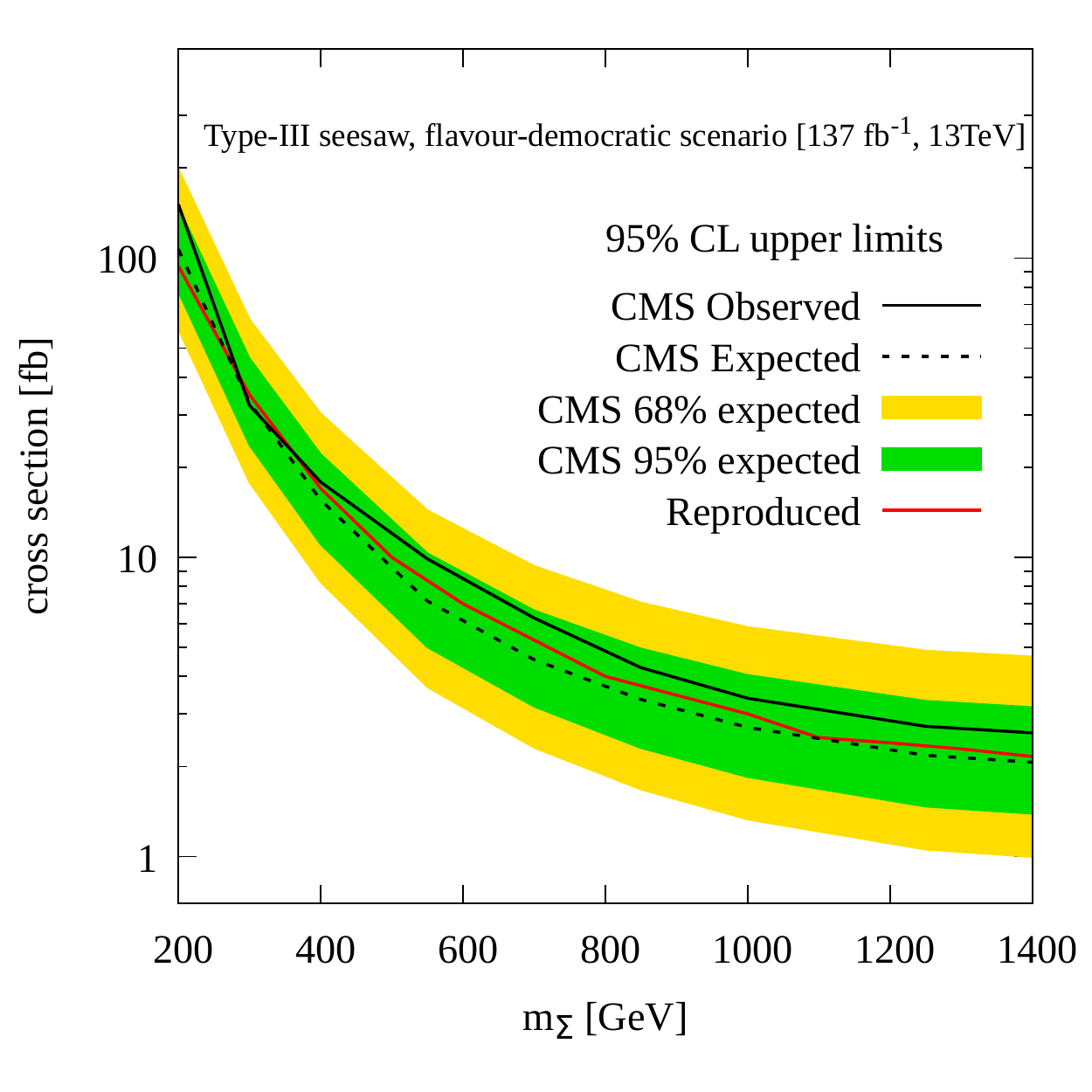}
\caption{The $95\%$ CL upper limits on the total production cross section of heavy fermionic pairs in the flavour-democratic scenario in type-III seesaw model using the CMS search \cite{cms_multilepton_137}. The reproduced $95\%$ CL upper limit is represented by red solid line.}
\label{fig:CL_typeIII}
\end{figure}

We use a hypothesis tester named `Profile Likelihood Number Counting Combination' \cite{bayes} which uses a library of C++ classes `RooFit' \cite{RooFit} in the ROOT environment to estimate CL. This package treats all the bins in different signal regions as independent channels for both signal and background events. The uncertainties are included via the Profile Likelihood Ratio. Before proceeding further, we try to validate our approach ({\it i.e.,} event simulation, object reconstruction, event selection for different signal bins and the statistical approach to obtain the bound) by reproducing the 95\% CL bound in Ref.~\cite{cms_multilepton_137} on the total triplet pair production cross-section in simplified flavor democratic type-III seesaw. The pair productions of type-III triplets for a given triplet mass are generated in MadGraph using the \texttt{FeynRules 2.0} \cite{feynrules} model file \cite{biggio2011} for the simplified type-III seesaw. The generated events are fed into our collider analysis packages like PYTHIA, Delphes, {\it etc.} for simulating the ISR, FSR, decays, hadronization, {\it etc.} and selecting signal events for different signal bins. The number of signal events in all the 40 signal bins along with the observed data points and the total (systematic and statistical) uncertainties are passed to the hypothesis tester \cite{bayes} for calculating the exclusion significance of the signal benchmark point. Figure \ref{fig:CL_typeIII} shows the CMS \cite{cms_multilepton_137} observed and expected $95\%$ CL upper limits on the total production cross section of heavy fermionic pairs in the flavour-democratic type-III seesaw model. The green and yellow bands, respectively, correspond to the regions containing $68\%$ and $95\%$ of the distribution of expected limits under the background-only hypothesis. The reproduced $95\%$ CL upper limit is represented by the red solid line. We see that our reproduced result matches quite well with the CMS one. We follow the same approach to estimate the $95\%$ CL upper limits on the total production cross sections of the exotic fermionic multiplets in the present model.

\begin{figure}[htb!]
\centering
\includegraphics[scale=0.39]{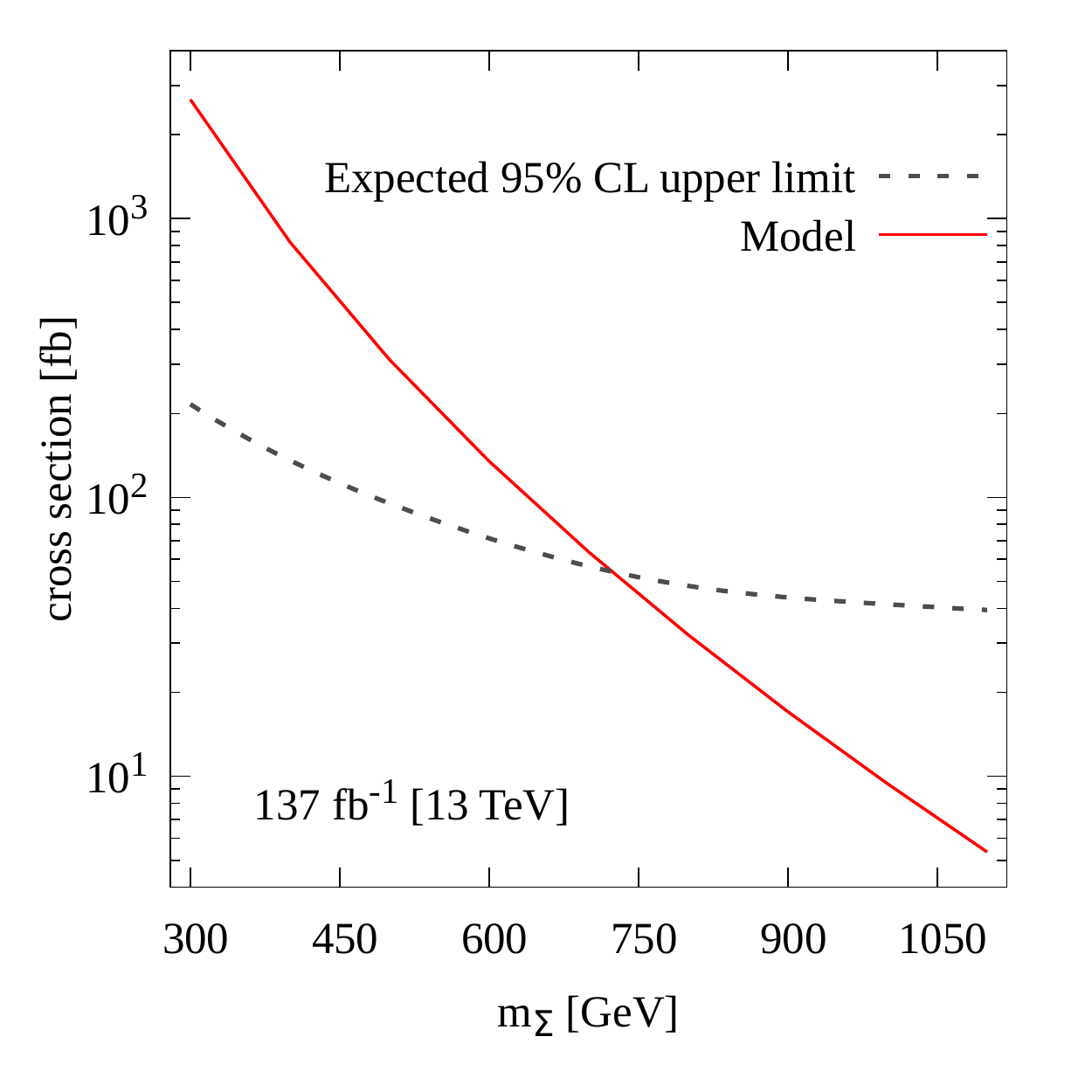}
\includegraphics[scale=0.39]{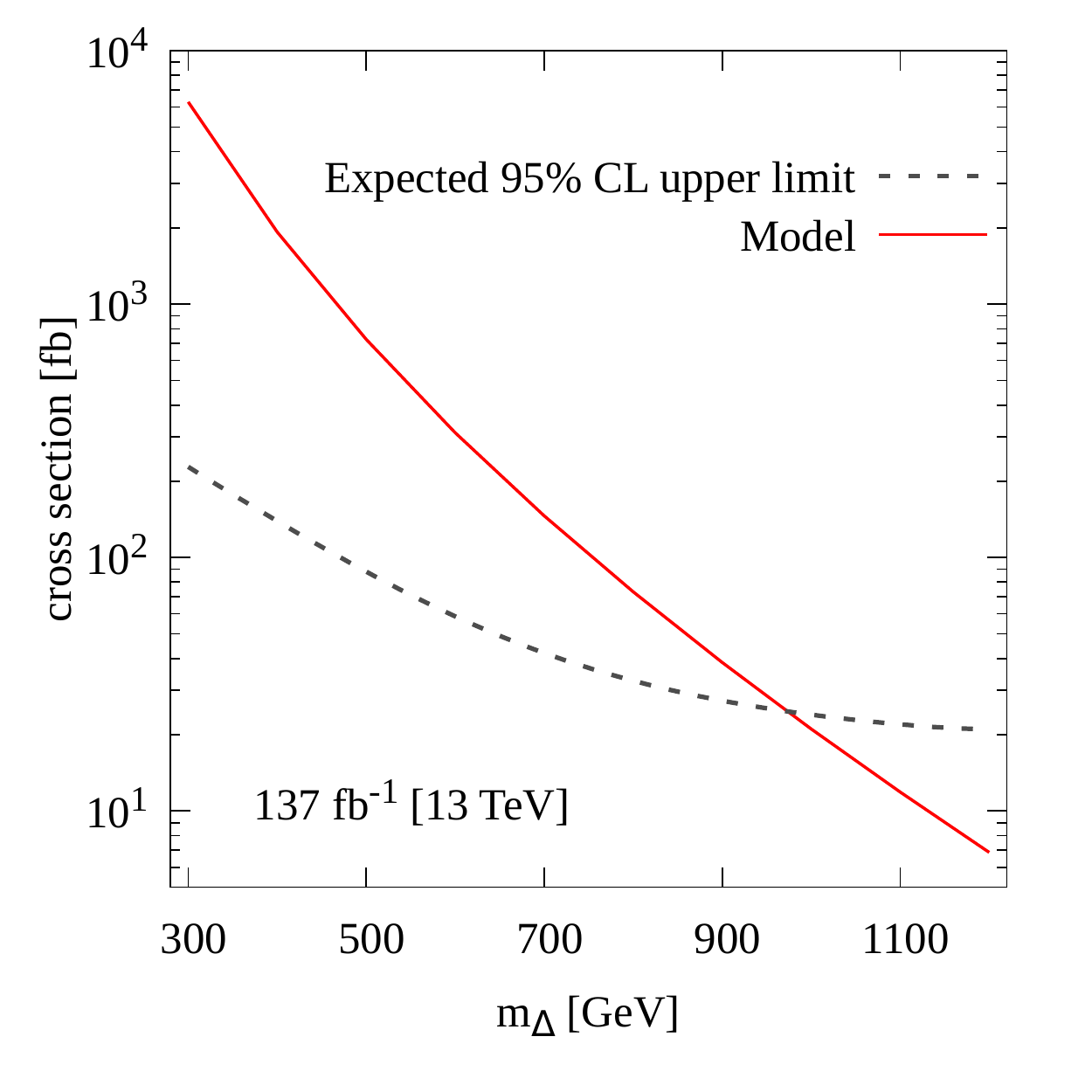}
\includegraphics[scale=0.39]{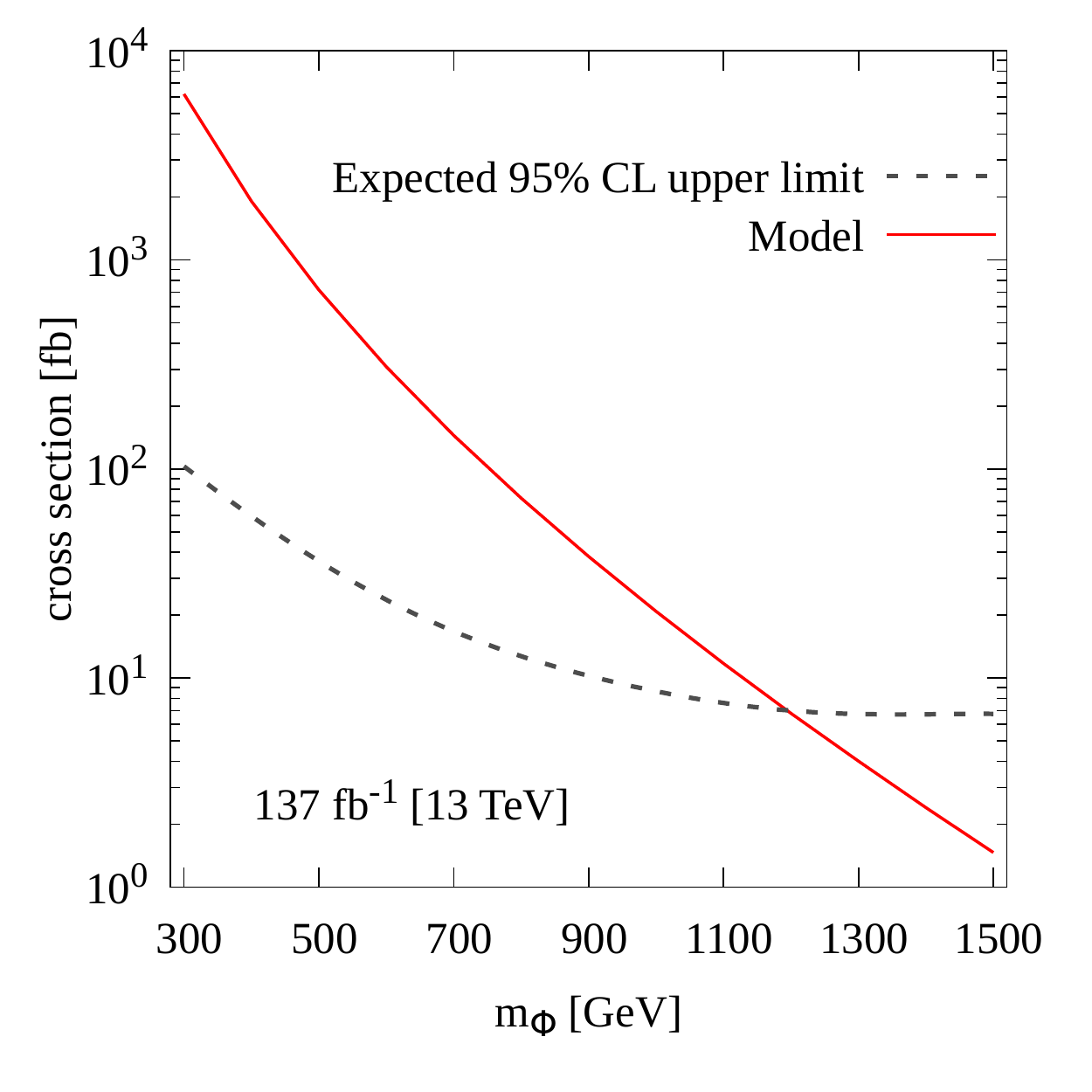}
\caption{\label{fig:CL} Solid line: $95\%$ CL upper limit on the total production cross section of triplet (left most panel), quadruplet (middle panel), quintuplet (right most panel) fermions. Dashed line: Model prediction for total production cross section of the exotics of a given multiplet. See text for details.}
\end{figure}
\begin{figure}[htb!]
\begin{center}
\includegraphics[scale=0.75]{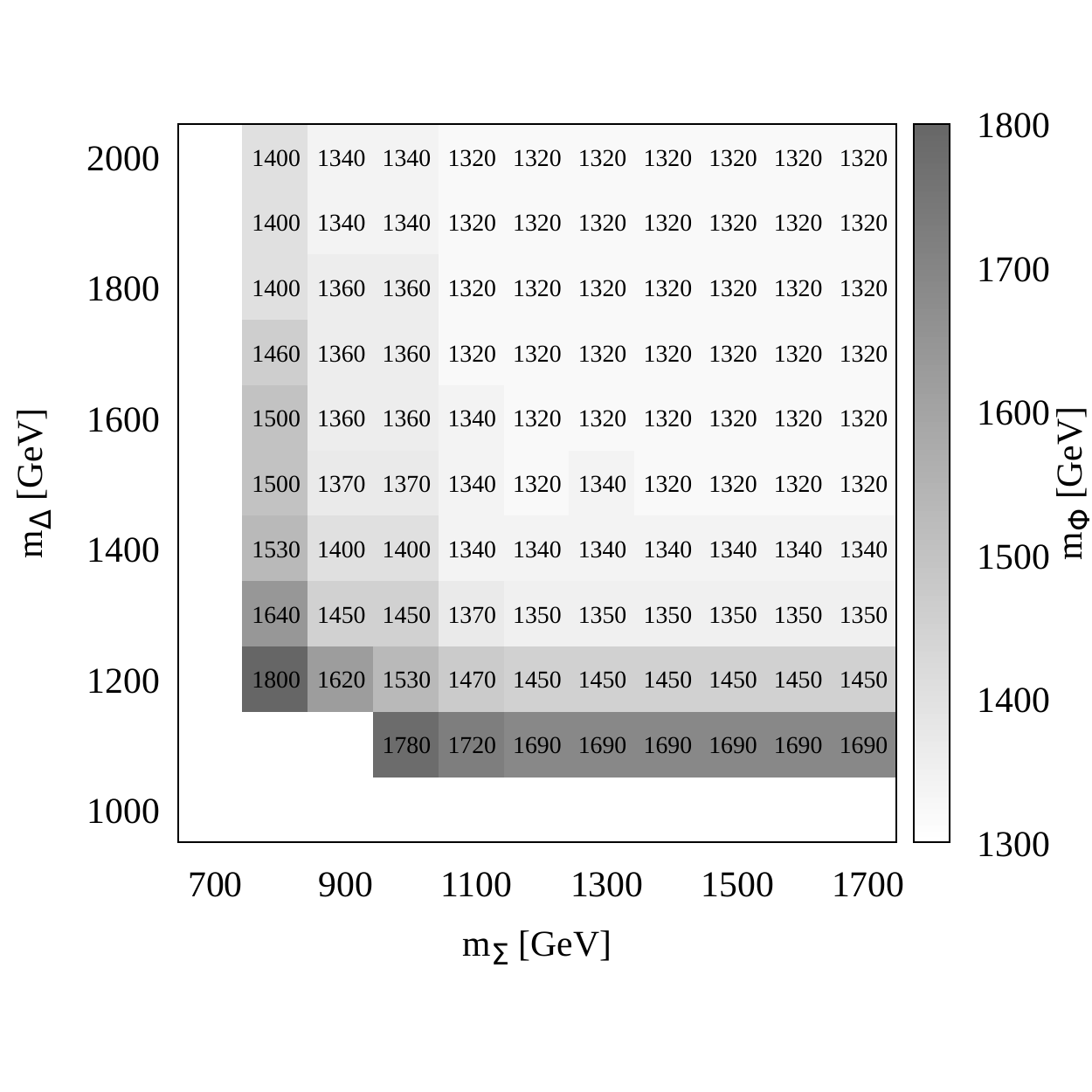}
\end{center}
\caption{\label{fig:CL345} Lower limit on $m_\Phi$ at $95\%$ CL as a function of  $M_\Sigma$ and $M_\Delta$. See text for details.}
\end{figure}

Red solid line in the left most (middle) [right most] panel in figure~\ref{fig:CL} show the $95\%$ CL upper limit on the total production cross section of exotic triplet (quadruplet) [quintuplet] fermionic pairs for a given set of Yukawa couplings: $y_{34} = y_{45} = y_{34}^\prime = y_{45}^\prime = 0.15$ in the simplified scenario assuming the lightest neutrino mass to be larger than $10^{-9} (10^{-7}) [10^{-5}]$ eV to ensure their prompt decays. The grey dashed line in the left most (middle) [right most] panel in figure~\ref{fig:CL} show theoretical prediction for the total pair production cross section of the triplet (quadruplet) [quintuplet] fermionic pairs. For each plot, the bare mass parameters in the Lagrangian are chosen such that the given multiplet (triplet, quadruplet or quintuplet) is the lightest and the other two multiplets are too heavy to contribute significantly in the signal bins. Figure~\ref{fig:CL} shows that the exotic fermion masses below 720, 970, and 1200 GeV are excluded for triplet, quadrupled and quintuplet, respectively.
As mentioned earlier, the contributions to the signal bins for the triplets are less compared to that of the quintuplets because of smaller triplet pair production cross-sections as well as the effective branching ratios of triplet pairs into multilepton final states. This explains the less stringent exclusion limit on the triplet mass than that on the quintuplet mass. Although the triplets in both realistic type-III seesaw model and the present model has similar (total) pair production cross section, the triplet pairs in the type-III seesaw model have larger effective branching ratios into multilepton final states than that in the present one. Thus, one would expect a more stringent\footnote{It can be seen from figure~\ref{fig:CL_typeIII} and \ref{fig:CL}(left most panel) that the 95\% CL upper limit on the total triplet production cross section in type-III seesaw is much more stringent than that in our model.} exclusion bound on the masses of the triplets in the type-III seesaw model than in the present model\footnote{The CMS analysis in Ref.~\cite{cms_multilepton_137} excludes triplet fermions with masses below 880 GeV at 95$\%$ CL in the flavour democratic scenario of type-III seesaw. Reinterpreting the same search in the context of a realistic type-III seesaw model with three generations of triplets \cite{asha_kirti} would yield an exclusion limit of 1110 GeV in the flavour democratic scenario.}. The lower bounds of 720, 970 and 1200 GeV ,respectively, on triplet, quadruplet and quintuplet masses resulting from figure~\ref{fig:CL} are based on the assumption that the triplet (quadruplet) [quintuplet] is the lightest with few hundred GeVs to a TeV mass while $m_\Delta$ and $m_\Phi$ ($m_\Sigma$ and $m_\Phi$) [$m_\Sigma$ and $m_\Delta$] are large ($\sim 3$ TeV). If we relax this assumption then all three multiplets of the type-V seesaw model start contributing to the multilepton final states. To display the bounds on the 3-dimensional parameter space defined by $m_\Sigma$--$m_\Delta$--$m_\Phi$, in figure~\ref{fig:CL345}, we present the lower limit on $m_\Phi$ at $95\%$ CL in the $m_\Sigma$--$m_\Delta$ plane.

\section{Displaced vertex}
\label{sec:displaced}
We have shown in section~\ref{sec:decay} that the total decay width for some of the lightest charged and neutral exotics can be small enough to result into displaced vertices or long-lived particles (LLPs) at the LHC. In the present model, displaced vertices, LLPs, vanishing charge track signature, {\it etc.} result for smaller values of the lightest neutrino mass, {\it i.e.} smaller $m_1(m_3)$ in NH(IH). Since last decade, the long-lived particles (LLPs), motivated by dark matter and tiny neutrino masses, are being discussed quite extensively in the literature \cite{high_dim_7,dis_heavy,dis_sterile,dis_sterile2,dis_lrsm,dis_rhn_u1x,dis_typei,dis_typeii_lrsm,dis_typeii,dis_typeiii}. Null results from the LHC searches for BSM particles at the electroweak scale have added up the existing speculations about the LLPs. LHC has been searching for the LLPs \cite{lhc_dis_1,lhc_dis_2,lhc_dis_3,lhc_dis_4,lhc_dis_5,lhc_dis_6}. A number of experimental proposals (MATHUSLA\footnote{The MATHUSLA detector is a proposed detector, anticipated to detect long-lived particles escaping prompt decay searches after being produced at the LHC, which will be placed at $\sim 100$ m away from the LHC at an angle of [0.23,0.8] away from the LHC beam. Being far away from the production center, it will have the advantage of an almost background free environment. Thus, it is well suited to detect long-lived particles with decay lengths of $\mathcal{O}(100)$ m or even longer.} \cite{mathusla,mathusla2,mathusla3}, LHeC \cite{lhec_dis}, FCC-he \cite{fcc-he_dis} \footnote{The absence of large hadronic background and pile-up events at higher luminosities allows $\it ep$ collider such as the LHeC and FCC-he to reliably reconstruct the long-lived particles with soft final states and/or very short lifetimes. Because of the asymmetric beam setup of $\it ep$ collider, the (long-lived) charged particles produced at the collider are significantly boosted along the beam direction increasing their lifetime in the lab frame. Thus, $\it ep$ collider has better sensitivity than $\it pp$ colliders to soft final states and/or very short lifetimes \cite{curtin_np}.}, etc.) to search for the same have come out in recent years. In this section, we discuss the possibility that the exotics of this model could also be long-lived\footnote{Authors of \cite{high_dim_7} have also discussed the same upto some extent.} leaving disappearing track signatures or displaced vertex at the detector.

The decay length of a given component with electric charge $Q$ of a multiplet $\chi$ is given by 
$$c\tau \approx 1 \rm mm \times 1.975 \times 10^{-13} \times \frac{1 \rm GeV}{\Gamma_{\rm total}^{\chi^Q}}~,$$
where $\Gamma_{\rm total}^{\chi^Q}$ is the total decay width of ${\chi^Q}$. For the estimation of decay length of the heavy fermions, we take $y_{34} = y_{45} = y_{34}^\prime = y_{45}^\prime = 0.15$, and the neutrino oscillation parameters are fixed at their best fit values (except the Majorana phases which are taken to be zero) assuming the mass spectrum to be normal hierarchical, and $Y_{23}$ is determined using the parametrisation Eq.~(\ref{eq:parametrisation}). The heavier multiplets can decay into lighter multiplets in association with an SM boson ($W,~Z$ or the Higgs boson). These decays are determined by the Yukawa couplings (which are ${\cal O}(0.1)$ in this model) involving different multiplets, and hence not suppressed. The only kinematically allowed decay modes for the lightest of the multiplets are the decays into the SM leptons in association with an SM boson or into pions/leptons in association with a lighter component of the same multiplet. While the former is suppressed by the small neutrino masses, the latter is suppressed due to the small radiative mass splitting within the components of a given multiplet. We see in section~\ref{sec:decay} that for the lightest among the triplet, quadrupled and quintuplet, the components of a particular generation which is associated to the lightest neutrino mass could have sizeable decay length ($c\tau$), and hence result in disappearing tracks or displaced vertex signatures at the detector. The components of other generations, being associated with second and third lightest neutrinos, have $c\tau$ always smaller than 0.1 mm, and hence no disappearing tracks or displaced vertex at the detector. So, we concentrate only on the components which are associated with the lightest neutrino mass ($m_1$ for NH and $m_3$ for IH). A detailed phenomenology for the LLPs of this model is beyond the scope of this work. However, we qualitatively discuss the possible characteristic signatures of LLPs resulting in different parts of the parameter-space in the following.

\begin{figure}[htb!]
\includegraphics[scale=0.39]{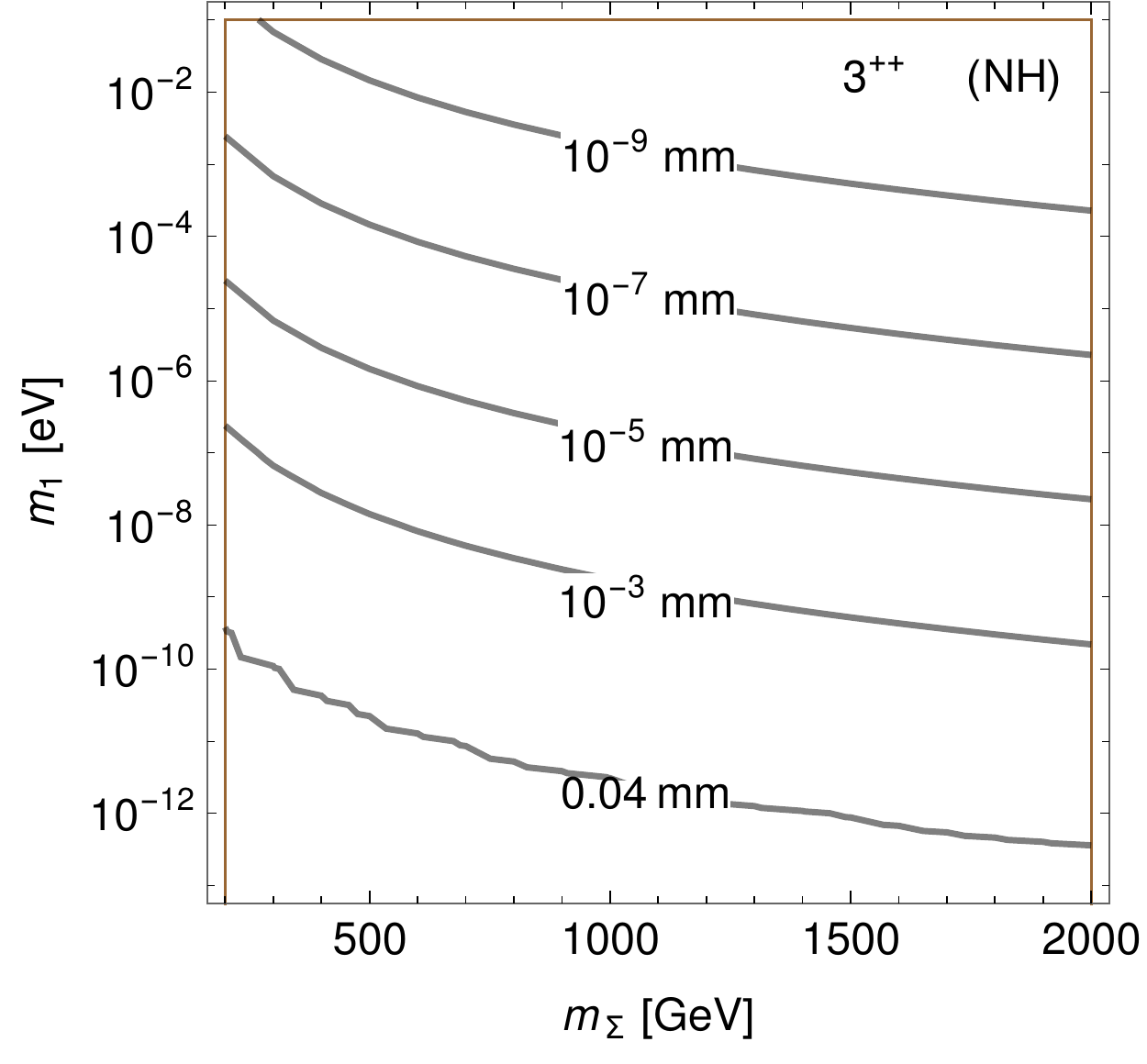}
\includegraphics[scale=0.39]{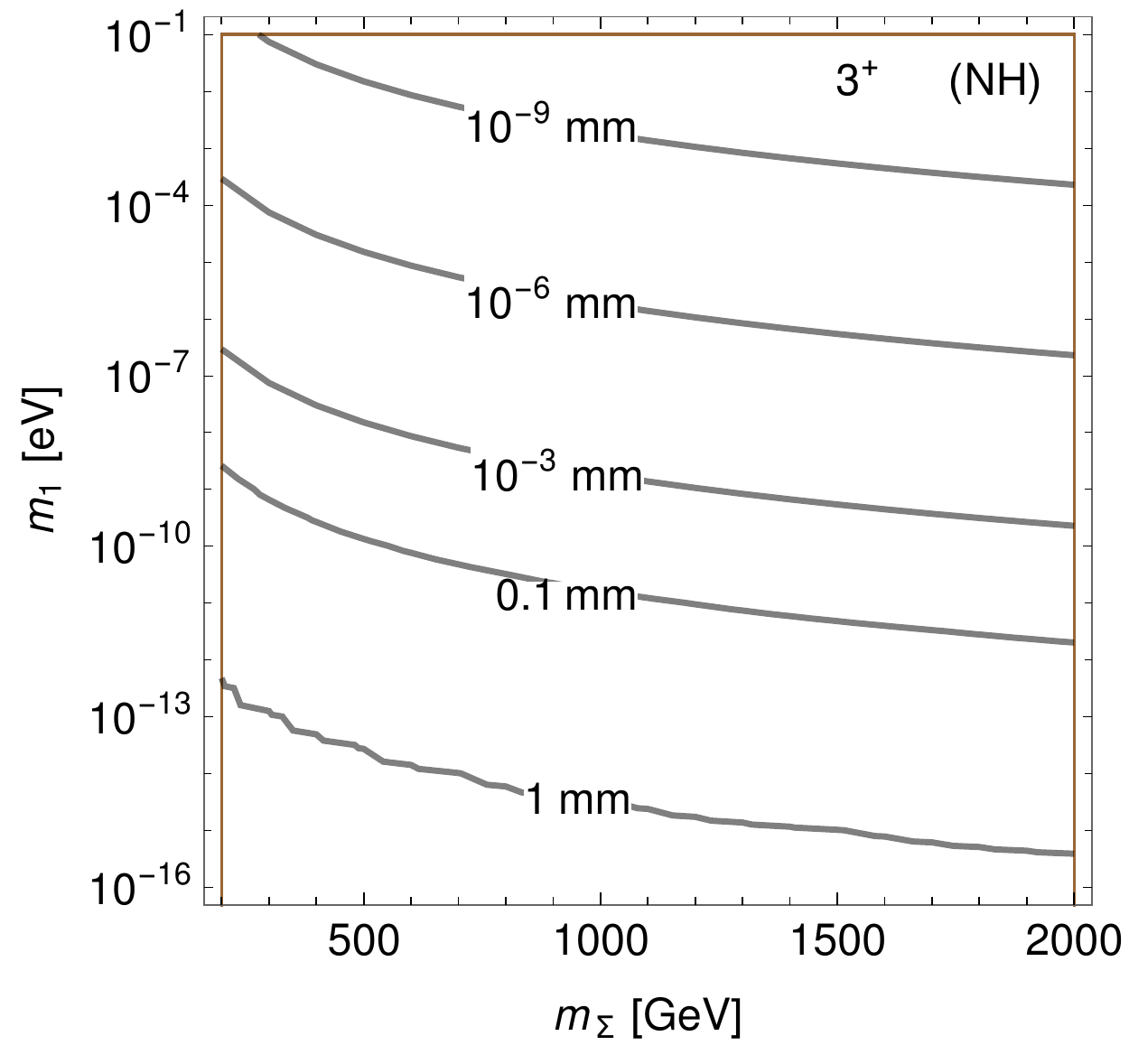}
\includegraphics[scale=0.39]{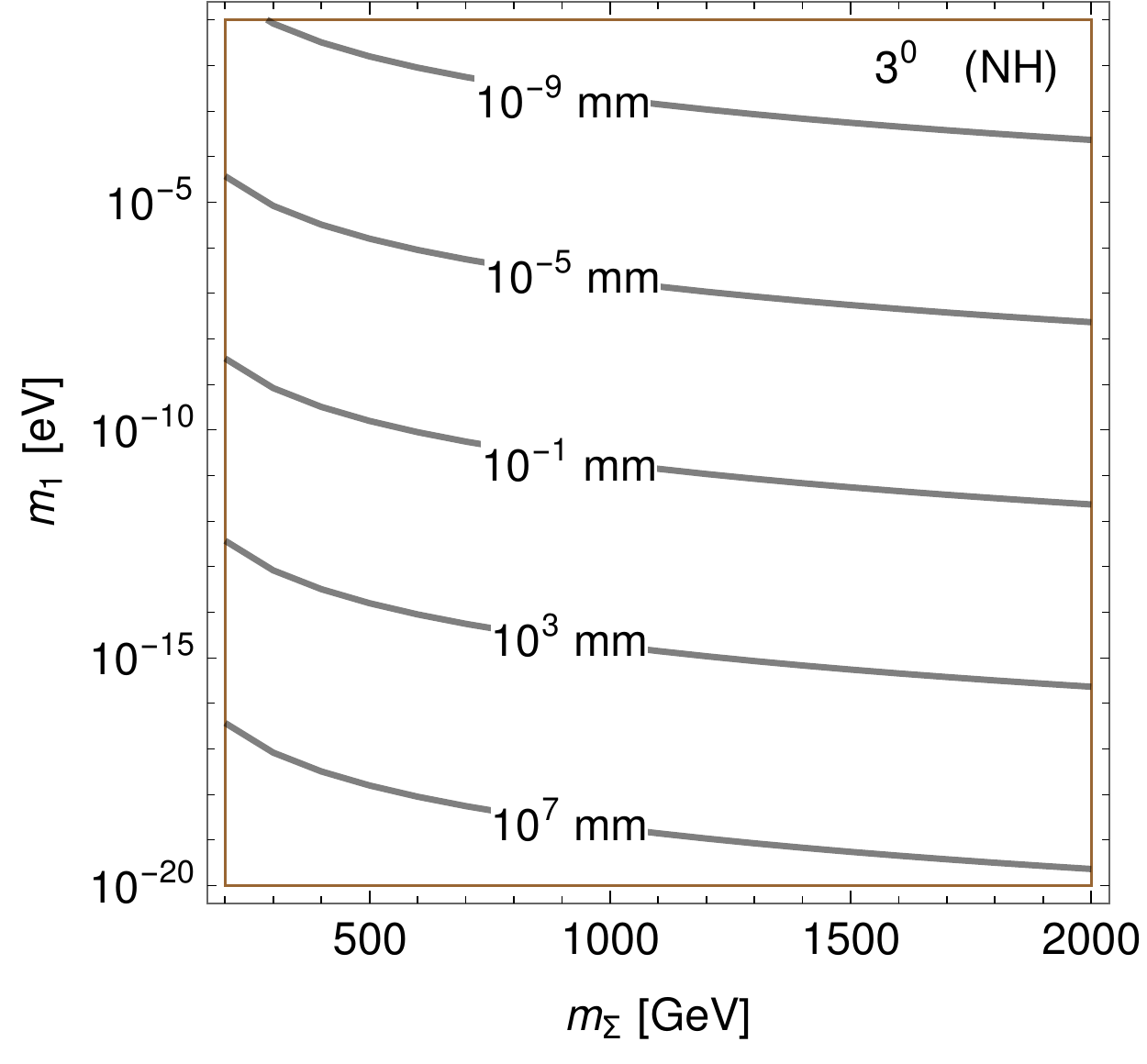}
\caption{Contour plots for decay length in $(m_1-m_\Sigma)$ plane for $3^{++},3^+$ and $3^0$, respectively, for normal hierarchical light neutrino mass spectrum.}
\label{fig:contour_3}
\end{figure}

\begin{figure}[htb!]
\centering
\includegraphics[scale=0.45]{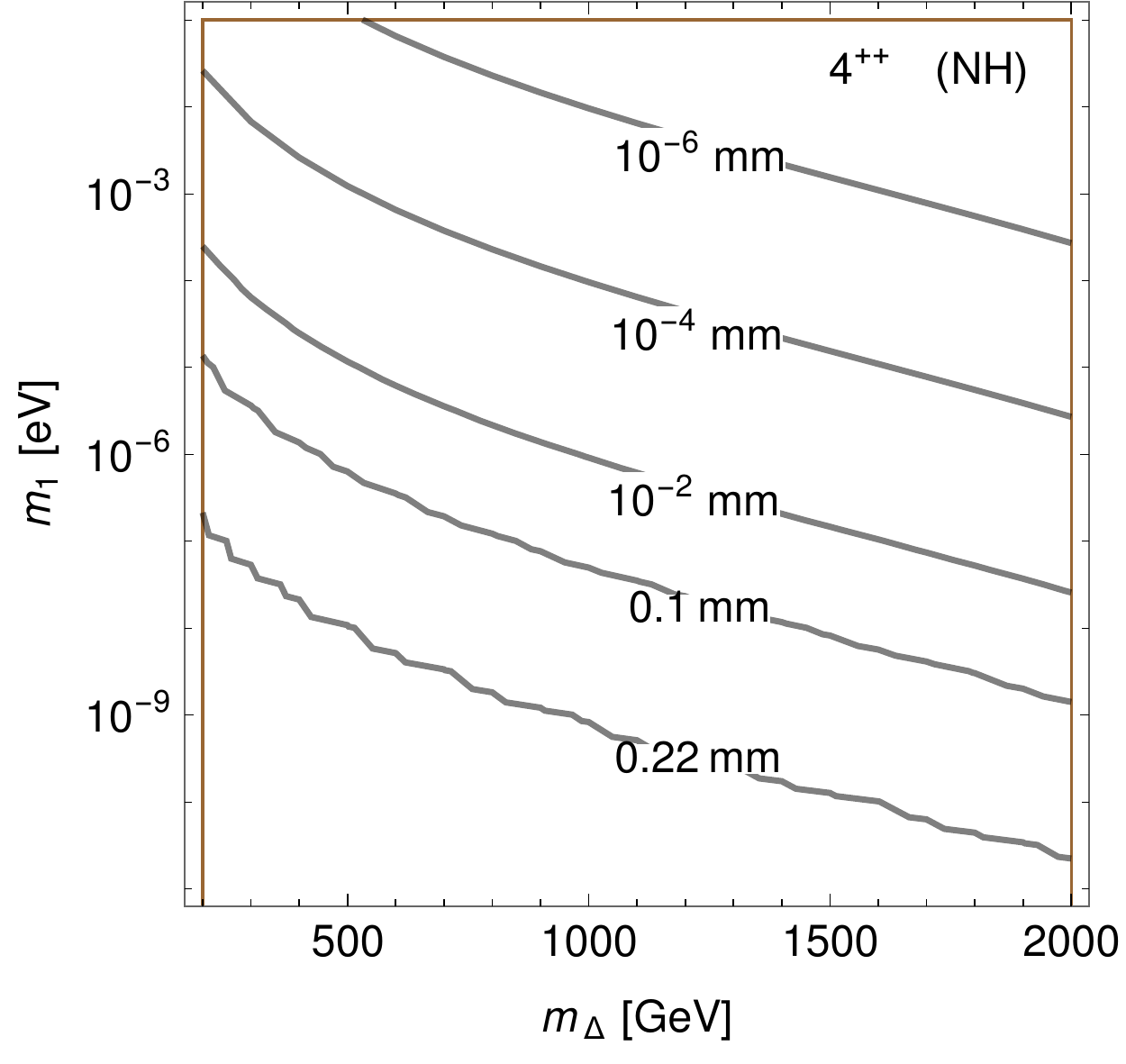}
\includegraphics[scale=0.45]{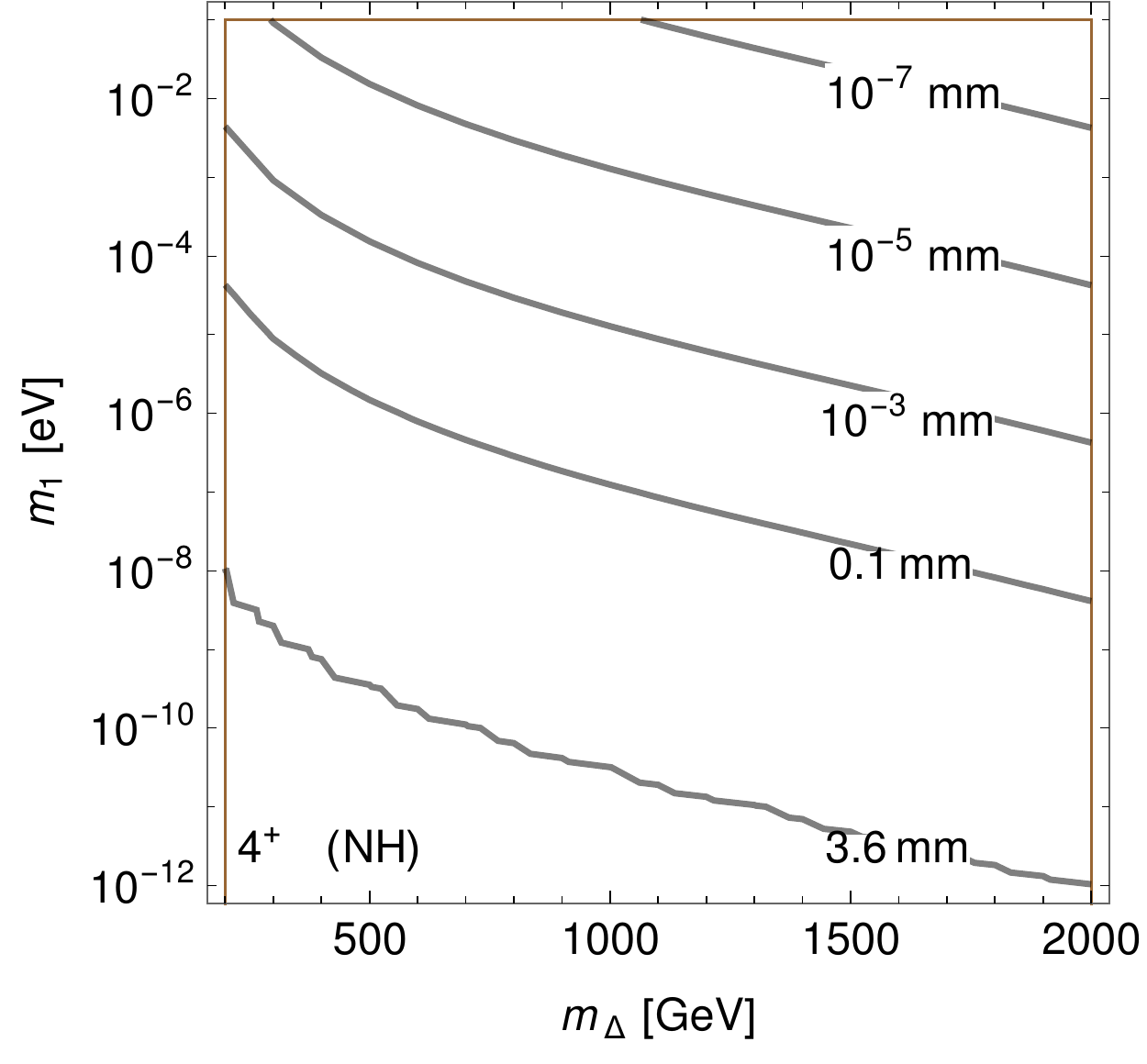}
\includegraphics[scale=0.45]{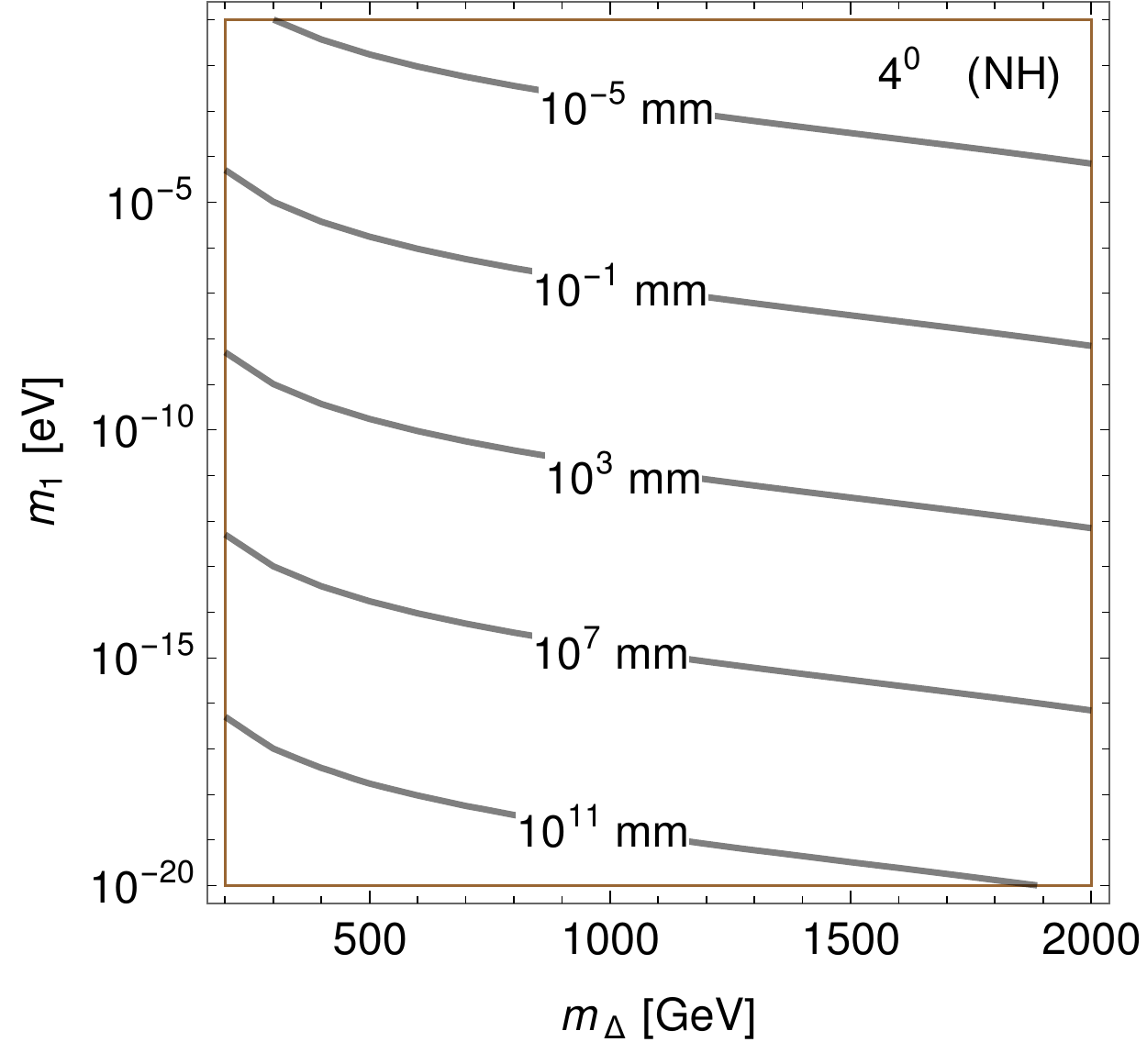}
\includegraphics[scale=0.45]{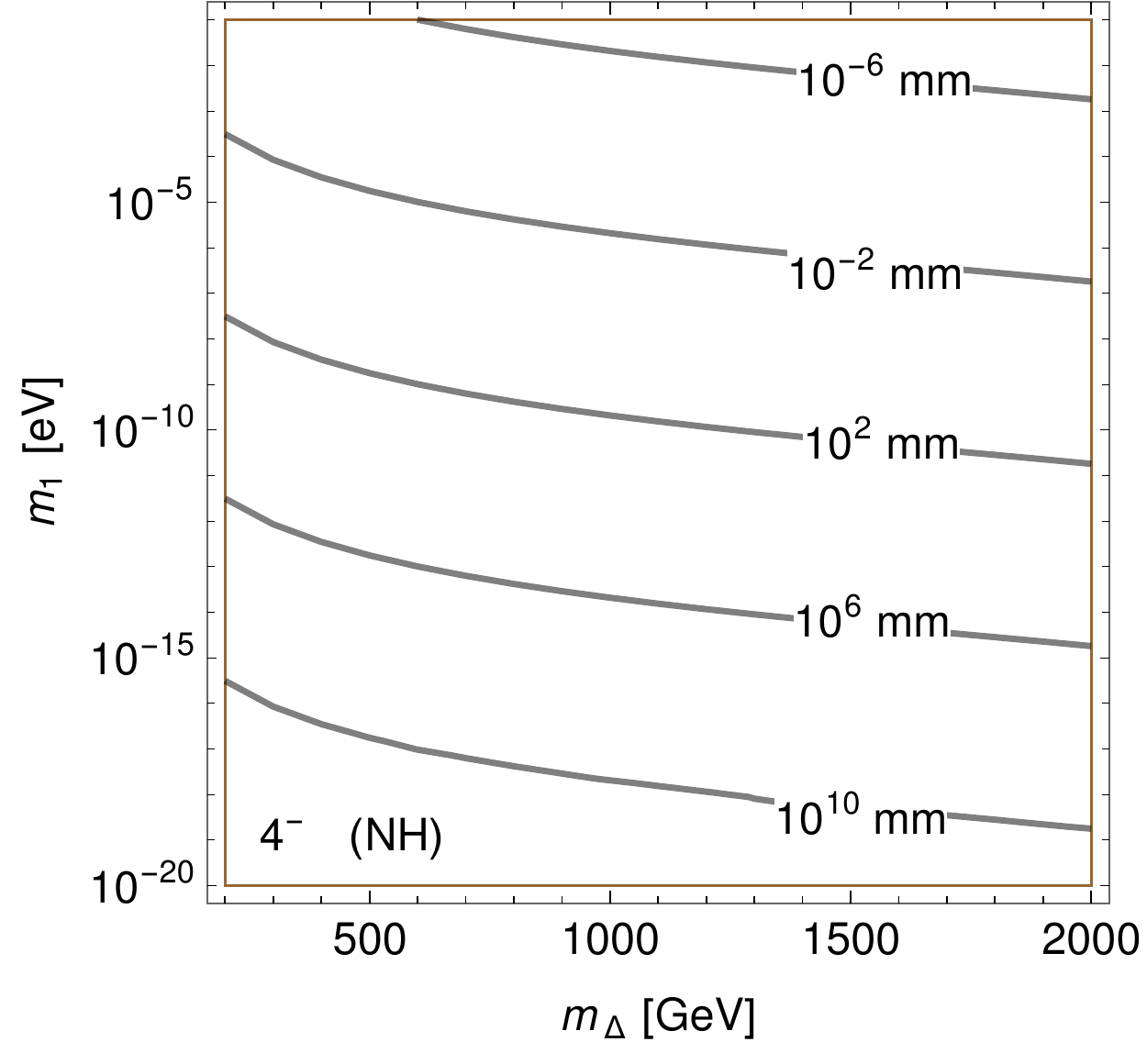}
\caption{Contour plots for decay length in $(m_1-M_\chi)$ plane for $4^{++},4^+,4^0$ and $4^-$, respectively, for normal hierarchical light neutrino mass spectrum.}
\label{fig:contour_4}
\end{figure}

\begin{figure}[htb!]
\includegraphics[scale=0.39]{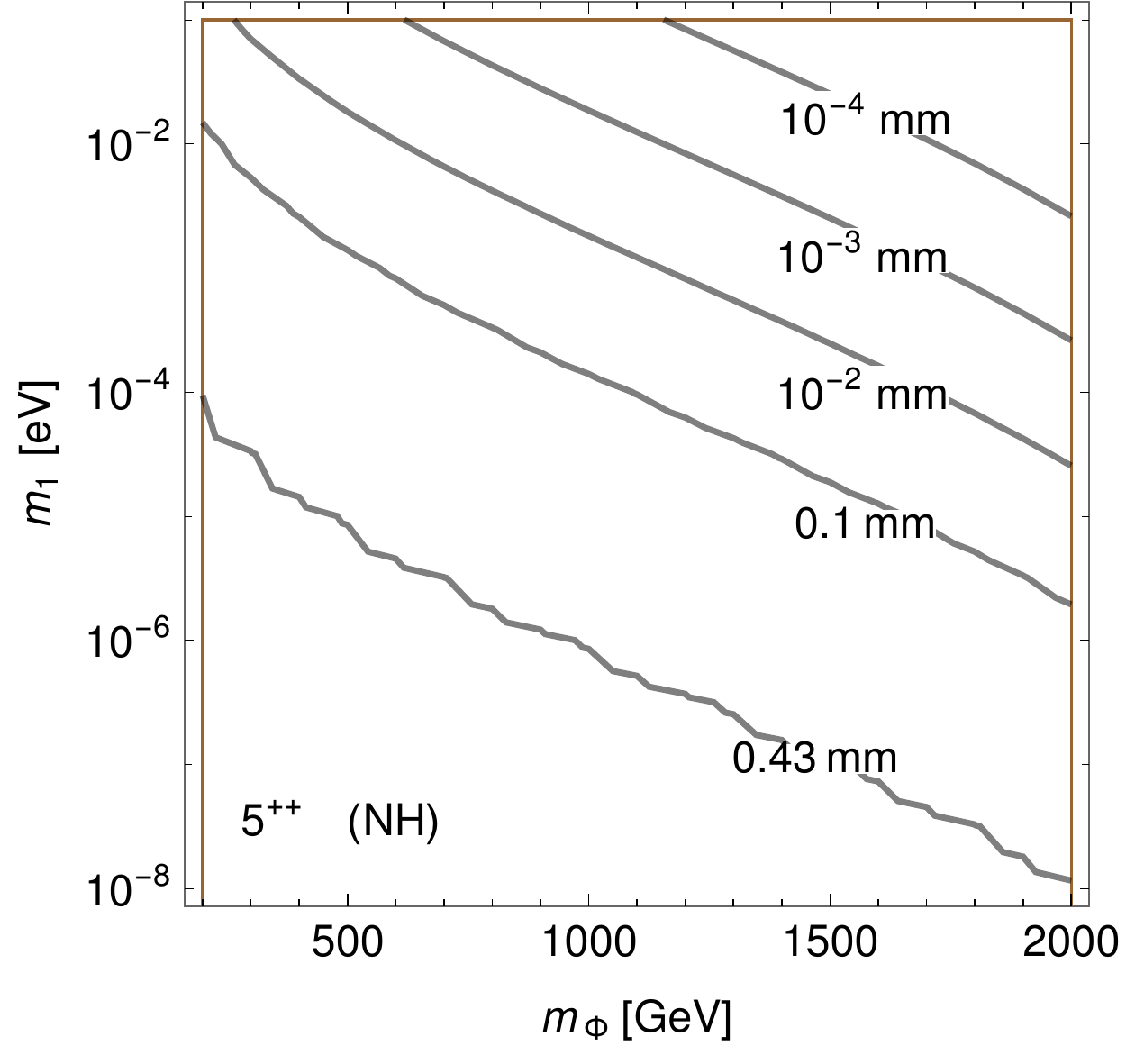}
\includegraphics[scale=0.39]{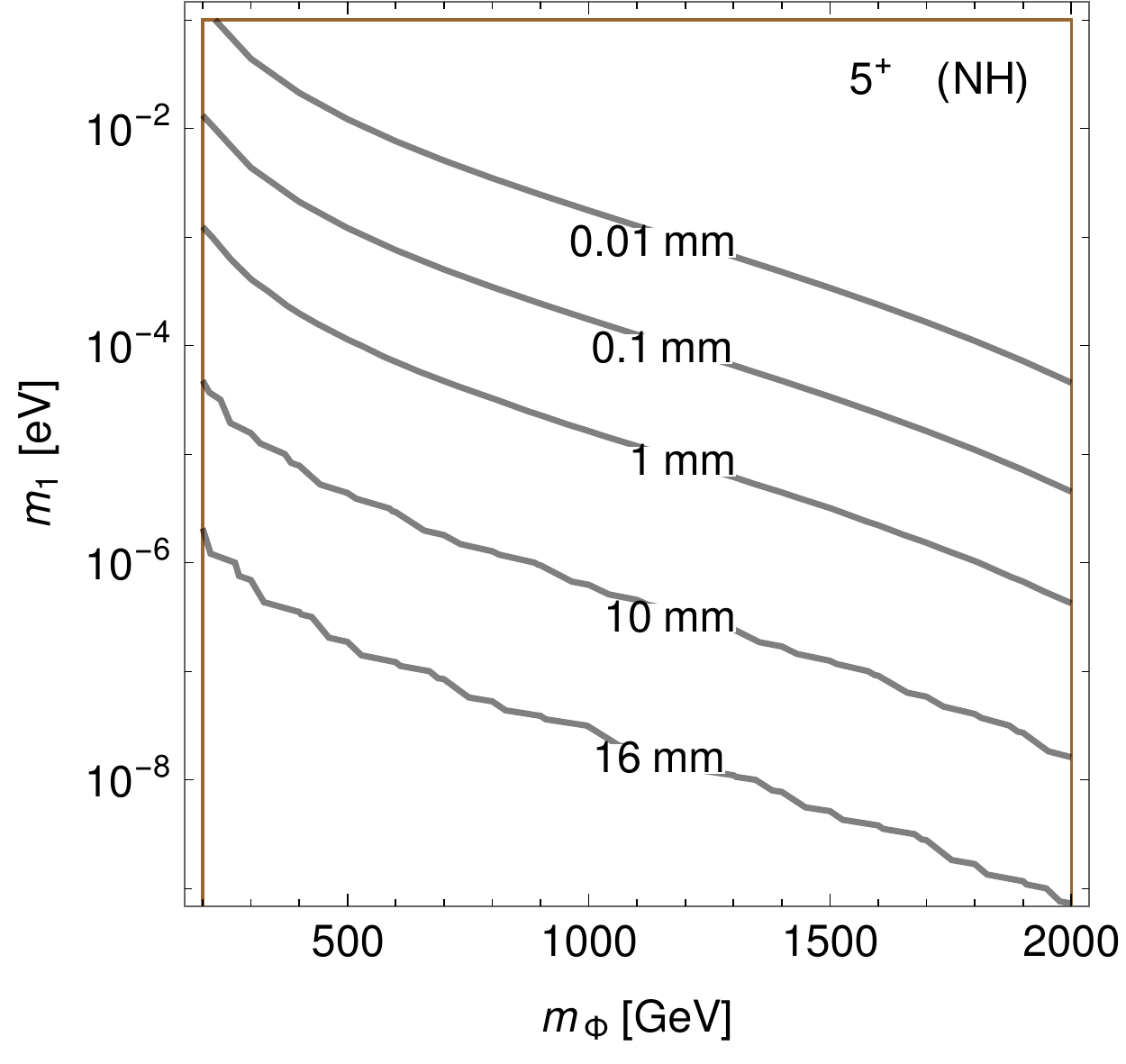}
\includegraphics[scale=0.39]{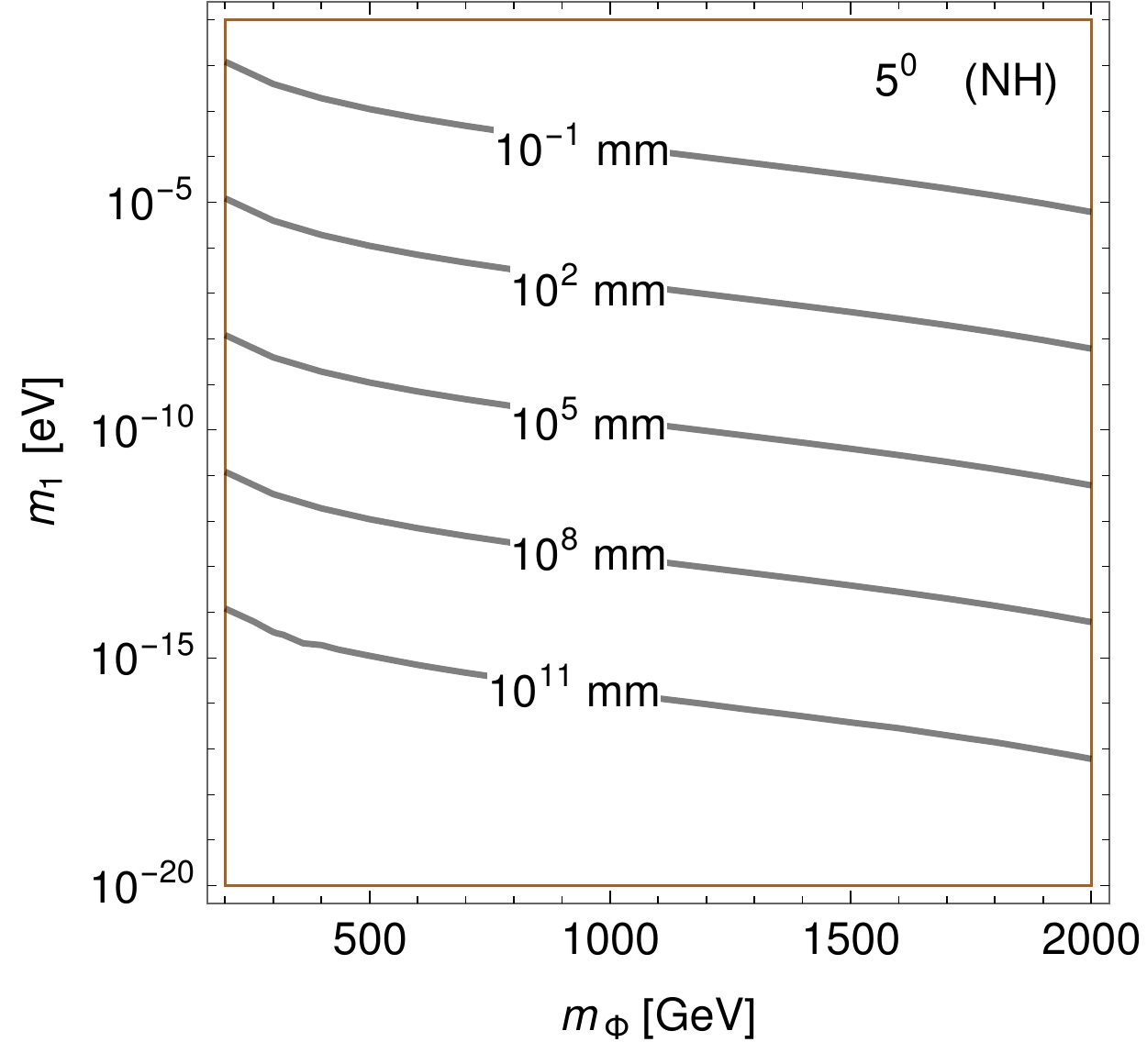}
\caption{Contour plots for decay lengths in $(m_1-m_\Delta)$ plane for $5^{++},5^+$ and $5^0$ for normal hierarchical light neutrino mass spectrum.}
\label{fig:contour_5}
\end{figure}

Figure \ref{fig:contour_3}, \ref{fig:contour_4} and \ref{fig:contour_5} show contour plots for $c\tau$ in $(m_1-m_\chi)$ plane assuming triplet, quadruplet and quintuplet, respectively, as the lightest\footnote{For each plot, once again, the bare mass parameters in the Lagrangian are chosen such that the corresponding multiplet is the lightest among all the multiplets.} multiplet. 
Figure \ref{fig:contour_3} shows contour plots for $c\tau$ in $(m_1-m_\Sigma)$ plane, respectively, for $3^{++},3^+$ and $3^0$ for normal hierarchical light neutrino mass spectrum. The decay lengths of $3^{++}$ and $3^+$ are almost constant ($\approx c\tau_{\rm max}$) for $m_1 \leq 10^{-9}$ and $10^{-12}$ eV, respectively. The reason for such constant $c\tau$ feature has been explained at the end of section \ref{sec:decay}. They have maximum decay length, $c\tau_{\rm max}$, of $\sim 0.04$ and 1 mm, respectively. So, one does not expect to observe the displaced vertex signatures of $3^{++}$ at the LHC (HL-LHC)\footnote{However, signals with $c\tau \in [0.01,100]$ mm can be observed by proposed $ep$-colliders like LHeC and FCC-he.}. Unlike $3^{++}$ and $3^+$, $c\tau$ for $3^0$ could be arbitrarily large as $m_1 (m_3) \to 0$ for NH (IH). So, a very limited region of $m_1-m_\Sigma$ parameters' space can be probed at MATHUSLA. For example, $m_\Sigma \sim 1$ TeV and $m_1 \in [\mathcal{O}(10^{-19})-\mathcal{O}(10^{-18})]$ eV give rise to ${\cal O}(100~{\rm m})$ decay length for $3^0$ and can be probed at MATHUSLA. On the other hand, $3^+$ with the maximum decay length 1 mm decays into very soft (too soft to be visible) poins and $3^0$ (which remains stable and invisible inside the LHC detector) and hence, gives rise to interesting disappearing track signature for $m_1 < 10^{-10}$ eV at future ${\it ep}$-colliders like LHeC and FCC-he. 

Figure \ref{fig:contour_4} shows contour plots for $c\tau$ in $(m_1-m_\Delta)$ plane for $4^{++},4^+,4^0$ and $4^-$. For $m_1 \leq 10^{-9}$ eV, $4^{++}$ and $4^+$ have constant $c\tau$. We see that $4^{++}$ and $4^+$ have $c\tau_{\rm max}$ of $\sim 0.22$ and $3.6$ mm, respectively. So, one would expect disappearing track signatures at future ${\it ep}$-colliders like LHeC and FCC-he for $m_1 \leq 10^{-9}$. Like $3^0$, $c\tau$ for $4^0$ and $4^-$ could be arbitrarily large as $m_1 (m_3) \to 0$ for NH (IH) \footnote{The mass splitting between $4^-$ and $4^0$ being smaller than $m_\pi,m_K$ and $m_\mu$, the only kinematically accessible decay for $4^-$ is $4^-\to 4^0 e^- \nu$. Though, in principle, $4^-$ undergoes heavy state transition to $4^0$, the corresponding decay width is negligibly small. This is why, like $4^0$, $c\tau$ for $4^-$ could be arbitrarily large as $m_1(m_3) \to 0$.}. Similar to the $3^0$ case, a very limited region of $m_1-m_\Delta$ parameters' space can be probed at MATHUSLA for both $4^0$ and $4^-$. For example, for $m_\Delta \sim 1$ TeV, $m_1 \in [\mathcal{O}(10^{-13})-\mathcal{O}(10^{-14})]$ eV can be probed at MATHUSLA.
Figure \ref{fig:contour_5} shows contour plots for $c\tau$ in $(m_1-m_\Phi)$ plane, respectively, for $5^{++},5^+$ and $5^0$ for normal hierarchical light neutrino mass spectrum. For $m_1 < 10^{-8}$ eV, $5^{++}$ and $5^+$ have almost constant decay length. They have sizeable maximum decay length of $\sim 0.43$ and $16$ mm, respectively. The disappearing track signatures produced by $5^{++}$ and $5^+$ can be probed at future {\it ep}-colliders for a significant part of the $m_1-m_\Phi$ plane. $c\tau$ for $5^0$ could be arbitrarily large as $m_1 (m_3) \to 0$ for NH (IH). Like $3^0$ and $4^0$, a very limited region of $m_1-m_\Phi$ parameters' space can be probed at MATHUSLA in case of $5^0$. For example, for $m_\Phi \sim 1$ TeV, $m_1 \in [\mathcal{O}(10^{-11})-\mathcal{O}(10^{-10})]$ eV can be probed at MATHUSLA.

Note that one would get similar disappearing track signatures for the charged fermions with $c\tau_{\rm max} \sim 5.6$ cm \cite{dis_typeiii,high_dim_7} in the type-III seesaw model. Also, the neutral fermion in the type-III seesaw model has a similar feature for the decay length. So, unless the lightest neutrino mass is known or very close to zero, distinguishing the triplet fermions in the type-III seesaw model from the other $SU(2)_L$ multiplets in this type of seesaw-inspired model using disappearing track signatures will be challenging.

In this section, a qualitative discussion on the possible characteristic signatures of LLPs has been presented. The numbers are to be taken with a grain of salt, none the less suggest that, in this model, for a limited region of parameters' space, the possible characteristic signatures of LLPs can be probed at future colliders. It would be prepossessing to undertake a detailed analysis for the same.

\section{Summary and conclusions}
\label{sec:conclusion}
We have studied a genuine model that generates neutrino masses at tree-level via low-energy effective operator with mass-dimension-9. A model is said to be `genuine' at a given dimension, if all the lower dimensional contributions to neutrino masses are automatically absent without the need for any additional $U(1)$ or $\mathbb{Z}_n$ symmetry. The introduction of such higher dimensional operator brings down the LNV mass scale to TeV making such model potentially testable, and hence falsifiable in the present or near-future collider. This model possesses several new $SU(2)_L$ fermionic multiplets --- vector-like $SU(2)_L$ triplet, $SU(2)_L$ quadruplet and chiral quintuplet --- three copies of each, and thus a rich phenomenology at the LHC. The observed neutrino masses and mixing angles can dexterously be accommodated in this model. The well-known Casas-Ibarra parameterization has been used to find the general texture for the Yukawa matrix $Y_{23}$ which couples the left-handed triplet with the SM lepton doublet and Higgs such that it reproduces the observed neutrino masses and mixing angles. LFV arises very naturally in such setup. Calculation of the decay rates of LFV magnetic transitions $\mu \to e\gamma$ and $\tau \to \ell \gamma$ ($\ell=e,\mu$) are presented. We obtained constraints on the Yukawa couplings and exotic fermion masses using the current experimental bounds on LFV processes. We have studied the production and subsequent decays of the new fermions in this model. Both Drell-Yan and photon-induced (photon-photon fusion) production processes of the exotic fermions have been considered. For a set of Yukawa couplings motivated from the current experimental bounds on LFV processes, $95\%$ CL upper limits on the total production cross section of exotic pairs of a heavy fermionic multiplet have been estimated using a recent CMS search based on multilepton final states. Pair production of exotic fermions for masses below 720, 970 and 1200 GeV are excluded for triplet, quadruplet and quintuplet, respectively. Note that these limits are not applicable if the exotics are long-lived. Although these limits are insensitive to the values of the Yukawa couplings $y_{34}^\prime$ and $y_{45}^\prime$ as long as exotic fermions decay promptly, a different texture for the Yukawa matrix $Y_{23}$ along with non-degenerate multiplet generations might lead to completely different limits. We have also briefly discussed the possibility that the exotics of the model could also be long-lived leaving disappearing track signatures or displaced vertex at the detector. The doubly charged fermions can have maximum decay length of 0.04, 0.22 and 0.43 mm for triplet, quadruplet and quintuplet, respectively. The singly charged fermions have maximum decay length of 1, 3.6 and 16 mm for triplet, quadruplet and quintuplet, respectively. Hence, one doesn't expect any displaced vertex/disappearing track signature for the charged fermions at the LHC, but one would expect disappearing track signatures resulting from the charged fermions at the future {\emph ep}-colliders like LHeC or FCC-he. We found that the decay lengths for neutral fermions can be arbitrarily large as $m_1$ approaches zero, and, hence, a limited region of $m_1-m_\chi$ parameters' space can be probed at MATHUSLA detector.

We mention in closing that the final states discussed in this work are very common to a large class of models. The multilepton final states or the LNV signatures arise in most of the seesaw-inspired models. Also, the disappearing track signatures and other displaced vertex signatures for fermions in this model could be quite similar to those in other seesaw models. Once a positive result from a search is found, one has to identify whether it corresponds to heavy neutrinos (type-I seesaw), scalar triplets (type-II seesaw), fermionic triplets (type-III seesaw) or any other seesaw-inspired (e.g., the model presented in this work or in \cite{high_dim_5,high_dim_6,high_dim_7,kumericki}) model. This can be done by analyzing the cross sections, phase space distributions \cite{aguila}, {\it etc}. This is beyond the scope of this article to address all these issues. However, this article will surely pave the way for such phenomenological studies.

\appendix
\section{Mass matrices for differently charged leptons and their diagonalisations}
\label{app:mass_matrices}
\paragraph{Neutral leptons}
The mass matrix for neutral leptons is given by
\small
\begin{equation*}
-\mathcal{L}_0 =\left(\overline{\tilde{\nu}_L} ~ \overline{\Sigma^0_R} ~\overline{\tilde{\Sigma}^0_L} ~ \overline{\Delta^0_R} ~ \overline{\tilde{\Delta}^0_L} ~ \overline{\Phi^0_R}\right)
\left(
\begin{array}{cccccc}
0 & 0 & \frac{v Y_{23}^\dagger}{\sqrt{2}} & 0 & 0 & 0 \\
0 & 0 & M_\Sigma & 0 & -\frac{v Y_{34}^{\prime \dagger}}{\sqrt{6}} & 0 \\
\frac{v Y_{23}^*}{\sqrt{2}} & M_\Sigma & 0 & -\frac{v Y_{34}^{T}}{\sqrt{6}} & 0 & 0 \\
0 & 0 & -\frac{v Y_{34}}{\sqrt{6}} & 0 & M_\Delta & \frac{v Y_{45}^\prime}{2} \\
0 & -\frac{v Y_{34}^{\prime *}}{\sqrt{6}} & 0 & M_\Delta & 0 & \frac{v Y_{45}^*}{2}\\
0 & 0 & 0 & \frac{v Y_{45}^{\prime T}}{2} & \frac{v Y_{45}^\dagger}{2} & M_\Phi
\end{array}
\right) \left( \begin{array}{c}
\nu_L \\ \tilde{\Sigma}^0_R \\ \Sigma^0_L \\ \tilde{\Delta}^0_R \\ \Delta^0_L \\ \tilde{\Phi}^0_R 
\end{array} \right)~.
\end{equation*}
\normalsize

\noindent The above symmetric mass matrix can be block diagonalised by the unitary transformation
\begin{equation}
\label{eq:inter_gauge_states}
\left( \begin{array}{c}
\nu_L \\ \tilde{\Sigma}^0_R \\ \Sigma^0_L \\ \tilde{\Delta}^0_R \\ \Delta^0_L \\ \tilde{\Phi}^0_R 
\end{array} \right)= U^0 \left( \begin{array}{c}
\nu_{iL} \\ \tilde{\Sigma}^0_{iR} \\ \Sigma^0_{iL} \\ \tilde{\Delta}^0_{iR} \\ \Delta^0_{iL} \\ \tilde{\Phi}^0_{iR} 
\end{array} \right)~,
\end{equation}
such that
\begin{equation*}
U^{0T} \left(\begin{array}{cccccc}
0 & 0 & \frac{v Y_{23}^\dagger}{\sqrt{2}} & 0 & 0 & 0 \\
0 & 0 & M_\Sigma & 0 & -\frac{v Y_{34}^{\prime \dagger}}{\sqrt{6}} & 0 \\
\frac{v Y_{23}^*}{\sqrt{2}} & M_\Sigma & 0 & -\frac{v Y_{34}^{T}}{\sqrt{6}} & 0 & 0 \\
0 & 0 & -\frac{v Y_{34}}{\sqrt{6}} & 0 & M_\Delta & \frac{v Y_{45}^\prime}{2} \\
0 & -\frac{v Y_{34}^{\prime *}}{\sqrt{6}} & 0 & M_\Delta & 0 & \frac{v Y_{45}^*}{2}\\
0 & 0 & 0 & \frac{v Y_{45}^{\prime T}}{2} & \frac{v Y_{45}^\dagger}{2} & M_\Phi
\end{array}\right) U^0 
\approx 
\left(\begin{array}{cccccc}
m_\nu & 0 & 0 & 0 & 0 & 0 \\
0 & 0 & M_\Sigma & 0 & 0 & 0 \\
0 & M_\Sigma & 0 & 0 & 0 & 0 \\
0 & 0 & 0 & 0 & M_\Delta & 0 \\
0 & 0 & 0 & M_\Delta & 0 & 0\\
0 & 0 & 0 & 0 & 0 & M_\Phi
\end{array}\right)~,
\end{equation*}
where
\begin{equation}
m_\nu \approx -\frac{v^6}{48} Y_{23}^\dagger M_\Sigma^{-1} Y_{34}^{\prime \dagger} M_\Delta^{-1} Y_{45}^\prime M_\phi^{-1} Y_{45}^{\prime T} M_\Delta^{-1} Y_{34}^{\prime *} M_\Sigma^{-1}  Y_{23}^*
\end{equation}
and
\begin{equation*}
U^0=\left( \begin{array}{cccccc}
U^0_{\nu_L \nu_L} & U^0_{\nu_L \tilde{\Sigma}^0_R} & U^0_{\nu_L \Sigma^0_L} & U^0_{\nu_L \tilde{\Delta}^{0}_R} & U^0_{\nu_L \Delta^0_L} & U^0_{\nu_L \tilde{\Phi}^{0}_R}
\\
U^0_{\tilde{\Sigma}^0_R \nu_L} & U^0_{\tilde{\Sigma}^0_R \tilde{\Sigma}^0_R} & U^0_{\tilde{\Sigma}^0_R \Sigma^0_L} & U^0_{\tilde{\Sigma}^0_R \tilde{\Delta}^0_R} & U^0_{\tilde{\Sigma}^0_R \Delta^0_L} & U^0_{\tilde{\Sigma}^0_R \tilde{\Phi}^0_R}
\\
U^0_{\Sigma^0_L \nu_L} & U^0_{\Sigma^0_L \tilde{\Sigma}^0_R} & U^0_{\Sigma^0_L \Sigma^0_L} & U^0_{\Sigma^0_L \tilde{\Delta}^0_R} & U^0_{\Sigma^0_L \Delta^0_L} & U^0_{\Sigma^0_L \tilde{\Phi}^0_R}
\\
U^0_{\tilde{\Delta}^0_R \nu_L} & U^0_{\tilde{\Delta}^0_R \tilde{\Sigma}^0_R} & U^0_{\tilde{\Delta}^0_R \Sigma^0_L} & U^0_{\tilde{\Delta}^0_R \tilde{\Delta}^0_R} & U^0_{\tilde{\Delta}^0_R \Delta^0_L} & U^0_{\tilde{\Delta}^0_R \tilde{\Phi}^0_R}
\\
U^0_{\Delta^0_L \nu_L} & U^0_{\Delta^0_L \tilde{\Sigma}^0_R} & U^0_{\Delta^0_L \Sigma^0_L} & U^0_{\Delta^0_L \tilde{\Delta}^0_R} & U^0_{\Delta^0_L \Delta^0_L} & U^0_{\Delta^0_L \tilde{\Phi}^0_R}
\\
U^0_{\tilde{\Phi}^0_R \nu_L} & U^0_{\tilde{\Phi}^0_R \tilde{\Sigma}^0_R} & U^0_{\tilde{\Phi}^0_R \Sigma^0_L} & U^0_{\tilde{\Phi}^0_R \tilde{\Delta}^0_R} & U^0_{\tilde{\Phi}^0_R \Delta^0_L} & U^0_{\tilde{\Phi}^0_R \tilde{\Phi}^0_R}
\end{array} \right)~.
\end{equation*}

\noindent  While diagonalising the mass matrices, corrections to the heavy lepton masses $M_\Sigma$, $M_\Delta$ and $M_\Phi$ are neglected for phenomenological considerations as they are relatively very small. Note that the eigenstates with subscript $i$ in equation \eqref{eq:inter_gauge_states} are neither mass eigenstates (denoted with subscript $m$) nor gauge eigenstates (denoted without any subscript) rather some kind of intermediate eigenstates. The elements of the mixing matrix $U^0$ is obtained following the procedure presented in Ref.\ \cite{Grimus_Lavoura} by expanding in powers of $M_\Sigma^{-1}$, $M_\Delta^{-1}$ and $M_\Phi^{-1}$. The following dimensionless quantities are used:
\begin{align*}
\epsilon_{23}^{(\prime)} &= \frac{v}{\sqrt{2}} Y_{23}^{(\prime)T} M_\Sigma^{-1}~,
\qquad
\epsilon_{34}^{(\prime)} = \frac{v}{\sqrt{6}} Y_{34}^{(\prime)T} M_\Delta^{-1}~,
\qquad
\delta_{34}^{(\prime)} = \frac{v}{\sqrt{6}} Y_{34}^{(\prime)} M_\Sigma^{-1}~,
\\
\epsilon_{45}^{(\prime)} &= \frac{v}{2} Y_{45}^{(\prime)*} M_\Phi^{-1}~,
\qquad \quad
\delta_{45}^{(\prime)} = \frac{v}{2} Y_{45}^{(\prime)\dagger} M_\Delta^{-1}~.
\end{align*}

\noindent  The elements of $U^0$ are:
\begin{multicols}{2}
\begin{subequations}
\begin{align*}
&U^0_{\nu_L \nu_L}=1-\frac{1}{2} \epsilon_{23} \epsilon_{23}^\dagger
\\
&U^0_{\nu_L \tilde{\Sigma}^0_R}=+\epsilon_{23}
\\
&U^0_{\nu_L \Sigma^0_L}=+\epsilon_{23} \epsilon^\prime_{34} \epsilon^\prime_{45} \delta^\prime_{45} \delta^\prime_{34}
\\
&U^0_{\nu_L \tilde{\Delta}^0_R}=+\epsilon_{23} \epsilon^\prime_{34}
\\
&U^0_{\nu_L \Delta^0_L}=+\epsilon_{23} \epsilon^\prime_{34} \epsilon^\prime_{45} \delta^\prime_{45}
\\
&U^0_{\nu_L \tilde{\Phi}^0_R}=-\epsilon_{23} \epsilon^\prime_{34} \epsilon^\prime_{45}
\\
&U^0_{\tilde{\Sigma}^0_R \nu_L}=-\epsilon_{23}^\dagger
\\
&U^0_{\Sigma^0_L \nu_L}=-\delta^{\prime \dagger}_{34} \delta^{\prime \dagger}_{45} \epsilon^{\prime \dagger}_{45} \epsilon^{\prime \dagger}_{34} \epsilon^\dagger_{23}
\\
&U^0_{\tilde{\Delta}^0_R \nu_L}=-\epsilon^{\prime \dagger}_{34} \epsilon^\dagger_{23}
\\
&U^0_{\Delta^0_L \nu_L}=-\delta^{\prime \dagger}_{45} \epsilon^{\prime \dagger}_{45} \epsilon^{\prime \dagger}_{34} \epsilon^\dagger_{23}
\\
&U^0_{\tilde{\Phi}^0_R \nu_L}=+\epsilon^{\prime \dagger}_{45} \epsilon^{\prime \dagger}_{34} \epsilon^\dagger_{23}
\\
&U^0_{\tilde{\Sigma}^0_R \tilde{\Sigma}^0_R}=1-\frac{1}{2}\epsilon^\dagger_{23} \epsilon_{23}
\\
&U^0_{\tilde{\Sigma}^0_R \Sigma^0_L}=-\frac{1}{2}\epsilon^\dagger_{23} \epsilon_{23} \epsilon^\prime_{34} \epsilon^\prime_{45} \delta^\prime_{45} \delta^\prime_{34}
\\
&U^0_{\tilde{\Sigma}^0_R \tilde{\Delta}^0_R}=-\frac{1}{2}\epsilon^\dagger_{23} \epsilon_{23} \epsilon^\prime_{34}
\\
&U^0_{\tilde{\Sigma}^0_R \Delta^0_L}=-\frac{1}{2}\epsilon^\dagger_{23} \epsilon_{23} \epsilon^\prime_{34} \epsilon^\prime_{45} \delta^\prime_{45}
\\
&U^0_{\tilde{\Sigma}^0_R \tilde{\Phi}^0_R}=+\frac{1}{2}\epsilon^\dagger_{23} \epsilon_{23} \epsilon^\prime_{34} \epsilon^\prime_{45}
\\
&U^0_{\Sigma^0_L \tilde{\Sigma}^0_R}=-\frac{1}{2} \delta^{\prime \dagger}_{34} \delta^{\prime \dagger}_{45} \epsilon^{\prime \dagger}_{45} \epsilon^{\prime \dagger}_{34} \epsilon^\dagger_{23} \epsilon_{23}
\\
&U^0_{\Sigma^0_L \Sigma^0_L}=1-\frac{1}{2} \delta^{\prime \dagger}_{34} \delta^{\prime \dagger}_{45} \epsilon^{\prime \dagger}_{45} \epsilon^{\prime \dagger}_{34} \epsilon^\dagger_{23} \epsilon_{23} \epsilon^\prime_{34} \epsilon^\prime_{45} \delta^\prime_{45} \delta^\prime_{34}
\\
&U^0_{\Sigma^0_L \tilde{\Delta}^0_R}=-\frac{1}{2} \delta^{\prime \dagger}_{34} \delta^{\prime \dagger}_{45} \epsilon^{\prime \dagger}_{45} \epsilon^{\prime \dagger}_{34} \epsilon^\dagger_{23} \epsilon_{23} \epsilon^\prime_{34}
\end{align*}
\begin{align*} 
\\
&U^0_{\Sigma^0_L \Delta^0_L}=-\frac{1}{2} \delta^{\prime \dagger}_{34} \delta^{\prime \dagger}_{45} \epsilon^{\prime \dagger}_{45} \epsilon^{\prime \dagger}_{34} \epsilon^\dagger_{23} \epsilon_{23} \epsilon^\prime_{34} \epsilon^\prime_{45} \delta^\prime_{45}
\\
&U^0_{\Sigma^0_L \tilde{\Phi}^0_R}=+\frac{1}{2} \delta^{\prime \dagger}_{34} \delta^{\prime \dagger}_{45} \epsilon^{\prime \dagger}_{45} \epsilon^{\prime \dagger}_{34} \epsilon^\dagger_{23} \epsilon_{23} \epsilon^\prime_{34} \epsilon^\prime_{45}
\\
&U^0_{\tilde{\Delta}^0_R \tilde{\Sigma}^0_R}=-\frac{1}{2} \epsilon^{\prime \dagger}_{34} \epsilon^\dagger_{23} \epsilon_{23}
\\
&U^0_{\tilde{\Delta}^0_R \Sigma^0_L}=-\frac{1}{2} \epsilon^{\prime \dagger}_{34} \epsilon^\dagger_{23} \epsilon_{23} \epsilon^\prime_{34} \epsilon^\prime_{45} \delta^\prime_{45} \delta^\prime_{34}
\\
&U^0_{\tilde{\Delta}^0_R \tilde{\Delta}^0_R}=1-\frac{1}{2} \epsilon^{\prime \dagger}_{34} \epsilon^\dagger_{23} \epsilon_{23} \epsilon^\prime_{34}
\\
&U^0_{\tilde{\Delta}^0_R \Delta^0_L}=-\frac{1}{2}\epsilon^{\prime \dagger}_{34} \epsilon^\dagger_{23} \epsilon_{23} \epsilon^\prime_{34} \epsilon^\prime_{45} \delta^\prime_{45}
\\
&U^0_{\tilde{\Delta}^0_R \tilde{\Phi}^0_R}=+\frac{1}{2} \epsilon^{\prime \dagger}_{34} \epsilon^\dagger_{23} \epsilon_{23} \epsilon^\prime_{34} \epsilon^\prime_{45}
\\
&U^0_{ \Delta^0_L \tilde{\Sigma}^0_R}=-\frac{1}{2} \delta^{\prime \dagger}_{45} \epsilon^{\prime \dagger}_{45} \epsilon^{\prime \dagger}_{34} \epsilon^\dagger_{23} \epsilon_{23}
\\
&U^0_{\Delta^0_L \Sigma^0_L}=-\frac{1}{2} \delta^{\prime \dagger}_{45} \epsilon^{\prime \dagger}_{45} \epsilon^{\prime \dagger}_{34} \epsilon^\dagger_{23} \epsilon_{23} \epsilon^\prime_{34} \epsilon^\prime_{45} \delta^\prime_{45} \delta^\prime_{34}
\\
&U^0_{ \Delta^0_L \tilde{\Delta}^0_R}=-\frac{1}{2} \delta^{\prime \dagger}_{45} \epsilon^{\prime \dagger}_{45} \epsilon^{\prime \dagger}_{34} \epsilon^\dagger_{23} \epsilon_{23} \epsilon^\prime_{34}
\\
&U^0_{\Delta^0_L \Delta^0_L}=1-\frac{1}{2}\delta^{\prime \dagger}_{45} \epsilon^{\prime \dagger}_{45} \epsilon^{\prime \dagger}_{34} \epsilon^\dagger_{23} \epsilon_{23} \epsilon^\prime_{34} \epsilon^\prime_{45} \delta^\prime_{45}
\\
&U^0_{\Delta^0_L \tilde{\Phi}^0_R}=+\frac{1}{2}\delta^{\prime \dagger}_{45} \epsilon^{\prime \dagger}_{45} \epsilon^{\prime \dagger}_{34} \epsilon^\dagger_{23} \epsilon_{23} \epsilon^\prime_{34} \epsilon^\prime_{45}
\\
&U^0_{\tilde{\Phi}^0_R \tilde{\Sigma}^0_R}=+\frac{1}{2} \epsilon^{\prime \dagger}_{45} \epsilon^{\prime \dagger}_{34} \epsilon^\dagger_{23} \epsilon_{23}
\\
&U^0_{\tilde{\Phi}^0_R \Sigma^0_L}=+\frac{1}{2} \epsilon^{\prime \dagger}_{45} \epsilon^{\prime \dagger}_{34} \epsilon^\dagger_{23} \epsilon_{23} \epsilon^\prime_{34} \epsilon^\prime_{45} \delta^\prime_{45} \delta^\prime_{34}
\\
&U^0_{\tilde{\Phi}^0_R \tilde{\Delta}^0_R}=+\frac{1}{2} \epsilon^{\prime \dagger}_{45} \epsilon^{\prime \dagger}_{34} \epsilon^\dagger_{23} \epsilon_{23} \epsilon^\prime_{34}
\\
&U^0_{\tilde{\Phi}^0_R \Delta^0_L}=+\frac{1}{2} \epsilon^{\prime \dagger}_{45} \epsilon^{\prime \dagger}_{34} \epsilon^\dagger_{23} \epsilon_{23} \epsilon^\prime_{34} \epsilon^\prime_{45} \delta^\prime_{45}
\\
&U^0_{\tilde{\Phi}^0_R \tilde{\Phi}^0_R}= 1-\frac{1}{2} \epsilon^{\prime \dagger}_{45} \epsilon^{\prime \dagger}_{34} \epsilon^\dagger_{23} \epsilon_{23} \epsilon^\prime_{34} \epsilon^\prime_{45}
\end{align*}
\end{subequations}
\end{multicols}

\noindent  Further, the unitary transformation
\begin{equation*}
\left( \begin{array}{c}
\nu_{iL} \\ \tilde{\Sigma}^0_{iR} \\ \Sigma^0_{iL} \\ \tilde{\Delta}^0_{iR} \\ \Delta^0_{iL} \\ \tilde{\Phi}^0_{iR} 
\end{array} \right)= U^1 \left( \begin{array}{c}
\nu_{mL} \\ \Sigma^1_{mL} \\ \Sigma^2_{mL} \\ \Delta^1_{mL} \\ \Delta^2_{mL} \\ \tilde{\Phi}^0_{mR} 
\end{array} \right)
\end{equation*}
completely diagonalises the (block diagonal) neutral mass matrix according to 
\begin{equation*}
U^{1T} \left(\begin{array}{cccccc}
m_\nu & 0 & 0 & 0 & 0 & 0 \\
0 & 0 & M_\Sigma & 0 & 0 & 0 \\
0 & M_\Sigma & 0 & 0 & 0 & 0 \\
0 & 0 & 0 & 0 & M_\Delta & 0 \\
0 & 0 & 0 & M_\Delta & 0 & 0 \\
0 & 0 & 0 & 0 & 0 & M_\Phi
\end{array}\right) U^1
=
\left(\begin{array}{cccccc}
\hat{m_\nu} & 0 & 0 & 0 & 0 & 0\\
0 & -M_\Sigma & 0 & 0 & 0 & 0\\
0 & 0 & M_\Sigma & 0 & 0 & 0\\
0 & 0 & 0 & -M_\Delta & 0 & 0\\
0 & 0 & 0 & 0 & M_\Delta & 0\\
0 & 0 & 0 & 0 & 0 & M_\Phi
\end{array} \right)
\end{equation*} 
with 
\begin{equation*}
U^1=\left(\begin{array}{cccccc}
V_{\rm PMNS} & 0 & 0 & 0 & 0 & 0\\
0 & -\frac{1}{\sqrt{2}} & \frac{1}{\sqrt{2}} & 0 & 0 & 0\\
0 & \frac{1}{\sqrt{2}} & \frac{1}{\sqrt{2}} & 0 & 0 & 0\\
0 & 0 & 0 & -\frac{1}{\sqrt{2}} & \frac{1}{\sqrt{2}} & 0\\
0 & 0 & 0 & \frac{1}{\sqrt{2}} & \frac{1}{\sqrt{2}} & 0\\
0 & 0 & 0 & 0 & 0 & 1
\end{array} \right)
\end{equation*}
and 
\begin{equation*}
\hat{m}_\nu=V_{\rm PMNS}^T m_\nu V_{\rm PMNS}~,
\end{equation*}
where $\hat{m}_\nu = {\rm diag}(m_1,m_2,m_3)$ is the diagonal light neutrino mass matrix and $V_{\rm PMNS}$ is the so-called Pontecorvo-Maki-Nakagawa-Sakata (light neutrino mixing) matrix. Note that $\Sigma^1_{mL}$ and $\Sigma^2_{mL}$ are not dominantly $\tilde{\Sigma}^0_{iR}$ or $\Sigma^0_{iL}$, rather equal-weight admixure of $\Sigma^0_{iL}$ and $\tilde{\Sigma}^0_{iR}$; same follows for $\Delta^1_{mL}$ and $\Delta^2_{mL}$:
\begin{align*}
\Sigma^1_{mL}&=-\frac{1}{\sqrt{2}} \tilde{\Sigma}^0_{iR} + \frac{1}{\sqrt{2}} \Sigma^0_{iL}~, \qquad \Sigma^2_{mL}=\frac{1}{\sqrt{2}} \tilde{\Sigma}^0_{iR} + \frac{1}{\sqrt{2}} \Sigma^0_{iL}~, \nonumber
\\
\Delta^1_{mL}&=-\frac{1}{\sqrt{2}} \tilde{\Delta}^0_{iR} + \frac{1}{\sqrt{2}} \Delta^0_{iL}~, \qquad \Delta^2_{mL}=\frac{1}{\sqrt{2}} \tilde{\Delta}^0_{iR} + \frac{1}{\sqrt{2}} \Delta^0_{iL}~.
\end{align*}
Finally, the gauge eingenstates and the mass eigenstates are related by the unitary transformation
\begin{equation*}
\left( \begin{array}{c}
\nu_L \\ \tilde{\Sigma}^0_R \\ \Sigma^0_L \\ \tilde{\Delta}^0_R \\ \Delta^0_L \\ \tilde{\Phi}^0_R 
\end{array} \right)= U \left( \begin{array}{c}
\nu_{mL} \\ \Sigma^1_{mL} \\ \Sigma^2_{mL} \\ \Delta^1_{mL} \\ \Delta^2_{mL} \\ \tilde{\Phi}^0_{mR} 
\end{array} \right),
\end{equation*}
where 
\begin{equation*}
U=U^0 U^1~.
\end{equation*}

\paragraph{Singly charged leptons}
The mass matrix for singly charged leptons is given by
\small
\begin{equation*}
-\mathcal{L}_1=\left( \begin{array}{ccccc}
\overline{\ell_R} & \overline{\tilde{\Sigma}^+_L} & \overline{\tilde{\Delta}^+_L} & \overline{\Delta^-_R} & \overline{\Phi^-_R}
\end{array} \right)
\left( \begin{array}{ccccc}
m_\ell & 0 & 0 & 0 & 0 \\
-\frac{v Y_{23}^*}{2} & M_\Sigma & -\frac{v Y_{34}^T}{\sqrt{3}} & 0 & 0 \\
0 & -\frac{v Y_{34}^{\prime *}}{\sqrt{3}} & M_\Delta & 0 & \frac{1}{2} \sqrt{\frac{3}{2}} v Y_{45}^*\\
0 & 0 & 0 & M_\Delta & -\frac{v Y_{45}^\prime}{2 \sqrt{2}} \\
0 & 0 & -\frac{1}{2} \sqrt{\frac{3}{2}} v Y_{45}^{\prime T} & \frac{v Y_{45}^\dagger}{2 \sqrt{2}} & -M_\Phi
\end{array} \right)
\left( \begin{array}{c}
\ell_L \\ \tilde{\Sigma}^+_R \\ \tilde{\Delta}^+_R \\ \Delta^-_L \\ \tilde{\Phi}^+_R
\end{array} \right)~,
\end{equation*} 
\normalsize
where $m_\ell=\frac{vY_\ell}{\sqrt{2}}$ is the SM charged lepton mass. The mass matrix for singly charged leptons can be diagonalised by the bi-unitary transformation
\begin{equation*}
\left( \begin{array}{c}
\ell_L \\ \tilde{\Sigma}^+_R \\ \tilde{\Delta}^+_R \\ \Delta^-_L \\ \tilde{\Phi}^+_R
\end{array} \right)=U^L \left( \begin{array}{c}
\ell_{mL} \\ \tilde{\Sigma}^+_{mR} \\ \tilde{\Delta}^+_{mR} \\ \Delta^-_{mL} \\ \tilde{\Phi}^+_{mR} \end{array} \right)~, \qquad \left( \begin{array}{c}
\ell_R \\ \tilde{\Sigma}^+_L \\ \tilde{\Delta}^+_L \\ \Delta^-_R \\ \Phi^-_{R}
\end{array} \right)=U^R \left( \begin{array}{c}
\ell_{mR} \\ \tilde{\Sigma}^+_{mL} \\ \tilde{\Delta}^+_{mL} \\ \Delta^-_{mR} \\ \Phi^-_{mR}
\end{array} \right)
\end{equation*}
according to
\begin{equation*}
U_R^\dagger \left( \begin{array}{ccccc}
m_\ell & 0 & 0 & 0 & 0 \\
-\frac{v Y_{23}^*}{2} & M_\Sigma & -\frac{v Y_{34}^T}{\sqrt{3}} & 0 & 0 \\
0 & -\frac{v Y_{34}^{\prime *}}{\sqrt{3}} & M_\Delta & 0 & \frac{1}{2} \sqrt{\frac{3}{2}} v Y_{45}^*\\
0 & 0 & 0 & M_\Delta & -\frac{v Y_{45}^\prime}{2 \sqrt{2}} \\
0 & 0 & -\frac{1}{2} \sqrt{\frac{3}{2}} v Y_{45}^{\prime T} & \frac{v Y_{45}^\dagger}{2 \sqrt{2}} & -M_\Phi
\end{array} \right) U_L 
\approx 
\left( \begin{array}{ccccc}
m_\ell & 0 & 0 & 0 & 0
\\ 
0 & M_\Sigma & 0 & 0 & 0 
\\
0 & 0 & M_\Delta & 0 & 0
\\
0 & 0 & 0 & M_\Delta & 0 
\\
0 & 0 & 0 & 0 & -M_\Phi
\end{array} \right)~.
\end{equation*}
where 
\begin{equation*}
U_L = \left( \begin{array}{ccccc}
U^L_{\ell \ell} & U^L_{\ell \tilde{\Sigma}^+_R} & U^L_{\ell \tilde{\Delta}^+_R} & U^L_{\ell \Delta^-_{L}} & U^L_{\ell \tilde{\Phi}^+_R} 
\\
U^L_{\tilde{\Sigma}^+_R \ell} & U^L_{\tilde{\Sigma}^+_R \tilde{\Sigma}^+_R} & U^L_{\tilde{\Sigma}^+_R \tilde{\Delta}^+_R} & U^L_{\tilde{\Sigma}^+_R \Delta^-_{L}} & U^L_{\tilde{\Sigma}^+_R \tilde{\Phi}^+_R}
\\
U^L_{\tilde{\Delta}^+_R \ell} & U^L_{\tilde{\Delta}^+_R \tilde{\Sigma}^+_R} & U^L_{\tilde{\Delta}^+_R \tilde{\Delta}^+_R} & U^L_{\tilde{\Delta}^+_R \Delta^-_{L}} & U^L_{\tilde{\Delta}^+_R \tilde{\Phi}^+_R}
\\
U^L_{\Delta^-_{L} \ell} & U^L_{\Delta^-_L \tilde{\Sigma}^+_R} & U^L_{\Delta^-_L \tilde{\Delta}^+_R} & U^L_{\Delta^-_L \Delta^-_L} & U^L_{\Delta^-_L \tilde{\Phi}^+_R}
\\
U^L_{\tilde{\Phi}^+_R \ell} & U^L_{\tilde{\Phi}^+_R \tilde{\Sigma}^+_R} & U^L_{\tilde{\Phi}^+_R \tilde{\Delta}^+_R} & U^L_{\tilde{\Phi}^+_R \Delta^-_L} & U^L_{\tilde{\Phi}^+_R \tilde{\Phi}^+_R}
\end{array} \right)
\end{equation*}
\begin{equation*}
U_R = \left( \begin{array}{ccccc}
U^R_{\ell \ell} & U^R_{\ell \tilde{\Sigma}^+_L} & U^R_{\ell \tilde{\Delta}^+_L} & U^R_{\ell \Delta^-_R} & U^R_{\ell \Phi^-_R} 
\\
U^R_{\tilde{\Sigma}^+_L \ell} & U^R_{\tilde{\Sigma}^+_L \tilde{\Sigma}^+_L} & U^R_{\tilde{\Sigma}^+_L \tilde{\Delta}^+_L} & U^R_{\tilde{\Sigma}^+_L \Delta^-_R} & U^R_{\tilde{\Sigma}^+_L \Phi^-_R}
\\
U^R_{\tilde{\Delta}^+_L \ell} & U^R_{\tilde{\Delta}^+_L \tilde{\Sigma}^+_L} & U^R_{\tilde{\Delta}^+_L \tilde{\Delta}^+_L} & U^R_{\tilde{\Delta}^+_L \Delta^-_R} & U^R_{\tilde{\Delta}^+_L \Phi^-_R}
\\
U^R_{\Delta^-_R \ell} & U^R_{\Delta^-_R \tilde{\Sigma}^+_L} & U^R_{\Delta^-_R \tilde{\Delta}^+_L} & U^R_{\Delta^-_R \Delta^-_R} & U^R_{\Delta^-_R \Phi^-_R}
\\
U^R_{\Phi^-_R \ell} & U^R_{\Phi^-_R \tilde{\Sigma}^+_L} & U^R_{\Phi^-_R \tilde{\Delta}^+_L} & U^R_{\Phi^-_R \Delta^-_R} & U^R_{\Phi^-_R \Phi^-_R}
\end{array} \right)
\end{equation*}
Here also, terms like $vY_{34}^{(\prime)}$ and $vY_{45}^{(\prime)}$ are neglected with respect to masses of heavy leptons. The elements of the mixing matrix $U^L$ and $U^R$ are obtained following the procedure presented in Ref.~\cite{Grimus_Lavoura}. Nonzero elements $U^L$ and $U^R$ are given in the following:
\begin{multicols}{2}
\begin{subequations}
\begin{align*}
&U^L_{\ell \ell}=1-\frac{1}{4} \epsilon_{23} \epsilon_{23}^\dagger
\\
&U^L_{\ell \tilde{\Sigma}^+_R}=-\frac{1}{\sqrt{2}} \epsilon_{23}
\\
&U^L_{\ell \tilde{\Delta}^+_R}=-\epsilon_{23} \epsilon_{34}^\prime
\\
&U^L_{\ell \Delta^-_{L}}~~~=\frac{\sqrt{3}}{2} \epsilon_{23}\epsilon_{34}^\prime \epsilon_{45}^\prime \delta_{45}^\prime
\\
&U^L_{\ell \tilde{\Phi}^+_R}=\sqrt{\frac{3}{2}} \epsilon_{23}\epsilon_{34}^\prime \epsilon_{45}^\prime
\\
&U^L_{\tilde{\Sigma}^+_R \ell}=\frac{1}{\sqrt{2}} \epsilon_{23}^\dagger
\\
&U^L_{\tilde{\Delta}^+_R \ell}=\epsilon_{34}^{\prime \dagger} \epsilon_{23}^\dagger
\\
&U^L_{\Delta^-_{L} \ell}~~~~=-\frac{\sqrt{3}}{2} \delta_{45}^{\prime \dagger} \epsilon_{45}^{\prime \dagger} \epsilon_{34}^{\prime \dagger} \epsilon_{23}^\dagger
\\
&U^L_{\tilde{\Phi}^+_R \ell}~=-\sqrt{\frac{3}{2}} \epsilon_{45}^{\prime \dagger} \epsilon_{34}^{\prime \dagger} \epsilon_{23}^\dagger
\\
&U^L_{\tilde{\Sigma}^+_R \tilde{\Sigma}^+_R}=1-\frac{1}{4} \epsilon_{23}^\dagger \epsilon_{23}
\\
&U^L_{\tilde{\Sigma}^+_R \tilde{\Delta}^+_R}=-\frac{1}{2\sqrt{2}} \epsilon_{23}^\dagger \epsilon_{23} \epsilon_{34}^\prime
\\
&U^L_{\tilde{\Sigma}^+_R \Delta^-_{L}}=\frac{\sqrt{3}}{4\sqrt{2}} \epsilon_{23}^\dagger \epsilon_{23}\epsilon_{34}^\prime \epsilon_{45}^\prime \delta_{45}^\prime
\\
&U^L_{\tilde{\Sigma}^+_R \tilde{\Phi}^+_R}=\frac{\sqrt{3}}{4} \epsilon_{23}^\dagger \epsilon_{23}\epsilon_{34}^\prime \epsilon_{45}^\prime
\\
&U^L_{\tilde{\Delta}^+_R \tilde{\Sigma}^+_R}=-\frac{1}{2\sqrt{2}} \epsilon_{34}^{\prime \dagger} \epsilon_{23}^\dagger \epsilon_{23}
\\
&U^L_{\tilde{\Delta}^+_R \tilde{\Delta}^+_R}=1-\frac{1}{2} \epsilon_{34}^{\prime \dagger} \epsilon_{23}^\dagger \epsilon_{23} \epsilon_{34}^\prime
\\
&U^L_{\tilde{\Delta}^+_R \Delta^-_{L}}=\frac{\sqrt{3}}{4} \epsilon_{34}^{\prime \dagger} \epsilon_{23}^\dagger \epsilon_{23}\epsilon_{34}^\prime \epsilon_{45}^\prime \delta_{45}^\prime
\\
&U^L_{\tilde{\Delta}^+_R \tilde{\Phi}^+_R}=\frac{\sqrt{3}}{2\sqrt{2}} \epsilon_{34}^{\prime \dagger} \epsilon_{23}^\dagger \epsilon_{23}\epsilon_{34}^\prime \epsilon_{45}^\prime
\\
&U^L_{\Delta^-_L \tilde{\Sigma}^+_R}=\frac{\sqrt{3}}{4\sqrt{2}} \delta_{45}^{\prime \dagger} \epsilon_{45}^{\prime \dagger} \epsilon_{34}^{\prime \dagger} \epsilon_{23}^\dagger \epsilon_{23}
\\
&U^L_{\Delta^-_L \tilde{\Delta}^+_R}=\frac{\sqrt{3}}{4} \delta_{45}^{\prime \dagger} \epsilon_{45}^{\prime \dagger} \epsilon_{34}^{\prime \dagger} \epsilon_{23}^\dagger \epsilon_{23} \epsilon_{34}^\prime
\\
&U^L_{\Delta^-_L \Delta^-_L}=1-\frac{3}{8} \delta_{45}^{\prime \dagger} \epsilon_{45}^{\prime \dagger} \epsilon_{34}^{\prime \dagger} \epsilon_{23}^\dagger \epsilon_{23}\epsilon_{34}^\prime \epsilon_{45}^\prime \delta_{45}^\prime
\end{align*}
\begin{align*}
\\
&U^L_{\Delta^-_L \tilde{\Phi}^+_R}=-\frac{3}{4\sqrt{2}} \delta_{45}^{\prime \dagger} \epsilon_{45}^{\prime \dagger} \epsilon_{34}^{\prime \dagger} \epsilon_{23}^\dagger \epsilon_{23}\epsilon_{34}^\prime \epsilon_{45}^\prime
\\
&U^L_{\tilde{\Phi}^+_R \tilde{\Sigma}^+_R}=\frac{\sqrt{3}}{4} \epsilon_{45}^{\prime \dagger} \epsilon_{34}^{\prime \dagger} \epsilon_{23}^\dagger \epsilon_{23}
\\
&U^L_{\tilde{\Phi}^+_R \tilde{\Delta}^+_R}=\frac{\sqrt{3}}{2\sqrt{2}} \epsilon_{45}^{\prime \dagger} \epsilon_{34}^{\prime \dagger} \epsilon_{23}^\dagger \epsilon_{23} \epsilon_{34}^\prime
\\
&U^L_{\tilde{\Phi}^+_R \Delta^-_L}=-\frac{3}{4\sqrt{2}} \epsilon_{45}^{\prime \dagger} \epsilon_{34}^{\prime \dagger} \epsilon_{23}^\dagger \epsilon_{23}\epsilon_{34}^\prime \epsilon_{45}^\prime \delta_{45}^\prime
\\
&U^L_{\tilde{\Phi}^+_R \tilde{\Phi}^+_R}=1-\frac{3}{4}\epsilon_{45}^{\prime \dagger} \epsilon_{34}^{\prime \dagger} \epsilon_{23}^\dagger \epsilon_{23}\epsilon_{34}^\prime \epsilon_{45}^\prime
\\
\\
\\
&U^R_{\ell \ell}=1
\\
&U^R_{\ell \tilde{\Sigma}^+_L}=-\frac{1}{\sqrt{2}} m_\ell \epsilon_{23} M_\Sigma^{-1}
\\
&U^R_{\ell \tilde{\Delta}^+_L}=-m_\ell \epsilon_{23} \left( \epsilon_{34}^\prime + \delta^T_{34} \right) M_\Delta^{-1}
\\
&U^R_{\ell \Delta^-_{R}}=\frac{\sqrt{3}}{2} m_\ell \epsilon_{23} \Big( \epsilon_{34}^\prime \epsilon_{45}^\prime \epsilon_{45}^T + \epsilon_{34}^\prime \epsilon_{45}^\prime \delta_{45}^\prime  
\\
&\qquad \qquad \qquad \qquad + \epsilon_{34}^\prime \delta_{45}^T \epsilon_{45}^T + \delta_{34}^T \delta_{45}^T \epsilon_{45}^T \Big) M_\Delta^{-1}
\\
&U^R_{\ell \Phi^-_{R}}=-\sqrt{\frac{3}{2}} m_\ell \epsilon_{23} \left(\delta_{34}^T \delta_{45}^T + \epsilon_{34}^\prime \epsilon_{45}^\prime \right) M_\Phi^{-1} 
\\
&U^R_{\tilde{\Sigma}^+_L \ell}=\frac{1}{\sqrt{2}} M_\Sigma^{-1} \epsilon_{23}^\dagger m_\ell 
\\
&U^R_{\tilde{\Delta}^+_L \ell}=M_\Delta^{-1}\left(\epsilon_{34}^{\prime \dagger} + \delta_{34}^*  \right) \epsilon_{23}^\dagger m_\ell 
\\
&U^R_{\Delta^-_{R} \ell}=-\frac{\sqrt{3}}{2} M_\Delta^{-1} \Big(\epsilon_{45}^* \epsilon_{45}^{\prime \dagger} \epsilon_{34}^{\prime \dagger} + \delta_{45}^{\prime \dagger} \epsilon_{45}^{\prime \dagger} \epsilon_{34}^{\prime \dagger} 
\\
& \qquad \qquad + \epsilon_{45}^* \delta_{45}^* \epsilon_{34}^{\prime \dagger} + \epsilon_{45}^* \delta_{45}^* \delta_{34}^* \Big) \epsilon_{23}^\dagger m_\ell 
\\
&U^R_{\Phi^-_{R} \ell}=\sqrt{\frac{3}{2}} M_\Phi^{-1} \big(\delta_{45}^* \delta_{34}^* + \epsilon_{45}^{\prime \dagger} \epsilon_{34}^{\prime \dagger} \big) \epsilon_{23}^\dagger m_\ell 
\\
&U^R_{\tilde{\Sigma}^+_L \tilde{\Sigma}^+_L}=1
\\
&U^R_{\tilde{\Delta}^+_L \tilde{\Delta}^+_L}=1
\\
&U^R_{\Delta^-_{R} \Delta^-_{R}}=1
\\
&U^R_{\Phi^-_{R} \Phi^-_{R}}=1
\end{align*}
\end{subequations}
\end{multicols}

\paragraph{Doubly charged leptons}
Finally, the mass marix for doubly charged leptons reads as
\begin{equation*}
-\mathcal{L}_2=\left( \begin{array}{ccc}
\overline{\tilde{\Sigma}^{++}_L} & \overline{\tilde{\Delta}^{++}_L} & \overline{\Phi^{--}_R}
\end{array} \right)
\left(
\begin{array}{ccccc}
M_\Sigma & \frac{v Y_{34}^{T}}{\sqrt{2}} & 0 \\
\frac{v Y_{34}^{\prime *}}{\sqrt{2}} & M_\Delta & \frac{v Y_{45}^*}{\sqrt{2}} \\
0 & \frac{v Y_{45}^{\prime T}}{\sqrt{2}} & M_\Phi \\
\end{array} \right)
\left( \begin{array}{c}
\tilde{\Sigma}^{++}_R \\ \tilde{\Delta}^{++}_R \\ \tilde{\Phi}^{++}_R
\end{array} \right)~.
\end{equation*}
The mass matrix for doubly charged leptons is already in almost diagonal form. For our phenemenological considerations, mixing among them can be neglected.

\section{Gauge Lagrangian}
\label{app:gauge_Lag}
For convenience, we define four-component Dirac spinors for charged fermionic fields and Majorana spinors for the neutral ones:
\begin{multicols}{2}
\begin{subequations}
\begin{align*}
\nu_{(m)} &= \nu_{(m)L} + \tilde{\nu}_{(m)L}
\\
\Sigma^0_c &= \tilde{\Sigma}^0_R + \Sigma^0_R
\\
\Sigma^0 &= \Sigma^0_L + \tilde{\Sigma}^0_L
\\
\Delta^0_c &= \tilde{\Delta}^0_R + \Delta^0_R
\\
\Delta^0 &= \Delta^0_L + \tilde{\Delta}^0_L
\\
\Phi^0 _{(m)}&= \tilde{\Phi}^0_{(m)R} +\Phi^0_{(m)R}
\end{align*}
\end{subequations}
\begin{subequations}
\begin{align*}
&\Sigma^{1,2}_m = \Sigma^{1,2}_{mL} + \tilde{\Sigma}^{1,2}_{mL}
\\
&\Delta^{1,2}_m = \Delta^{1,2}_{mL} + \tilde{\Delta}^{1,2}_{mL}
\\
&\Phi^{+,++}_{(m)} = \tilde{\Phi}^{-,--}_{(m)R} + \Phi^{+,++}_{(m)R}
\\
&\ell_{(m)} = \ell_{(m)L}+\ell_{(m)R}
\\
&\Sigma^{+,++}_{(m)} = \Sigma^{+,++}_{(m)L}+\Sigma^{+,++}_{(m)R}
\\
&\Delta^{-,+,++}_{(m)} = \Delta^{-,+,++}_{(m)L}+\Delta^{-,+,++}_{(m)R}
\end{align*}
\end{subequations}
\end{multicols}

\noindent The relevant parts of the gauge Lagrangian in the gauge-eigenstates' basis are given by
\begin{subequations}
\begin{align*}
\mathcal{L}^{\rm SM}_{\rm gauge} \supset &~ \frac{g}{\sqrt{2}} \overline{\nu} \gamma^\mu P_L \ell W^+_\mu + h.c. +\frac{g}{2c_w} \Big\{ (2s_w^2-1) \overline{\ell} \gamma^\mu P_L \ell + 2s^2_w \overline{\ell} \gamma^\mu P_R \ell + \overline{\nu} \gamma^\mu P_L \nu  \nonumber
\\
&-\overline{\nu} \gamma^\mu P_R \nu \Big\} Z_\mu -e \overline{\ell} \gamma^\mu \ell A_\mu~,
\\
\mathcal{L}^\Sigma_{\rm gauge} \supset & g \left\{\overline{\Sigma^+} \gamma^\mu P_R \Sigma^0_c + \overline{\Sigma^+} \gamma^\mu P_L \Sigma^0 - \overline{\Sigma^{++}} \gamma^\mu \Sigma^+ \right\} W^+_\mu + h.c. + \frac{g}{c_w}\Big\{
(1-2s^2_w) \overline{\Sigma^{++}} \gamma^\mu \Sigma^{++} \nonumber
\\
& -s^2_w \overline{\Sigma^+} \gamma^\mu \Sigma^+ +2 \overline{\Sigma^0_c} \gamma^\mu P_L \Sigma^0_c -2 \overline{\Sigma^0} \gamma^\mu P_L \Sigma^0 \Big\} Z_\mu + e\left\{2\overline{\Sigma^{++}} \gamma^\mu \Sigma^{++} + \overline{\Sigma^+} \gamma^\mu \Sigma^+ \right\} A_\mu~,
\\
\mathcal{L}^\Delta_{\rm gauge} \supset &~ g\Big\{ \sqrt{\frac{3}{2}} \overline{\Delta^{++}} \gamma^\mu \Delta^+ +\sqrt{2} \overline{\Delta^+} \gamma^\mu P_R \Delta^0_c +\sqrt{2} \overline{\Delta^+} \gamma^\mu P_L \Delta^0 +\sqrt{\frac{3}{2}} \overline{\Delta^0} \gamma^\mu P_L \Delta^- + \nonumber
\\
&\sqrt{\frac{3}{2}} \overline{\Delta^0_c} \gamma^\mu P_R \Delta^- \Big\} W^+_\mu + h.c. + \frac{g}{2c_w} \Big\{(3-4s_w^2) \overline{\Delta^{++}} \gamma^\mu \Delta^{++} + (1-2s_w^2) \overline{\Delta^{+}} \gamma^\mu \Delta^{+} \nonumber
\\
& +2 \overline{\Delta^0_c} \gamma^\mu P_L \Delta^0_c -2 \overline{\Delta^0} \gamma^\mu P_L \Delta^0 - (3-2s_w^2) \overline{\Delta^-} \gamma^\mu \Delta^- \Big\} Z_\mu + e\Big\{2\overline{\Delta^{++}} \gamma^\mu \Delta^{++} + \nonumber
\\
& \overline{\Delta^+} \gamma^\mu \Delta^+ - \overline{\Delta^-} \gamma^\mu \Delta^- \Big\} A_\mu~,
\\
\mathcal{L}^\Phi_{\rm gauge} \supset &~ g\left\{\sqrt{2} \left( \overline{\Phi^{++}} \gamma^\mu P_R \Phi^+ -\overline{\Phi^{++}} \gamma^\mu P_L \Phi^+ \right) + \sqrt{3} \left( \overline{\Phi^+} \gamma^\mu P_R \Phi^0 - \overline{\Phi^+} \gamma^\mu P_L \Phi^0 \right) \right\}W^+_\mu + \nonumber
\\
& h.c.+\left\{2\overline{\Phi^{++}} \gamma^\mu \Phi^{++} + \overline{\Phi^+} \gamma^\mu \Phi^+ \right\} \left(eA_\mu + gc_w Z_\mu \right)~.
\end{align*}
\end{subequations}
The parts of the Lagrangian relevant for gauge production of the exotic fermions in the mass-eigenstates' basis are given by
\begin{subequations}
\begin{align}
\label{eq:gauge_prod_3}
\mathcal{L}^{\Sigma \overline{\Sigma}}_{\rm prod} = &~ \frac{g}{\sqrt{2}} \left\{\overline{\Sigma^+_m} \gamma^\mu P_L \Sigma^1_m - \overline{\Sigma^+_m} \gamma^\mu P_R \Sigma^1_m + \overline{\Sigma^+_m} \gamma^\mu \Sigma^2_m -\sqrt{2} \overline{\Sigma^{++}_m} \gamma^\mu \Sigma^+_m \right\} W^+_\mu + h.c.~ + \nonumber
\\
& \frac{g}{c_w}\Big\{(1-2s^2_w) \overline{\Sigma^{++}_m} \gamma^\mu \Sigma^{++}_m -s^2_w \overline{\Sigma^+_m} \gamma^\mu \Sigma^+_m + 2\overline{\Sigma^1_m} \gamma^\mu P_R \Sigma^2_m - 2\overline{\Sigma^1_m} \gamma^\mu P_L \Sigma^2_m \Big\} Z_\mu \nonumber 
\\&+ e\left\{2\overline{\Sigma^{++}_m} \gamma^\mu \Sigma^{++}_m + \overline{\Sigma^+_m} \gamma^\mu \Sigma^+_m \right\} A_\mu~,
\\
\label{eq:gauge_prod_4}
\mathcal{L}^{\Delta \overline{\Delta}}_{\rm prod} = &~ g \Big\{ \sqrt{\frac{3}{2}} \overline{\Delta^{++}_m} \gamma^\mu \Delta^+_m + \overline{\Delta^+_m} \gamma^\mu P_L \Delta^1_m - \overline{\Delta^+_m} \gamma^\mu P_R \Delta^1_m + \overline{\Delta^+_m} \gamma^\mu \Delta^2_m +\frac{\sqrt{3}}{2} \overline{\Delta^1_m} \gamma^\mu P_L \Delta^-_m \nonumber
\\
&-\frac{\sqrt{3}}{2} \overline{\Delta^1_m} \gamma^\mu P_R \Delta^-_m +\frac{\sqrt{3}}{2} \overline{\Delta^2_m} \gamma^\mu \Delta^-_m \Big\} W^+_\mu + h.c. + \frac{g}{2c_w} \Big\{(3-4s_w^2) \overline{\Delta^{++}_m} \gamma^\mu \Delta^{++}_m \nonumber
\\
&+(1-2s_w^2) \overline{\Delta^+_m} \gamma^\mu \Delta^+_m + 2\overline{\Delta^1_m} \gamma^\mu P_R \Delta^2_m - 2\overline{\Delta^1_m} \gamma^\mu P_L \Delta^2_m -(3-2s_w^2) \overline{\Delta^-_m} \gamma^\mu \Delta^-_m \Big\} Z_\mu \nonumber
\\
&+e\Big\{2\overline{\Delta^{++}_m} \gamma^\mu \Delta^{++}_m + \overline{\Delta^+_m} \gamma^\mu \Delta^+_m - \overline{\Delta^-_m} \gamma^\mu \Delta^-_m \Big\} A_\mu~,
\\
\label{eq:gauge_prod_5}
\mathcal{L}^{\Phi \overline{\Phi}}_{\rm prod} = &~ g\left\{\sqrt{2} \left( \overline{\Phi^{++}_m} \gamma^\mu P_R \Phi^+_m -\overline{\Phi^{++}_m} \gamma^\mu P_L \Phi^+_m \right) + \sqrt{3} \left( \overline{\Phi^+_m} \gamma^\mu P_R \Phi^0_m - \overline{\Phi^+_m} \gamma^\mu P_L \Phi^0_m \right) \right\}W^+_\mu  \nonumber
\\
&+ h.c. +\left\{2\overline{\Phi^{++}_m} \gamma^\mu \Phi^{++}_m + \overline{\Phi^+}_m \gamma^\mu \Phi^+_m \right\} \left(eA_\mu + gc_w Z_\mu \right)~.
\end{align}
\end{subequations}
The parts of the Lagrangian relevant for the decays of the exotics to $W^{\pm}/Z$-boson and a SM lepton in the mass-eigenstates' basis are given by
\begin{subequations}
\begin{align}
\label{eq:dec_CC_3}
\mathcal{L}_{\rm CC}^\Sigma \approx &~+\frac{g}{2} \overline{\Sigma^1_m} \left\{\left(\sqrt{2}U^L_{\tilde{\Sigma}^+_R \ell} + U^0_{\tilde{\Sigma}^0_R \nu_L} \right) \gamma^\mu P_L -\sqrt{2} U^R_{\tilde{\Sigma}^+_L \ell} \gamma^\mu P_R \right\} \ell^-_m W^+_\mu \nonumber
\\
&-\frac{g}{2} \overline{\Sigma^2_m} \left\{\left(\sqrt{2}U^L_{\tilde{\Sigma}^+_R \ell} + U^0_{\tilde{\Sigma}^0_R \nu_L} \right) \gamma^\mu P_L +\sqrt{2} U^R_{\tilde{\Sigma}^+_L \ell} \gamma^\mu P_R \right\} \ell^-_m W^+_\mu \nonumber
\\
&+ \frac{g}{\sqrt{2}} \overline{\Sigma^+_m} \left\{\sqrt{2} U^0_{\Sigma^0_L \nu_L} V_{\rm PMNS} \gamma^\mu P_L + \left(\sqrt{2} U^{0*}_{\tilde{\Sigma}^0_R \nu_L}+ U^{L*}_{\tilde{\Sigma}^+_R \ell} \right) V_{\rm PMNS}^* \gamma^\mu P_R \right\} \nu_m W^+_\mu \nonumber
\\
&-g \overline{\Sigma^{++}_m} \left(U^{R*}_{\tilde{\Sigma}^+_L \ell} \gamma^\mu P_L + U^{L*}_{\tilde{\Sigma}^+_R \ell} \gamma^\mu P_R \right) \ell^+_m W^+_\mu~,
\\
\label{eq:dec_CC_4}
\mathcal{L}_{\rm CC}^\Delta \approx &+\frac{g}{2} \overline{\Delta^1_m} \left\{\left(2U^L_{\tilde{\Delta}^+_R \ell} + U^0_{\tilde{\Delta}^0_R \nu_L} \right) \gamma^\mu P_L -2 U^R_{\tilde{\Delta}^+_L \ell} \gamma^\mu P_R \right\} \ell^-_m W^+_\mu \nonumber
\\
&-\frac{g}{2} \overline{\Delta^2_m} \left\{\left(2U^L_{\tilde{\Delta}^+_R \ell} + U^0_{\tilde{\Delta}^0_R \nu_L} \right) \gamma^\mu P_L +2 U^R_{\tilde{\Delta}^+_L \ell} \gamma^\mu P_R \right\} \ell^-_m W^+_\mu \nonumber
\\
&+\frac{g}{\sqrt{2}} \overline{\Delta^+_m} \left\{2 U^0_{\Delta^0_L \nu} V_{\rm PMNS} \gamma^\mu P_L + \left(2 U^{0*}_{\tilde{\Delta}^0_R \nu}+ U^{L*}_{\tilde{\Delta}^+_R \ell} \right) V_{\rm PMNS}^* \gamma^\mu P_R \right\} \nu_m W^+_\mu \nonumber
\\
&-\frac{g}{\sqrt{2}} \overline{\Delta^-_m} \left\{\left(\sqrt{3} U^0_{\Delta^0_L \nu} + U^L_{\Delta^-_L \ell} \right) V_{\rm PMNS} \gamma^\mu P_L - \sqrt{3} U^{0*}_{\tilde{\Delta}^0_R \nu} V_{\rm PMNS}^* \gamma^\mu P_R  \right\} \nu_m W^-_\mu \nonumber
\\
& +\sqrt{\frac{3}{2}}g \overline{\Delta^{++}_m} \left(U^{R*}_{\tilde{\Delta}^+_L \ell} \gamma^\mu P_L + U^{L*}_{\tilde{\Delta}^+_R \ell} \gamma^\mu P_R \right) \ell^+_m W^+_\mu~,
\\
\label{eq:dec_CC_5}
\mathcal{L}_{\rm CC}^\Phi \approx &-\frac{g}{\sqrt{2}} \overline{\Phi^0_m} \left\{\left(\sqrt{6} U^L_{\tilde{\Phi}^+_R \ell} + U^0_{\tilde{\Phi}^0_R \nu_L} \right)\gamma^\mu P_L -\sqrt{6} U^R_{\Phi^-_R \ell} \gamma^\mu P_R \right\} \ell^-_m W^+_\mu \nonumber
\\
& +\frac{g}{\sqrt{2}} \overline{\Phi^+_m} \left\{-\sqrt{6} U^0_{\tilde{\Phi}^0_R \nu} V_{\rm PMNS} \gamma^\mu P_L + \left( \sqrt{6} U^{0*}_{\tilde{\Phi}^0_R \nu} + U^{L*}_{\tilde{\Phi}^+_R \ell} \right) V_{\rm PMNS}^* \gamma^\mu P_R \right\} \nu_m W^+_\mu \nonumber
\\
&-\sqrt{2}g \overline{\Phi^{++}_m} \left(U^{R*}_{\Phi^-_R \ell} \gamma^\mu P_L + U^{L*}_{\tilde{\Phi}^+_R \ell} \gamma^\mu P_R \right) \ell^+_m W^+_\mu
\end{align}
\end{subequations}
\begin{subequations}
\begin{align}
\label{eq:dec_NC_3}
\mathcal{L}_{\rm NC}^\Sigma \approx & \frac{g}{2\sqrt{2}\cos\theta} (\overline{\Sigma^2_m}-\overline{\Sigma^1_m}) \left\{U^0_{\tilde{\Sigma}^0_R \nu_L}V_{\rm PMNS} \gamma^\mu P_L - U^{0*}_{\tilde{\Sigma}^0_R \nu_L} V_{\rm PMNS}^* \gamma^\mu P_R \right\} \nu_m Z_\mu \nonumber
\\
& -\frac{g}{2\cos\theta} \overline{\Sigma^+_m} U^{L*}_{\tilde{\Sigma}^+_R \ell} \gamma^\mu P_R \ell^+_m Z_\mu~,
\\
\label{eq:dec_NC_4}
\mathcal{L}_{\rm NC}^\Delta \approx & - \frac{g}{2\sqrt{2}\cos\theta} (\overline{\Delta^1_m} + \overline{\Delta^2_m}) \left\{U^0_{\Delta^0_L \nu} V_{\rm PMNS} \gamma^\mu P_L - U^{0*}_{\Delta^0_L \nu_L} V_{\rm PMNS}^* \gamma^\mu P_R \right\} \nu_m Z_\mu \nonumber
\\
& +\frac{g}{2\cos\theta} \left\{ \overline{\Delta^+_m} U^{R*}_{\tilde{\Delta}^+_L \ell} \gamma^\mu P_L \ell^+_m - \overline{\Delta^-_m} \left(2U^L_{\Delta^-_L \ell} \gamma^\mu P_L +3 U^R_{\Delta^-_R \ell} \gamma^\mu P_R \right) \ell^-_m \right\} Z_\mu~,
\\
\label{eq:dec_NC_5}
\mathcal{L}_{\rm NC}^\Phi \approx & -\frac{g}{2\cos\theta} \overline{\Phi^0_m} \left\{U^0_{\tilde{\Phi}^0_R \nu} V_{\rm PMNS} \gamma^\mu P_L - U^{0*}_{\tilde{\Phi}^0_R \nu} V_{\rm PMNS}^* \gamma^\mu P_R \right\} \nu_m Z_\mu \nonumber
\\
& + \frac{g}{2\cos\theta} \overline{\Phi^+_m} \left(2U^{R*}_{\Phi^-_R \ell} \gamma^\mu P_L +U^{L*}_{\tilde{\Phi}^+_R \ell} \gamma^\mu P_R \right) \ell^+_m Z_\mu~.
\end{align}
\end{subequations}
The elements of the mixing matrices $U^0$,$U^L$ and $U^R$ are given in appendix \ref{app:mass_matrices}.
The Lagrangian involving the interactions of two fermions with the SM Higgs can be obtained by expanding the Yuakawa Lagrangian in equation (\ref{eq:yukawa_int}). In the gauge-eigenstates' basis, the same is given by
\begin{eqnarray}
\mathcal{L}_h &=& h \Bigg\{ \overline{\ell_L} \frac{Y_\ell}{\sqrt{2}} \ell_R + \overline{\Sigma^0_L} \frac{Y_{23}}{\sqrt{2}} \tilde{\nu_L} - \overline{\Sigma^+_L} \frac{Y_{23}}{2} \tilde{\ell}_L - \overline{\Delta^+_R} \frac{Y_{34}}{\sqrt{3}} \Sigma^+_L - \overline{\Delta^0_R} \frac{Y_{34}}{\sqrt{6}} \Sigma^0_L - \overline{\Delta^+_L} \frac{Y_{34}^\prime}{\sqrt{3}} \Sigma^+_R -\nonumber
\\
&& \overline{\Delta^0_L} \frac{Y_{34}^\prime}{\sqrt{6}} \Sigma^0_R + 3\overline{\Delta^+_L} \frac{Y_{45}}{\sqrt{24}} \Phi^+_R + \overline{\Delta^0_L} \frac{Y_{45}}{2} \Phi^0_R + \overline{\Delta^-_L} \frac{Y_{45}}{\sqrt{8}} \Phi^-_R -3\overline{\Delta^+_R} \frac{Y_{45}^\prime}{\sqrt{24}} \tilde{\Phi}^-_R + \overline{\Delta^0_R} \frac{Y_{45}^\prime}{2} \tilde{\Phi}^0_R \nonumber
\\
&& -\overline{\Delta^-_R} \frac{Y_{45}^\prime}{\sqrt{8}} \tilde{\Phi}^+_R \Bigg\} + h.c. \nonumber
\end{eqnarray}
The parts of the Lagrangian relevant for the decays of the exotic fermions to the SM Higgs and a SM lepton in the mass-eigenstates' basis are procured in the following:
\begin{subequations}
\begin{align}
\label{eq:dec_h_3}
\mathcal{L}^\Sigma_h &\approx h~\overline{\Sigma^+_m} \left[-\frac{1}{\sqrt{2}} U^{L*}_{\tilde{\Sigma}^+_R \ell} Y_\ell P_L -\frac{Y_{23}}{2} P_R \right] \ell^+_m + h~\overline{\Sigma^1_m} \left[\frac{Y_{23}^*}{2} V_{\rm PMNS} P_L + \frac{Y_{23}}{2} V_{\rm PMNS}^* P_R \right] \nu_m \nonumber
\\
&+ h~\overline{\Sigma^2_m} \left[\frac{Y_{23}^*}{2} V_{\rm PMNS} P_L + \frac{Y_{23}}{2} V_{\rm PMNS}^* P_R \right] \nu_m~,
\\
\label{eq:dec_h_4}
\mathcal{L}^\Delta_h &\approx h~\overline{\Delta^+_m} \left[ \left\{-\frac{Y_{34}}{\sqrt{3}} U^{R*}_{\tilde{\Sigma}^+_L \ell} -\frac{1}{\sqrt{2}} U^{L*}_{\tilde{\Delta}^+_R \ell} Y_\ell \right\} P_L -\frac{Y_{34}^\prime}{\sqrt{3}} U^{L*}_{\tilde{\Sigma}^+_R \ell} P_R \right]\ell^+_m \nonumber
\\ 
&+ h~\overline{\Delta^-_m} \left[-\frac{1}{2\sqrt{2}} Y_{45}^\prime U^L_{\tilde{\Phi}^+_R \ell} P_L + \left\{-\frac{1}{\sqrt{2}} U^L_{\Delta^-_L \ell} Y_\ell + \frac{1}{2\sqrt{2}} Y_{45} U^R_{\Phi^-_R \ell} \right\} P_R \right] \ell^-_m \nonumber
\\
&+ h~\overline{\Delta^1_m} \left[-\frac{Y^{\prime *}_{34}}{\sqrt{12}} U^0_{\tilde{\Sigma}^0_R \nu_L} V_{\rm PMNS} P_L -\frac{Y^\prime_{34}}{\sqrt{12}} U^{0*}_{\tilde{\Sigma}^0_R \nu_L} V_{\rm PMNS}^* P_R \right] \nu_m \nonumber
\\
&+ h~\overline{\Delta^2_m} \left[-\frac{Y^{\prime *}_{34}}{\sqrt{12}} U^0_{\tilde{\Sigma}^0_R \nu_L} V_{\rm PMNS} P_L -\frac{Y^\prime_{34}}{\sqrt{12}} U^{0*}_{\tilde{\Sigma}^0_R \nu_L} V_{\rm PMNS}^* P_R \right] \nu_m~,
\\
\label{eq:dec_h_5}
\mathcal{L}^\Phi_h &\approx h~\overline{\Phi^+_m} \left[ \left\{ -\frac{1}{\sqrt{2}} U^{L*}_{\tilde{\Phi}^+_R \ell} Y_\ell + \sqrt{\frac{3}{8}} Y_{45}^\dagger U^R_{\tilde{\Delta}^+_L \ell} \right\} P_L -\sqrt{\frac{3}{8}} Y_{45}^\dagger U^{L*}_{\tilde{\Delta}^+_R \ell} P_R \right] \ell^+_m \nonumber
\\
&+ h~\overline{\Phi^0_m} \left[\frac{Y^{\prime T}_{45}}{2} U^0_{\tilde{\Delta}^0_R \nu_L} V_{\rm PMNS} P_L + \frac{Y_{45}^{\prime \dagger}}{2} U^{0*}_{\tilde{\Delta}^0_R \nu_L} V_{\rm PMNS}^* P_R \right] \nu_m~.
\end{align}
\end{subequations}

\acknowledgments
We thank Debajyoti Choudhury for some useful discussions. SA thanks Avelino Vicente for his help regarding implentation of the model in SARAH. KG acknowledges the support from the DST/INSPIRE Research Grant [DST/INSPIRE/04/2014/002158] and SERB Core Research Grant [CRG/2019/006831].


\begin{thebibliography}{49}%

\bibitem{nu_osci_expt_1}
SNO Collaboration,
\newblock Q.~R.~Ahmad et al.,
\newblock \emph{Direct Evidence for Neutrino Flavor Transformation from Neutral-Current Interactions in the Sudbury Neutrino Observatory},
\newblock \emph{Phys. Rev. Lett.} {\bf 89} (2002) 011301.

\bibitem{nu_osci_expt_2}
SNO Collaboration,
\newblock Q.~R.~Ahmad et al.,
\newblock \emph{Measurement of Day and Night Neutrino Energy Spectra at SNO and Constraints on Neutrino Mixing Parameters},
\newblock \emph{Phys. Rev. Lett.} {\bf 89} (2002) 011302.

\bibitem{nu_osci_expt_3}
KamLAND Collaboration,
\newblock K.~Eguchi et al.,
\newblock \emph{First Results from KamLAND: Evidence for Reactor Antineutrino Disappearance},
\newblock \emph{Phys. Rev. Lett.} {\bf 90} (2003) 021802.

\bibitem{nu_osci_expt_4}
K2K Collaboration,
\newblock M.~H.~Ahn et al.,
\newblock \emph{Indications of Neutrino Oscillation in a 250 km Long-Baseline Experiment},
\newblock \emph{Phys. Rev. Lett.} {\bf 90} (2003) 041801.

\bibitem{nu_osci_expt_5}
Super-Kamiokande Collaboration,
\newblock J.~Hosaka et al.,
\newblock \emph{Three flavor neutrino oscillation analysis of atmospheric neutrinos in Super-Kamiokande},
\newblock \emph{Phys. Rev.} {\bf D 74} (2006) 032002.

\bibitem{nu_osci_expt_6}
T2K Collaboration,
\newblock K.~Abe et al.,
\newblock \emph{Indication of Electron Neutrino Appearance from an Accelerator-Produced Off-Axis Muon Neutrino Beam},
\newblock \emph{Phys. Rev. Lett.} {\bf 107} (2011) 041801.

\bibitem{nu_osci_param}
I.~Esteban et al.,
\newblock \emph{Global analysis of three-flavour neutrino oscillations: synergies and tensions in the determination of $\theta_{23}$, $\delta_{CP}$, and the mass ordering},
\newblock \emph{JHEP} {\bf 01} (2019) 106, \emph{NuFIT} {\bf 4.1} (2019), \emph{www.nu-fit.org}.

\bibitem{seesaw_gut_2}
S.~Weinberg,
\newblock \emph{Baryon- and Lepton-Nonconserving Processes},
\newblock \emph{Phys. Rev. Lett.} {\bf 43} (1979) 1566.

\bibitem{seesaw_gut_5}
E.~Ma,
\newblock \emph{Pathways to Naturally Small Neutrino Masses},
\newblock \emph{Phys. Rev. Lett.} {\bf 81} (1998) 1171.

\bibitem{seesaw1_1}
P. Minkowski,
\newblock \emph{$\mu \to e\gamma$ at a rate of one out of $10^9$ muon decays?},
\newblock \emph{Phys. Lett.} {\bf B 67} (1977) 421-428.

\bibitem{seesaw1_2}
T.~Yanagida,
\newblock \emph{Horizontal Symmetry and Masses of Neutrinos},
\newblock \emph{Conf. Proc.} {\bf C7902131} (1979) 95-99.

\bibitem{seesaw1_3}
M.~Gell-Mann, P.~Ramond, and R.~Slansky,
\newblock \emph{Complex Spinors and Unified Theories},
\newblock \emph{Conf.Proc.} {\bf C790927} (1979) 315-321.

\bibitem{seesaw1_4}
S.~L.~Glashow,
\newblock \emph{The future of elementary particle physics}, 
\newblock \emph{NATO Sci. Ser.} {\bf B 61} (1980) 687-713.

\bibitem{seesaw1_5}
R.~N.~Mohapatra and G.~Senjanovi$\acute{\rm c}$, 
\newblock \emph{Neutrino Mass and Spontaneous Parity Violation},
\newblock \emph{Phys. Rev. Lett.} {\bf 44} (1980) 912.

\bibitem{seesaw2_1}
W.~Konetschny and W.~Kummer,
\newblock \emph{Nonconservation of total lepton number with scalar bosons},
\newblock \emph{Phys. Lett.} {\bf B 70} (1977) 433.

\bibitem{seesaw2_2}
T.~P.~Cheng and L.-F.~Li, 
\newblock \emph{Neutrino Masses, Mixings and Oscillations in $\rm{SU(2)\times U(1)}$ Models of Electroweak Interactions},
\newblock \emph{Phys. Rev.} {\bf D 22} (1980) 2860.

\bibitem{seesaw2_3}
G.~Lazarides, Q.~Shafi, and C.~Wetterich, 
\newblock \emph{Proton Lifetime and Fermion Masses in an $\rm{SO(10)}$ Model},
\newblock \emph{Nucl. Phys.} {\bf B 181} (1981) 287-300.

\bibitem{seesaw2_4}
J. Schechter and J. W. F. Valle, 
\newblock \emph{Neutrino Masses in $\rm{SU(2)\times U(1)}$ Theories},
\newblock \emph{Phys. Rev.} {\bf D 22} (1980) 2227.

\bibitem{seesaw2_5}
M.~Magg and C.~Wetterich, 
\newblock \emph{Neutrino Mass Problem and Gauge Hierarchy},
\newblock \emph{Phys. Lett.} {\bf B 94} (1980) 61.

\bibitem{seesaw2_6}
R.~N.~Mohapatra and G.~Senjanovi$\acute{\rm c}$, 
\newblock \emph{Neutrino Masses and Mixings in Gauge Models with Spontaneous Parity Violation},
\newblock \emph{Phys. Rev.} {\bf D 23} (1981) 165.

\bibitem{seesaw3_1}
R.~Foot, H.~Lew, X.-G.~He, and G.~C.~Joshi,
\newblock \emph{Seesaw Neutrino Masses Induced by a Triplet of Leptons},
\newblock \emph{Z. Phys.} {\bf C} 44 (1989) 441.

\bibitem{seesaw_gut_1}
P.~Ramond,
\newblock \emph{The Family Group in Grand Unified Theories},
\newblock \emph{Conf.Proc.} {\bf C790225} (1979) 265-280.

\bibitem{seesaw_gut_3}
E.~Witten,
\newblock \emph{Neutrino masses in the minimal $\rm{O(10)}$ theory},
\newblock \emph{Phys. Lett.} {\bf B 91} (1980) 81.

\bibitem{seesaw_gut_4}
J.~C.~Pati and A.~salam,
\newblock \emph{Lepton number as the fourth ``color''},
\newblock \emph{Phys. Rev.} {\bf D 10} (1974) 275.

\bibitem{seesaw_gut_6}
B.~Bajc and G.~Senjanovi$\acute{\rm c}$,  
\newblock \emph{Seesaw at LHC},
\newblock \emph{JHEP} {\bf 0708} (2007) 014.

\bibitem{seesaw_gut_7}
P.~F.~P$\acute{\rm e}$rez, 
\newblock \emph{Renormalizable Adjoint $\rm{SU(5)}$},
\newblock \emph{Phys. Lett.} {\bf B 654} (2007) 189.

\bibitem{seesaw_gut_8}
P.~F.~P$\acute{\rm e}$rez,
\newblock \emph{Supersymmetric adjoint $\rm{SU(5)}$},
\newblock \emph{Phys. Rev.} {\bf D 76} (2007) 071701.

\bibitem{high_dim_1}
F.~Bonnet, D.~Hernandez, T.~Ota and W.~Winter,
\newblock \emph{Neutrino masses from higher than $d=5$ effective operators},
\newblock \emph{JHEP} {\bf 10} (2009) 76 .

\bibitem{high_dim_2}
S.~Kanemura and T.~Ota,
\newblock \emph{Neutrino Masses from Loop-induced $d\geq 7$ Operators},
\newblock \emph{Phys.Lett.} {\bf B 694} (2010) 233-237.

\bibitem{high_dim_3}
Y.~Liao,
\newblock \emph{Unique neutrino mass operator at any mass dimension},
\newblock \emph{Phys.Lett.} {\bf B 694} (2011) 346-348.

\bibitem{high_dim_4}
F.~Bonnet, M.~Hirsch, T.~Ota and W.~Winter,
\newblock \emph{Systematic study of the $d=5$ Weinberg operator at one-loop order},
\newblock \emph{JHEP} {\bf 07} (2012) 153.

\bibitem{high_dim_5}
R.~Cepedello, M.~Hirscha and J.C.~Helo,
\newblock \emph{Lepton number violating phenomenology of $d=7$ neutrino mass models},
\newblock \emph{JHEP} {\bf 01} (2018) 009.

\bibitem{high_dim_6}
G.~Anamiati, O.~Castillo-Felisola, R.M.~Fonseca, J.C.~Helo,and M.~Hirsch,
\newblock \emph{High-dimensional neutrino masses},
\newblock \emph{JHEP} {\bf 12} (2018) 066.

\bibitem{high_dim_7}
C.~Arbel$\acute{\rm a}$ez, J.~C.~Helo and M.~Hirsch,
\newblock \emph{Long-lived heavy particles in neutrino mass models},
\newblock \emph{Phys. Rev.} {\bf D 100} (2019) 055001.

\bibitem{picek}
I.~Picek and B.~Radov$\check{\rm c}$i$\acute{\rm c}$,
\newblock \emph{Novel TeV-scale seesaw mechanism with Dirac mediators},
\newblock \emph{Phys. Lett.} {\bf B 687} (2010) 338-341.

\bibitem{kumericki}
K.~Kumeri$\check{\rm c}$ki, I.~Picek and B.~Radov$\check{\rm c}$i$\acute{\rm c}$,
\newblock \emph{Exotic seesaw-motivated heavy leptons at the LHC},
\newblock \emph{Phys. Rev.} {\bf D 84} (2011) 093002.

\bibitem{kumericki1}
K.~Kumeri$\check{\rm c}$ki, I.~Picek and B.~Radov$\check{\rm c}$i$\acute{\rm c}$,
\newblock \emph{TeV-scale Seesaw with Quintuplet Fermions},
\newblock \emph{Phys. Rev.} {\bf D 86} (2012) 013006.

\bibitem{cms_multilepton_137}
CMS Collaboration,
\newblock \emph{Search for physics beyond the standard model in multilepton final states in proton-proton collisions at $\sqrt{s}$=13 TeV},
\newblock \emph{JHEP} {\bf 03} (2020) 051.

\bibitem{Casas_Ibarra}
J.~Casas and A.~Ibarra,
\newblock \emph{Oscillating neutrinos and $\mu \to e \gamma$},
\newblock \emph{Nucl. Phys.} {\bf B 618} (2001) 171-204.

\bibitem{Ibarra_Ross}
A.~Ibarra and G.G.~Ross,
\newblock \emph{Neutrino Phenomenology---the case of two right-handed neutrinos},
\newblock \emph{Phys. Lett.} {\bf B 591} (2004) 285-296.

\bibitem{cirelli}
M.~Cirelli, N.~Fornengo and A.~Strumia,
\newblock \emph{Minimal dark matter},
\newblock \emph{Nucl. Phys.} {\bf B 753} (2006) 178-194.

\bibitem{biggio2011}
C.~Biggio and F.~Bonnet,
\newblock \emph{Implementation of the type III seesaw model in FeynRules/MadGraph and prospects for discovery with early LHC data},
\newblock \emph{Eur. Phys. J.} {\bf C 72} (2012) 1899.

\bibitem{lhc_1}
CMS Collaboration,
A.M.~Sirunyan et al.,
\newblock \emph{Search for Evidence of the Type-III Seesaw Mechanism in Multilepton Final States in Proton-Proton Collisions at $\sqrt{s}=13$ TeV},
\newblock \emph{Phys. Rev. Lett.} {\bf 119} (2017) 221802.

\bibitem{lhc_2}
ATLAS Collaboration,
\newblock \emph{Search for type-III seesaw heavy leptons in proton-proton collisions at $\sqrt{s}=13$ TeV with the ATLAS detector},
\newblock \emph{ATLAS-CONF}-2018-020.

\bibitem{lhc_3}
ATLAS Collaboration,
G.~Aad et al.,
\newblock \emph{Search for type-III seesaw heavy leptons in $pp$ collisions at $\sqrt{s}=8$ TeV with the ATLAS detector}
\newblock \emph{Phys. Rev.} {\bf D 92} (2015) 032001.

\bibitem{lhc_4}
CMS collaboration,
\newblock \emph{Search for heavy lepton partners of neutrinos in proton-proton collisions in the context of the type III seesaw mechanism}
\newblock \emph{Phys. Lett.} {\bf B 718} (2012) 348-368.

\bibitem{Lavoura}
L.~Lavoura, 
\newblock \emph{General formulae for $f_1 \to f_2 \gamma$}, 
\newblock \emph{Eur. Phys. J.} {\bf C 29} (2003) 191-195.

\bibitem{Abada_et_al}
A.~Abada, C.~Biggio, F.~Bonnet, M.B.~Gavela and T.~Hambye,
\newblock \emph{$\mu \to e\gamma$ and $\tau \to e\gamma$ decays in the fermion triplet seesaw model},
\newblock \emph{Phys. Rev.} {\bf D 78} (2008) 033007.

\bibitem{muegamma}
MEG Collaboration,
\newblock A.~M.~Baldini et al.,
\newblock \emph{Search for the lepton flavour violating decay $\mu^+ \to e^+ \gamma$ with the full dataset of the MEG experiment},
\newblock \emph{Eur. Phys. J} {\bf C 76} (2016) 434.

\bibitem{tauellgamma}
The BABAR Collaboration,
\newblock B.~Aubert et al.,
\newblock \emph{Searches for Lepton Flavor Violation in the Decays $\tau^\pm \to e^\pm \gamma$ and $\tau^\pm \to \mu^\pm \gamma$},
\newblock \emph{Phys. Rev. Lett.} {\bf 104} (2010) 021802. 

\bibitem{mu3e}
SINDRUM Collaboration,
\newblock U.~Bellgardt et al.,
\newblock \emph{Search for the decay $\mu^+ \to e^+ e^+ e^-$},
\newblock \emph{Nucl. Phys.} {\bf B 299} (1988) 1-6.

\bibitem{mueau}
SINDRUM II Collaboration,
\newblock W.~H.~Bertl et al.,
\newblock \emph{A search for muon to electron conversion in muonic gold},
\newblock \emph{Eur. Phys. J} {\bf C 47} (2006) 337-346.

\bibitem{mueal}
COMET Collaboration,
\newblock Y.~Kuno,
\newblock \emph{A search for muon-to-electron conversion at J-PARC: The COMET experiment},
\newblock \emph{PTEP} {\bf 2013} (2013) 022C01.

\bibitem{mueal2}
Mu2e Collaboration, 
\newblock G.~Pezzullo,
\newblock \emph{The Mu2e experiment at Fermilab: a search for lepton avor violation},
\newblock \emph{Nucl. Part. Phys. Proc.} {\bf 285-286} (2017) 3.

\bibitem{tau3ell}
Belle Collaboration,
\newblock K.~Hayasaka et al.,
\newblock \emph{Search for lepton-flavor-violating $tau$ decays into three leptons with 719 million produced $\tau^+ \tau^-$ pairs},
\newblock \emph{Phys. Lett.} {\bf B 687} (2010) 139-143.

\bibitem{muepb}
\newblock SINDRUM II Collaboration,
\newblock W.~Honeckeret al.,
\newblock \emph{Improved limit on the branching ratio of $\mu \to e\gamma$ conversion on lead},
\emph{Phys. Rev. Lett.} {\bf 76} (1996) 200-203.

\bibitem{mueti}
\newblock SINDRUM II Collaboration,
\newblock J.~Kaulardet al.,
\newblock \emph{Improved limit on the branching ratio of $\mu^- \to e^+$ conversion on titanium},
\newblock \emph{Phys. Lett.} {\bf B 422} (1998) 334-338.

\bibitem{sarah}
F.~Staub,
\newblock \emph{SARAH 4: A tool for (not only SUSY) model builders},
\newblock \emph{Comput. Phys. Commun} {\bf 185} (2014) 1773-1790.

\bibitem{sarah2}
F.~Staub,
\newblock \emph{Exploring new models in all detail with SARAH},
\newblock \emph{Adv. High Energy Phys.} {\bf 2015} (2015) 840780.

\bibitem{spheno}
W.~Porod,
\newblock \emph{SPheno, a program for calculating supersymmetric spectra, SUSY particle decays and SUSY particle production at $e^+e^-$ colliders},
\newblock \emph{Comput. Phys. Commun.} {\bf 153} (2003) 275-315.

\bibitem{spheno2}
W.~Porod and F.~Staub,
\newblock \emph{SPheno 3.1: extensions including flavour, CP-phases and models beyond the MSSM},
\newblock \emph{Comput. Phys. Commun.} {\bf 183} (2012) 2458-2469.

\bibitem{babu_photon}
K.~S.~Babu and S.~Jana,
\newblock \emph{Probing doubly charged Higgs bosons at the LHC through photon initiated processes},
\newblock \emph{Phys. Rev.} {\bf D 95} (2017) 055020.

\bibitem{kirti_photon}
K.~Ghosh, S.~Jana and S.~Nandi,
\newblock \emph{Neutrino mass generation at TeV scale and new physics signatures from charged Higgs at the LHC for photon initiated processes},
\newblock \emph{JHEP} {\bf 03} (2018) 180.

\bibitem{agar_photon}
S.~K.~Agarwalla, K.~Ghosh, N.~Kumar and A~Patra,
\newblock \emph{Same-sign multilepton signatures of an ${\rm SU(2)_R}$ quintuplet at the LHC},
\newblock \emph{JHEP} {\bf 01} (2019) 080.

\bibitem{avnish_photon}
Avnish and K.~Ghosh,
\newblock \emph{Multi-charged TeV scale scalars and fermions in the framework of a radiative seesaw model},
\newblock \emph{arXiv}:2007.01766 [hep-ph].

\bibitem{mg5}
J.~Alwall, M.~Herquet, F.~Maltoni, O.~Mattelaer and T.~Stelzer,
\newblock \emph{MadGraph 5: Going Beyond},
\newblock \emph{JHEP} {\bf 06} (2011) 128.

\bibitem{mg52}
J.~Alwall et al.,
\newblock \emph{The automated computation of tree-level and next-to-leading order differential cross sections and their matching to parton shower simulations},
\newblock \emph{JHEP} {\bf 07} (2014) 079.

\bibitem{nnpdf}
NNPDF collaboration, 
\newblock R.D. Ball et al.,
\newblock \emph{Parton distributions for the LHC Run II},
\newblock \emph{JHEP} {\bf 04} (2015) 040.

\bibitem{nnpdf2}
NNPDF collaboration, 
\newblock R.D. Ball et al.,
\newblock \emph{Parton distributions with QED corrections},
\newblock \emph{Nucl. Phys.} {\bf B 877} (2013) 290.

\bibitem{ibe}
M.~Ibe, T.~Moroi and T.T.~Yanagida, 
\newblock \emph{Possible Signals of Wino LSP at the Large Hadron Collider},
\newblock \emph{Phys. Lett.} {\bf B 644} (2007) 355.

\bibitem{mcdonald}
K.L.~McDonald,
\newblock \emph{Probing exotic fermions from a seesaw/radiative model at the LHC},
\newblock \emph{JHEP} {\bf 11} (2013) 131.

\bibitem{Vud}
J.~Hardy and I.~S.~Towner,
\newblock \emph{$\rm V_{ud}$ from Nuclear Beta Decays},
\newblock \emph{PoS} (CKM2016) 028.

\bibitem{fpi}
S.~Aoki, Y.~Aoki, D.~Be$\check{\rm c}$irevi$\acute{\rm c}$ et al. (Flavour Lattice Averaging Group),
\newblock \emph{FLAG Review 2019},
\newblock \emph{Eur. Phys. J. C} {\bf 80} (2020) 113.

\bibitem{Vus}
N.~Cabibbo, E.~C.~Swallow, and R.~Winston,
\newblock \emph{Semileptonic Hyperon Decays and Cabibbo-Kobayashi-Maskawa Unitarity},
\newblock \emph{Phys. Rev. Lett.} {\bf 92} (2004) 251803.

\bibitem{frho}
Qin Chang et al.,
\newblock \emph{Decay constants of pseudoscalar and vectormesons with improved holographic wavefunction},
\newblock \emph{Chinese Phys.} {\bf C 42} (2018) 073102.

\bibitem{aguila}
F.~del Aguila and J.~A.~Aguilar-Saavedra,
\newblock \emph{Distinguishing seesaw models at LHC with multi-lepton signals},
\newblock \emph{Nucl. Phys} {\bf B 813} (2009) 22-90.

\bibitem{aguila2}
F.~del Aguila and J.~A.~Aguilar-Saavedra,
\newblock \emph{Electroweak scale seesaw and heavy Dirac neutrino signals at LHC},
\newblock \emph{Phys. Lett.} {\bf B 672} (2009) 158-165.

\bibitem{li}
T.~Li and X.-G.~He,
\newblock \emph{Neutrino Masses and Heavy Triplet Leptons at the LHC: Testability of Type III Seesaw},
\newblock \emph{Phys. Rev.} {\bf D 80} (2009) 093003.

\bibitem{asha_kirti}
S.~Ashanujjaman and K.~Ghosh,
\newblock \emph{Type-III Seesaw: Phenomenological Implications of the Information Lost in Decoupling from High-Energy to Low-Energy},
\newblock \emph{in preparation}. 

\bibitem{cms137hepdata}
HEPData for Search for physics beyond the standard model in multilepton final states in proton-proton collisions at $\sqrt{s}$=13 TeV,
\newblock \emph{https://www.hepdata.net/record/91969}.

\bibitem{pythia}
T.~Sj\"{o}strand et al.,
\newblock \emph{An introduction to PYTHIA 8.2},
\newblock \emph{Comput. Phys. Commun.} {\bf 191} (2015) 159-177.

\bibitem{delphes}
The DELPHES 3 collaboration,
\newblock J.~de Favereau,C.~Delaere,P.~Demin et al.,
\newblock \emph{DELPHES 3, A modular framework for fast simulation of a generic collider experiment},
\newblock \emph{JHEP} {\bf 02} (2014) 057.

\bibitem{anti-kt}
M.~Cacciari, G.P.~Salam, and G.~Soyez,
\newblock \emph{The anti-kT jet clustering algorithm},
\newblock \emph{JHEP} {\bf 04} (2008) 063.

\bibitem{bayes}
K.~Cranmer,
\newblock \emph{Profile Likelihood Number Counting Combination},
\newblock \emph{http://phystat.org/phystat/packages/0803001.1.html}.

\bibitem{RooFit}
W.~Verkerke and D.P.~Kirkby,
\newblock \emph{The RooFit toolkit for data modeling},
\newblock \emph{eConf C0303241 (2003) MOLT007}.

\bibitem{feynrules}
A.~Alloul, N.D.~Christensen, C.~Degrande et al.,
\newblock \emph{FeynRules 2.0 --- A complete toolbox for tree-level phenomenology}, 
\newblock \emph{Comput. Phys. Commun.} {\bf 185} (2014) 2250-2300.

\bibitem{dis_heavy}
J.C.~Helo, M.~Hirsch and S.~Kovalenko,
\newblock \emph{Heavy neutrino searches at the LHC with displaced vertices},
\newblock \emph{Phys. Rev.} {\bf D 89} (2014) 073005 [Erratum: \emph{Phys. Rev.} {\bf D 93} (2016) 099902].

\bibitem{dis_sterile}
G.~Cottin, J.C.~Helo and M.~Hirsch,
\newblock \emph{Searches for light sterile neutrinos with multitrack displaced vertices},
\newblock \emph{Phys. Rev.} {\bf D 97} (2018) 055025. 

\bibitem{dis_sterile2}
G.~Cottin, J.C.~Helo and M.~Hirsch,
\newblock \emph{Displaced vertices as probes of sterile neutrino mixing at the LHC},
\newblock \emph{Phys. Rev.} {\bf D 98} (2018) 035012.

\bibitem{dis_lrsm}
G.~Cottin, J.C.~Helo, M.~Hirsch and D.~Silva,
\newblock \emph{Revisiting the LHC reach in the displaced region of the minimal left-right symmetric model},
\newblock \emph{Phys. Rev.} {\bf D 99} (2019) 115013.

\bibitem{dis_rhn_u1x}
A.~Das, P.S.~Bhupal Dev and N.~Okada,
\newblock \emph{Long-lived TeV-scale right-handed neutrino production at the LHC in gauged $U(1)_X$ model}, 
\newblock \emph{Phys. Lett.} {\bf B 799} (2019) 135052.

\bibitem{dis_typei}
S.~Jana, N.~Okada and D.~Raut,
\newblock \emph{Displaced vertex signature of type-I seesaw model},
\newblock \emph{Phys. Rev.} {\bf D 98} (2018) 035023.

\bibitem{dis_typeii_lrsm}
P.S.~Bhupal Dev and Y.~Zhang,
\newblock \emph{Displaced vertex signatures of doubly charged scalars in the type-II seesaw and its left-right extensions},
\newblock \emph{JHEP} {\bf 10} (2018) 199.

\bibitem{dis_typeii}
Y.~Du, A.~Dunbrack, M.J.~Ramsey-Musolf and J.-H.~Yu,
\newblock \emph{Type-II seesaw scalar triplet model at a 100 TeV pp collider: discovery and higgs portal coupling determination},
\newblock \emph{JHEP} {\bf 01} (2019) 101.

\bibitem{dis_typeiii}
S.~Jana, N.~Okada and D.~Raut,
\newblock \emph{Displaced Vertex and Disappearing Track Signatures in type-III Seesaw},
\newblock \emph{arXiv:}1911.09037 [hep-ph].

\bibitem{lhc_dis_1}
ATLAS collaboration,
\newblock M.~Aaboud et al.,
\newblock \emph{Search for metastable heavy charged particles with large ionization energy loss in pp collisions at $\sqrt{s}=13$ TeV using the ATLAS experiment},
\newblock \emph{Phys. Rev.} {\bf D 93} (2016) 112015.

\bibitem{lhc_dis_2}
ATLAS collaboration,
\newblock M.~Aaboud et al.,
\newblock \emph{Search for heavy long-lived charged R-hadrons with the ATLAS detector in 3.2 fb$^{-1}$ of proton–proton collision data at $\sqrt{s}=13$ TeV},
\newblock \emph{Phys. Lett.} {\bf B 760} (2016) 647.

\bibitem{lhc_dis_3}
ATLAS collaboration,
\newblock \emph{Search for long-lived neutral particles decaying into displaced lepton jets in proton–proton collisions at $\sqrt{s}=13$ TeV with the ATLAS detector},
\newblock \emph{ATLAS-CONF}-2016-042.

\bibitem{lhc_dis_4}
ATLAS collaboration,
\newblock \emph{Search for long-lived neutral particles decaying in the hadronic calorimeter of ATLAS at $\sqrt{s}=13$ TeV in 3.2 fb$^{-1}$ of data},
\newblock \emph{ATLAS-CONF}-2016-103.

\bibitem{lhc_dis_5}
ATLAS collaboration,
\newblock M.~Aaboud et al.,
\newblock \emph{Search for long-lived, massive particles in events with displaced vertices and missing transverse momentum in $\sqrt{s}=13$ TeV pp collisions with the ATLAS detector},
\newblock \emph{Phys. Rev.} {\bf D 97} (2018) 052012.

\bibitem{lhc_dis_6}
ATLAS collaboration,
\newblock M.~Aaboud et al.,
\newblock \emph{Search for long-lived charginos based on a disappearing-track signature in pp collisions at $\sqrt{s}=13$ TeV with the ATLAS detector},
\newblock \emph{JHEP} {\bf 06} (2018) 022.

\bibitem{mathusla}
\newblock J.P.~Chou, D.~Curtin and H.J.~Lubatti,
\newblock \emph{New Detectors to Explore the Lifetime Frontier},
\newblock \emph{Phys. Lett.} {\bf B 767} (2017) 29.

\bibitem{mathusla2}
D.~Curtin et al.,
\newblock \emph{Long-Lived Particles at the Energy Frontier: The MATHUSLA Physics Case},
\newblock \emph{Rept. Prog. Phys.} {\bf 82} (2019) 116201.

\bibitem{mathusla3}
D.~Curtin and M.E.~Peskin,
\newblock \emph{Analysis of Long Lived Particle Decays with the MATHUSLA Detector},
\newblock \emph{Phys. Rev.} {\bf D 97} (2018) 015006.

\bibitem{lhec_dis}
\newblock LHeC Study Group,
\newblock J.L.~Abelleira Fernandez et al.,
\newblock \emph{A Large Hadron Electron Collider at CERN: Report on the Physics and Design Concepts for Machine and Detector},
\newblock \emph{J. Phys.} {\bf G 39} (2012) 075001.

\bibitem{fcc-he_dis}
\newblock M.~Kuze,
\newblock \emph{Energy-Frontier Lepton-Hadron Collisions at CERN: the LHeC and the FCC-eh},
\newblock \emph{Int. J. Mod. Phys. Conf. Ser.} {\bf 46} (2018) 1860081.

\bibitem{curtin_np}
\newblock D.~Curtin, K.~Deshpande, O.~Fischer et al.,
\newblock \emph{New physics opportunities for long-lived particles at electron-proton colliders},
\newblock \emph{JHEP} {\bf 07} (2018) 024.

\bibitem{srubabati_2019}
S.~Goswami, Vishnudath.~K.N. and N.~Khan,
\newblock \emph{Constraining minimal Type-III Seesaw Model from naturalness, Lepton Flavor Violation nad Electroweak Vacuum stability},
\newblock \emph{Phys. Rev.} {\bf D 99} (2019) 075012. 

\bibitem{Grimus_Lavoura}
W.~Grimus and L.~Lavoura,
\newblock \emph{The seesaw mechanism at arbitrary order: disentangling the small scale from the large scale},
\newblock \emph{JHEP} {\bf 11} (2000) 042.

\end{thebibliography}
\end{document}